\documentclass[
  nofootinbib,
 twocolumn, 
  showpacs,superscriptaddress,preprintnumbers]{revtex4-2}

\usepackage{color} 
\definecolor{mydarkred}{RGB}{233,20,35}
\definecolor{mypurple}{RGB}{120, 35, 160}
\definecolor{mydarkpurple}{RGB}{128, 100, 162}
\definecolor{mybrown}{RGB}{255, 195, 0}
\definecolor{myaqua}{RGB}{29, 153, 168}
\definecolor{myblue}{RGB}{91, 129, 184}  
\definecolor{mygreen}{RGB}{0, 125, 0}  

\definecolor{brickred}{rgb}{0.8, 0.25, 0.33}
\usepackage[utf8]{inputenc}
\usepackage[english]{babel}
\usepackage{natbib}
\usepackage{hyperref}
\hypersetup{
    colorlinks=true,
    linkcolor=black,
    filecolor=black,
    urlcolor=black,
    citecolor=black
}

\usepackage{amsmath,amssymb}

\usepackage{graphicx} 
\graphicspath{{fig/}} 

\usepackage{sidecap}

\newcommand{\p}{\partial}
\newcommand{\non}{\nonumber}
\newcommand{\al}{\alpha}

\begin{document}

\title{Dynamic sampling bias and overdispersion induced by skewed offspring distributions}

\author{Takashi Okada}
\affiliation{Departments of Physics and Integrative Biology, University of California, Berkeley, California 94720}
\affiliation{RIKEN iTHEMS, Wako, Saitama 351-0198, Japan}

\author{Oskar Hallatschek}
\affiliation{Departments of Physics and Integrative Biology, University of California, Berkeley, California 94720}

\begin{abstract}
Natural populations often show enhanced genetic drift consistent with a strong skew in their offspring number distribution. The skew arises because the variability of family sizes is either inherently strong or amplified by population expansions, leading to so-called `jackpot' events. The resulting allele frequency fluctuations are large and, therefore, challenge standard models of population genetics, which assume sufficiently narrow offspring distributions. While the neutral dynamics backward in time can be readily analyzed using coalescent approaches, we still know little about the effect of broad offspring distributions on the dynamics forward in time, especially with selection. Here, we employ an exact asymptotic analysis combined with a scaling hypothesis to demonstrate that over-dispersed frequency trajectories emerge from the competition of conventional forces, such as selection or mutations, with an emerging time-dependent sampling bias against the minor allele. The sampling bias arises from the characteristic time-dependence of the largest sampled family size within each allelic type. Using this insight, we establish simple scaling relations for allele frequency fluctuations, fixation probabilities, extinction times, and the site frequency spectra that arise when offspring numbers are distributed according to a power law $~n^{-(1+\alpha)}$. To demonstrate that this coarse-grained model captures a wide variety of non-equilibrium dynamics, we validate our results in traveling waves, where the phenomenon of 'gene surfing' can produce any exponent $1<\alpha <2$.  We argue that the concept of a dynamic sampling bias is useful generally to develop both intuition and statistical tests for the unusual dynamics of populations with skewed offspring distributions, which can confound commonly used tests for selection or demographic history.
\end{abstract}

\maketitle
{
\hypersetup{linkcolor=black}
}





\vspace{33pt}
\section{Introduction}
Interpreting the genetic differences between and within populations we observe today requires a robust understanding of how allele frequencies change over time. Most theoretical and statistical advancements have been based on the Wright-Fisher model~\citep{fisher1930genetical,wright1931evolution}, which has  shaped the intuition of generations of population geneticists for how evolutionary dynamics works \citep{crow1970introduction}. The Wright-Fisher model assumes that the genetic makeup of a generation results from resampling the gene pool of the previous generation, whereby biases are introduced to account for most relevant evolutionary forces, such as selection, migration, or variable population sizes. For large populations, the resulting dynamics can be approximated by a biased diffusion process, which simplifies the statistical modeling of the genetic diversity. More importantly, Wright-Fisher diffusion is the limiting allele frequency process of a wide variety of microscopic models, as long as they satisfy seemingly mild assumptions (see below). This flexibility has made Wright-Fisher diffusion the standard model of choice to infer the demographic history of a species, loci of selection or the strength of polygenic selection~\citep{Bollback2008-wv,Feder2014-vx,berg2014population,foll2015wfabc,Schraiber2016-jt,tataru2017statistical}.  

Despite its versatility, Wright-Fisher diffusion can be a poor approximation when the population dynamics is driven by rare but strong number fluctuations. It is increasingly recognized that number fluctuations can be inflated for very different reasons. First, the considered species may have a broad offspring distribution, which occurs for marine species and plants with a Type III survivorship curve~\citep{hedgecock1994does,Eldon2006-ko} as well as  viruses and fungi (reviewed in \citep{Tellier2014-zs}). Broad offspring distributions also arise in infectious disease, when relatively few super-spreaders are responsible for the majority of the disease transmissions~\citep{lloyd2005superspreading}. In the recent SARS-CoV-2 pandemic, for example, a strongly skewed offspring distributions were consistently inferred from both contact tracing data and infection cluster size distributions~\citep{laxminarayan2020epidemiology,adam2020clustering}. Understanding allele frequency trajectories in these systems is extremely challenging, as statistical inference based on the Wright-Fisher model is often misleading (see e.g.~ \cite{sackman2019inferring}).

A second mechanism for strong number fluctuations are so-called jackpot events, which can occur in any species no matter the actual offspring distribution. Jackpot events are population bottlenecks that arise when the earliest, the most fit or the most advanced individuals have an unusual large number of descendants. Temporal jackpot events ("earliest") were first discovered by Luria and Delbr\"{u}ck~\citep{Luria1943-ke} and studied as a signal of spontaneous mutations in an expanding population. They observed that a phage resistant mutant clone can grow exceptionally large if the resistance mutation by chance occurs early in an expansion event. Despite being rare, these jackpot events are easily detectable in large populations because they strongly inflate the variance of the expected number of mutants and lead to power-law descendant distributions. 

The very same descendant distribution arises in models of rampant adaptation and of background selection. In these models, mutations generate jackpot events when they arise within the few fittest individuals~\citep{Neher2013-sl}. Jackpot events also arise in range expansions, where the most advanced individuals in the front of the population have a good chance to leave many descendants over the next few generations. This phenomenon of gene surfing can produce a wide range of scale-free descendant distributions~\citep{hallatschek2008gene,fusco2016excess,Birzu2018-wq,birzu2020genealogical}.

 \begin{figure*}[t]
\centering
\includegraphics[width=18cm]{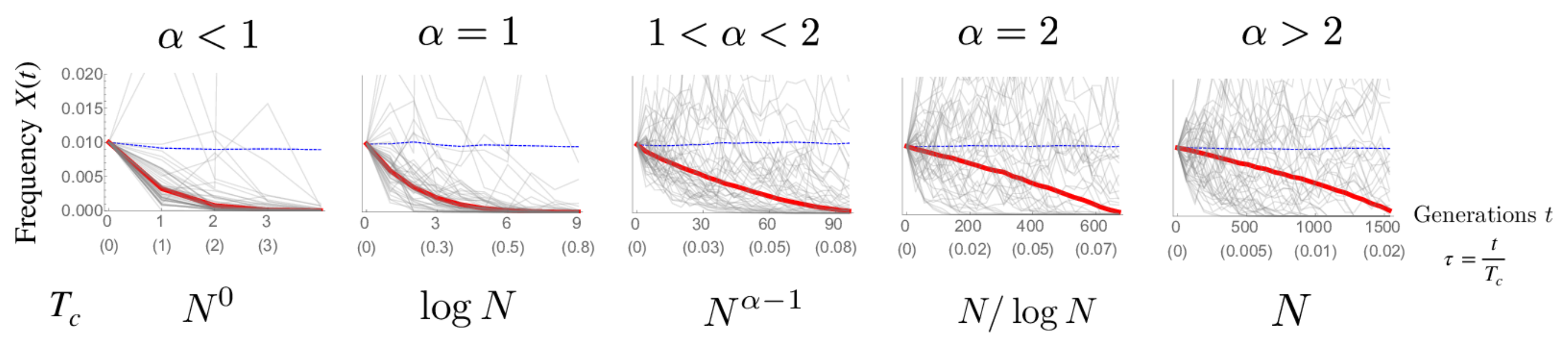}
\caption{
The mean (blue dashed curve) and median (red solid curve) of allele frequency trajectories for $\alpha=0.8, 1, 1.5, 2,$ and $2.5$. 
For each $\alpha$, $10^4$  trajectories are generated  with the  initial  frequency $x_0=0.01$.  For ease of viewing,  only 50 trajectories  are shown in gray in each panel. The  time $t$  in units of generations and  the one  $\tau=\frac{t}{T_c}$ re-scaled by the coalescent timescale $T_c$  are shown in the horizontal axes.  The dependence of  the coalescent time  on the population size $N$ is written below each panel. The population size is $N=10^5$.
}\label{fig:traj}
\end{figure*}

To account for skewed  offspring  distributions,  a number of theoretical studies have been conducted in the context of the coalescent framework. Based on this backward-in-time, a striking feature of broad offspring  distributions is the simultaneous merging of multiple lineages.  One of the most widely studied models is the beta-coalescent \citep{Schweinsberg2003-fj}, which is a subclass of the $\Lambda$-coalescent and corresponds to the population dynamics with a power-law offspring number distribution $\propto u^{-1+\alpha}$. The case $\alpha=1$, called Bolthausen-Sznitman coalescent~\citep{Bolthausen1998-as}, has been shown to be the limiting coalescent in models of so-called "pulled" traveling waves, which describe the most basic scenarios of range expansions~\citep{brunet2007effect} and of rampant adaptation~\citep{Neher2013-sl,desai2013genetic,kosheleva2013dynamics,schweinsberg2017rigorous}. Moreover, so-called "semi-pushed" traveling waves that contain some level of co-operativity, induced e.g. by an Allee effect, generate power-law offspring distributions with $1<\alpha<2$~\citep{Birzu2018-wq}, indicating that their coalescent is intermediate between the Bolthausen-Sznitman and Kingman coalescents. 

 
The tractability of coalescent approaches make it particularly useful for  inferring demographic histories and detecting outlier behaviors~\citep{basdevant2008asymptotics, eldon2009structured, eldon2011estimation}. However, as it is notoriously difficult to integrate selection in coalescent frameworks, there is also a strong need for forward-in-time approaches that capture the competition between genetic drift and selection. While for $\al\geq 2$, the limiting allele frequency dynamics is given by the well-understood Wright-Fisher process, much less is known for the case $\al < 2$. This is unfortunate because, as mentioned above, any exponent $1\leq \alpha \leq  2$ can arise dynamically. 


Recently, the forward-dynamics of the special case $\al=1$ was studied by one of the authors~\citep{hallatschek2018selection}, finding that an emergent sampling bias generates strong deviations from Wright-Fisher dynamics. The sampling bias arises because, in each generation, an allele with high frequency can sample more often and, hence, deeper into the tail of the offspring distribution than an allele with small frequency. The major allele of a biallelic site, therefore, has with high probability a greater number of offspring per individual than the minority type. This sampling bias acts like a selective advantage of the major allele, but its average effect is compensated by rare frequency hikes of the minor allele so that the expected change in frequency only changes in the presence of genuine selection. 

Here, we focus on the understudied case $1<\al<2$ intermediate between the known cases of $\alpha\geq 2$, corresponding to Wright-Fisher diffusion, and $\alpha=1$ described by jumps and sampling bias but vanishing diffusion. Similarly to the $\alpha=1$ borderline case, we find that a minor-allele-suppressing sampling bias arises but that it is fading over time as the offspring distributions are sampled more and more thoroughly. This time-dependent sampling bias determines the scaling of the fixation probability, extinction time, stationary distribution, and site frequency spectrum. The combination of jumps and bias generates a so-called Levy-flight which controls the variability of allele frequency trajectories, for instance between unlinked genes or between populations. The flexibility of our model should enable to fit wide range cases that deviate from Wright-Fisher diffusion.


\section{Sampling allele frequencies across generations}

To study the impact of broad offspring numbers, we consider an idealized, panmictic, haploid population of constant size $N$ that produces non-overlapping generations in the following way. First, we associate with each individual $i$ a "reproductive value" \citep{fisher1930genetical, barton2011relation} $U_i$, which represents its \textit{expected} contribution to the population of the next generation. The random numbers $U_i$ are drawn from a specified distribution $P_U$. In a second step, we sample each individual in proportion to its reproductive value until we have obtained $N$ new individuals representing the next generation. 

Our model belongs to the general class of Cannings models~\citep{cannings1974latent}. The Wright-Fisher model is obtained if we choose $P_U$ to be a Dirac delta function, such that all individuals have the same reproductive value.

We focus most of our analysis on the dynamics of two mutually exclusive alleles, a wild type and a mutant allele. The dynamics of the two alleles is captured by the time-dependent frequency $X(t)$ ($0\leq X \leq 1$) of mutants. The wild type frequency is given by $1-X(t)$. 
The total reproductive values $M$ and $W$ of the mutant population and the the wild type population, respectively, are given by
\begin{align}
M \equiv \sum_{i=1}^{N X}U^{(\text{M})}_i,\quad W  \equiv \sum_{i=1}^{N (1-X)}U^{(\text{W})}_i\;.\label{MW}
\end{align}
Here, $U^{(\text{M})}_i$ and $U^{(\text{W})}_i$ are the individual reproductive values of mutants and wild types and sampled from the distribution $P_U$. The population at the next generation is generated by binomially sampling $N$ individuals with success probability 
$\frac{M}{M+W}$. Mutations and selection are included as in the Wright-Fisher model. If the fitness of the mutant relative to the wild-type is $1+s$, where $s$ is the selection coefficient, and the forward- and back- mutation rates are $\mu_{1}$ and $\mu_2$ respectively, then the success probability is given by $
\frac{ (1-\mu_2) (1+s)M+ (1-\mu_1)W}{(1+s)M+W}$.

For the offspring  distribution $P_U$, we consider a family of fat-tailed distributions, which asymptotically behave as $P_U \sim \frac{1}{u^{\alpha+1}}$ with $\alpha$ being a positive constant. To make our presentation concrete, we choose  $P_U(u)=\alpha /u^{\alpha+1}\, (u\geq 1)$, which is known as the Pareto distribution.  In the large population size limit, the neutral allele-frequency dynamics is known to only depend on the asymptotic power law exponent $\alpha$ provided we measure time in units of the coalescence time~\citep{schweinsberg2003coalescent}.

\section{Simulation results}

Our goal is to understand the asymptotic dynamics of our model for large $N$, where the frequency becomes continuous over time \citep{kimura1954stochastic, Gardiner2009-nq}  provided that $\alpha \geq 1$ \citep{schweinsberg2003coalescent}.
 We first present  simulation results regarding  relevant measures in the population genetics. 
 Later, we provide a heuristic argument to explain them. Many separate observations (the fixation probability, extinction time, allele frequency fluctuations, stationary distribution, and site frequency spectrum) can be matched up with a unifying scaling picture.
 
Below, $t$ and $\tau=t/T_c$  denote a time in units of generations and one normalized by the characteristic (coalescent) timescale $T_c$, respectively. $T_c$ depends on the population size and the exponent $\alpha$ as follows: $T_c =N$ when $\al>2$, $T_c =N/\log N$ when $\al=2$,  $T_c = N^{\al-1}$ when $1<\al <2$, and  by $T_c = \log N $ when $\al=1$.  These timescales were originally derived in the coalescent framework \citep{schweinsberg2003coalescent}. Later, we explain how they can be rationalized within the forward-in-time approach. 
 
 To understand the frequency dynamics when $1\leq \alpha<2$, it is essential  to distinguish between average and typical trajectories. As a proxy for typical trajectories,  we use the median of the frequencies, denoted by $X^{\text{med}}(\tau)$, throughout this paper. 
 
\subsection{Neutral dynamics: typical trajectories and extinction time}
First, we characterize the allele frequency dynamics in the absence of selection $s=0$. 
In this neutral limit, the expected value of the allele frequency does not change over time, i.e., $ \langle  X(t)\rangle =X(0)$. Yet, despite the overall neutrality, a typical trajectory experiences a bias against the minority allele. This can be seen in Figure~\ref{fig:traj}, where the mean and median are plotted across many realizations that start from the same frequency $X(0)=0.01$. While the mean does not change over time, as required from  neutrality,  the median decays to zero in an $\alpha$--dependent manner. By symmetry, the median increases towards fixation if the starting frequency is larger than 50$\%$. Thus, the median experiences a bias against the minor allele. Note also that, when $1<\al<2$, the velocity of the median approaching extinction  decreases as it approaches the extinction boundary (see the red curve in Figure~\ref{fig:traj}). As we will show later, an uptick of the site frequency spectrum at the  boundaries originates from this slowing.

\begin{figure}[th]
\centering
  \includegraphics[width=9cm]{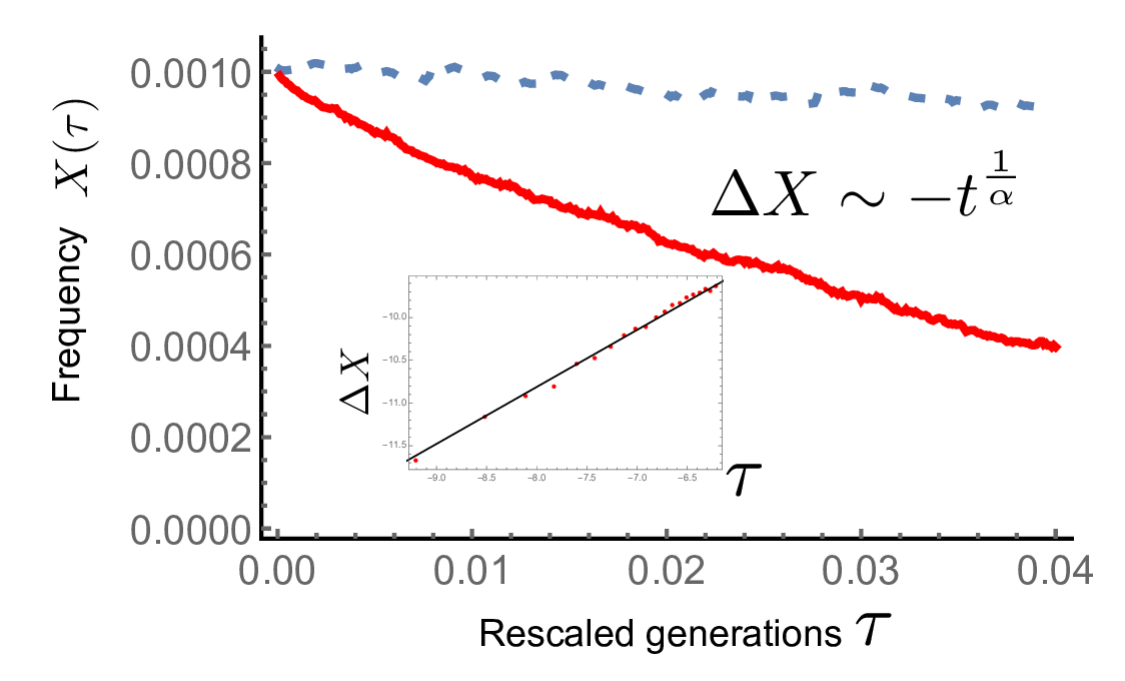}
  \caption{Trajectories of the median (red thick line) and the mean (blue dashed) of allele frequency when $\alpha=1.5$ and $N=10^8$. Inset: The trajectory (red points) of $|\Delta X^{\text{med}} |= X(0) -X^{\text{med}}(\tau)$ is shown in log-log plot.  The median agrees well with the expectation from the scaling argument $\Delta X^{\text{med}} \sim  - \tau^{\frac{1}{\alpha}}$  (black solid line). 
  }
 \label{fig:dx_scaling}
\end{figure}

Numerical simulations of the early part of trajectories show that time-dependent median displacement follows a simple power law, 
\begin{align}
    \Delta X^{\text{med}} \equiv X^{\text{med}}(\tau)-X(0) \sim -\tau^{\frac{1}{\alpha}}, \label{dx0}
\end{align}
up to a frequency-independent prefactor. 
 Figure~\ref{fig:dx_scaling} shows the numerical result  for  $\al=1.5$. The red curve represents the median of trajectories, which agrees well with 
$ \Delta X^{\text{med}}\sim -\tau^{\frac{2}{3}}$.

\begin{figure}[th]\centering
  \includegraphics[width=9cm]{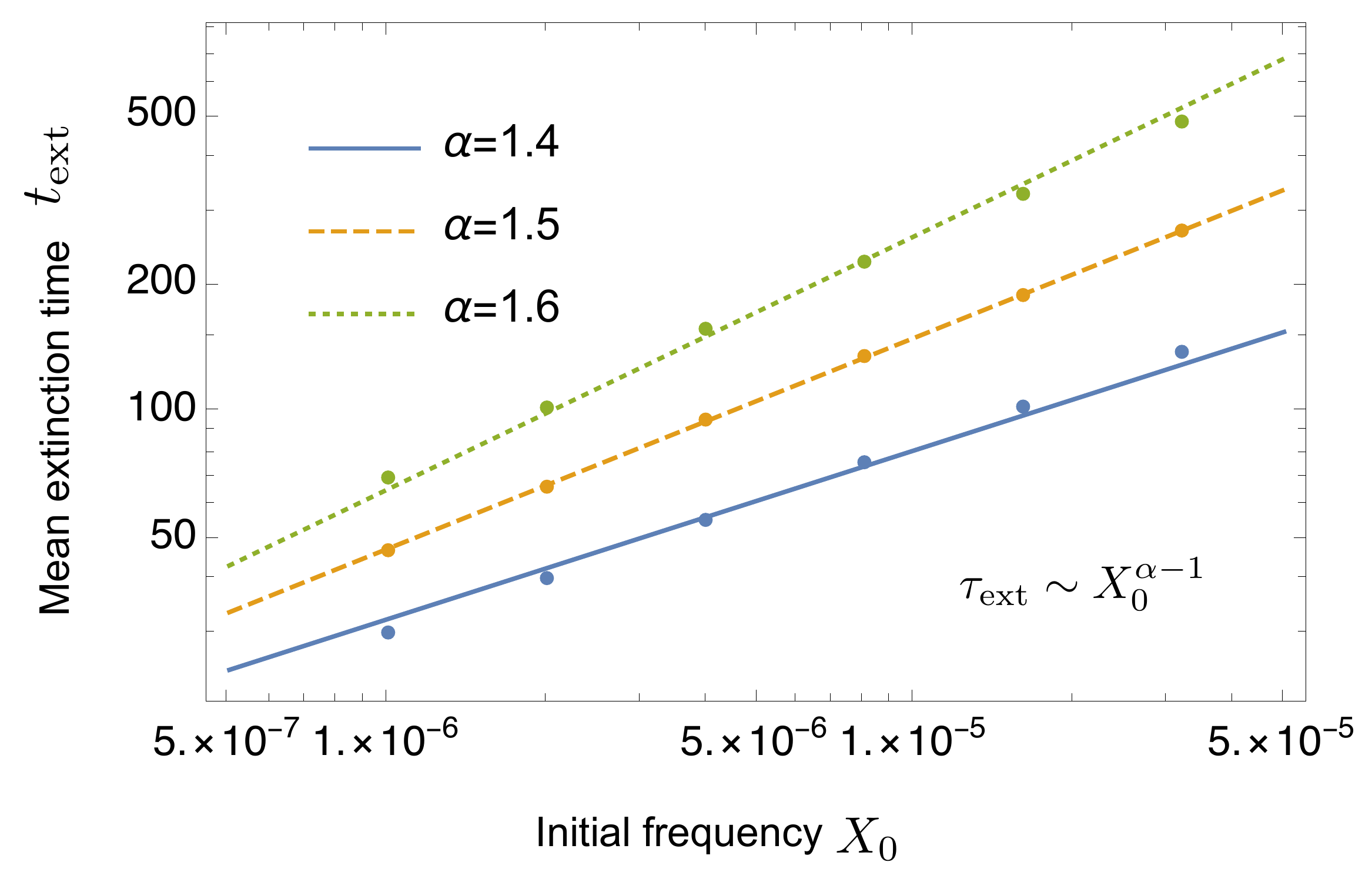}
  \caption{The mean extinction time $ t_{\text{ext}}$ (in units of generations) as a function of initial allele frequency $X(0)$ is plotted for $\alpha=1.4,\, 1.5,\, 1.6$. Each of the straight lines has the slope $\alpha-1$.  $t_{\text{ext}}$ can be  fitted well by Equation~\ref{tauext0}. The population size is $N=10^8$.
  }
 \label{fig:Text}
\end{figure}

Next we quantify the time to extinction, which turns out to be driven by the above minor-allele suppressing bias. Numerical results of the mean extinction time are  consistent with  
\begin{align}
\tau_{\text{ext}} \sim X(0)^{\al -1}\label{tauext0},
\end{align}
as shown in Figure~\ref{fig:Text}. Hence, in units of the coalescence time, the mean extinction time $\tau_{\text{ext}}$ becomes larger as $\al$ decreases (namely, for a broader offspring distribution). Note, however, that if one measures time  in  units of generations, Equation~\ref{tauext0}  can be rewritten as  $ t_{\text{ext}} = \tau_{\text{ext}} T_c \sim  (N X(0))^{\al-1}$, which becomes smaller as $\al$ decreases since $N X(0)\geq 1$.

\subsection{Allele frequency fluctuations as a signature of broad offspring distributions
}
\begin{figure*}[ht]%
\includegraphics[width=.6\textwidth]{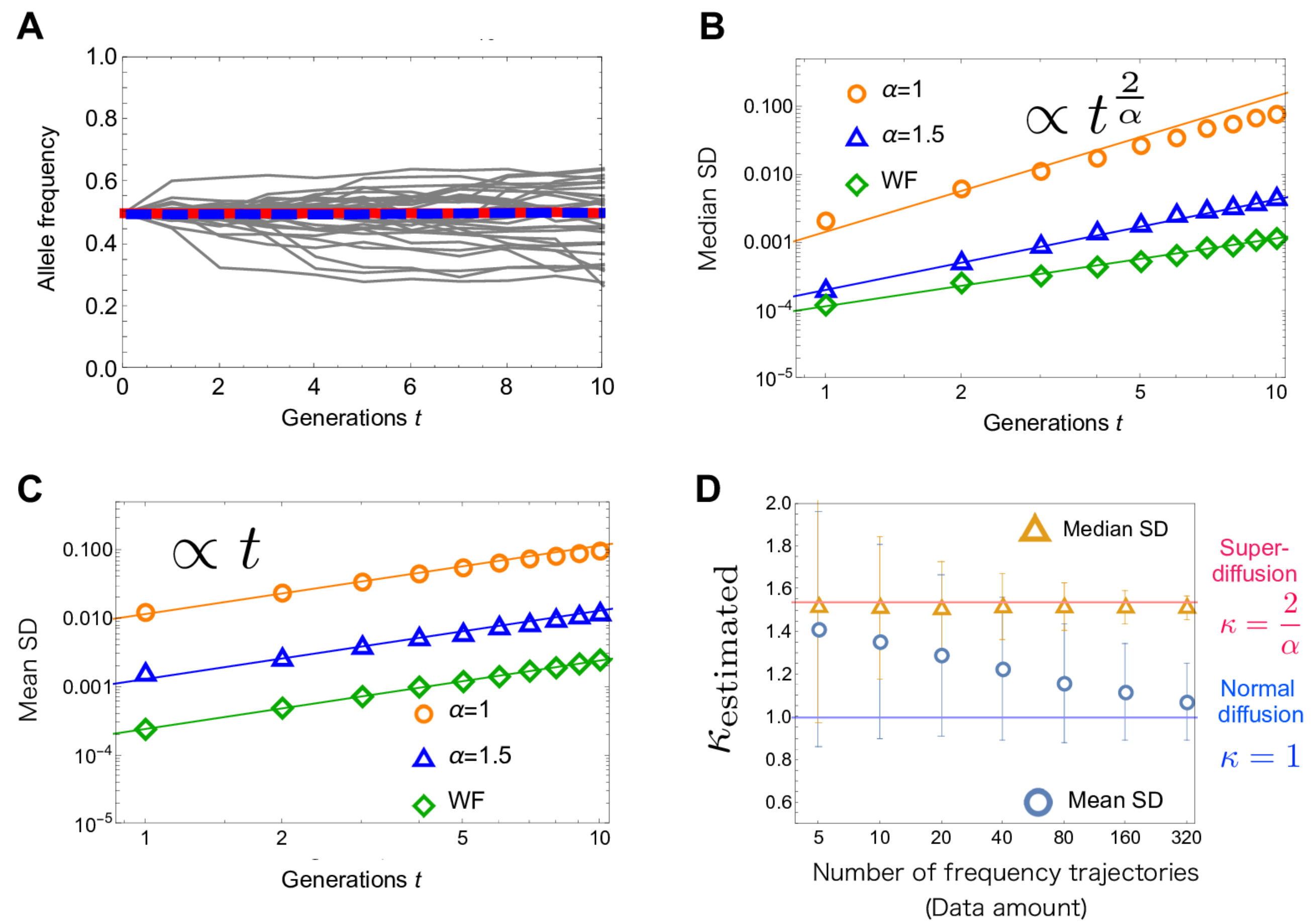}%
\hspace{0.5cm}  
\begin{minipage}[top]{.33\textwidth}%
\vspace{-7cm}
\caption{ ({\bf A}) Fluctuations of neutral allele frequencies when $\alpha=1.5$ and $N=10^5$. For $X(0)=0.5$, the median $X^{\text{med}}(t)$ (red) is constant as well as the mean $\langle X(t)\rangle $ (blue). ({\bf B}) The median square displacement computed from a data set of 1000 trajectories. For $\alpha=1,1.5$ and the Wright-Fisher model, $N= 10^8, 10^4,$ and $10^3$ are used  respectively. The straight lines represent the scaling in Equation~\ref{MedSD_main}. For $\alpha=1$, the fitting after $t\gtrsim 5$ is not perfect, since $t/T_c=t/\ln N\ll 1$ is not satisfied. ({\bf C}) The mean square displacement (mean SD) for different values of $\alpha$. Solid lines represent linear scaling, which is expected for a regular diffusion process. ({\bf D}) Data-size dependence of the estimated diffusion exponent $\kappa_{\text {estimated}}$ for the mean SD (blue circle) and that for the median SD (orange triangle). See the main text for the detailed explanation.  The horizontal lines show $\kappa=\frac{2}{\alpha}$  and $\kappa=1$.  The bars  represent  the standard deviations of $\kappa_{\text {estimated}}$.  $\alpha=1.3$ and $N=10^8$ are used. }
\label{fig:medsd}
\end{minipage}%
\end{figure*}
 
Next, we explore to what extent the spectrum of allele frequency fluctuations can provide a clue for identifying the exponent $\alpha$ of the offspring  distribution. A deviation from the Wright-Fisher diffusion is  most clearly revealed by measuring the median square displacement (median SD), 
\begin{align}
{\rm Median\ SD}\equiv {\mathbb M}\left[\Bigl(X(t)-{\mathbb M}[X(t)]\Bigr)^2\right],
\end{align}
where ${\mathbb M}[\, \cdot\,]$  denotes taking the median (e.g.~${\mathbb M}[X(t)]=X^{\text{med}}(t)$). 
To measure  the median SD,  we simulate 1000   neutral allele frequency trajectories with initial condition  $X(0)=0.5$,  for  $\alpha=1,\ 1.5$ and the Wright-Fisher model (Figure~\ref{fig:medsd}A). As shown in Figure~\ref{fig:medsd}B, the median SD  computed from this data set  
 is consistent with the scaling, 
 \begin{align}
\text {Median SD} \sim t^{\frac{2}{\alpha}},  \label{MedSD_main}
\end{align}
 when $t/T_c\ll 1$. Noting $1\leq \alpha<2$, this scaling  means that typical fluctuations characterized by the median SD  exhibit  super-diffusion.

Usually, allele frequency fluctuations are quantified by using the mean SD $ \equiv \langle (X(t)-X(0))^2\rangle$, rather than the median SD. For the Wright-Fisher diffusion, the distinction between these two measures is irrelevant since both of them increase linearly with time, except with differing prefactors. However, for $1 \leq \alpha < 2$, the $\alpha$-dependence in Equation~\ref{MedSD_main} can be detected by measuring the median SD. As shown in Figure~\ref{fig:medsd}C, the mean SD (computed from a large data set) grows linearly in time even when $\alpha$ is less than 2, as if the underlying process was diffusive. 

That the dynamics is not diffusive also impacts the mean SD, but somewhat subtly in that its value depends on the size of the data set  
(i.e., the number of frequency trajectories) used to measure it.
 This is because while  rare large jumps contribute the mean SD in a large data set, these jumps are not observed in a small data set (with high probability). To demonstrate  this data-size dependence,  we  prepare an ensemble of data sets, where each data set consists of a given number of allele-frequency trajectories. Then, for each data set, we measure the diffusion exponent $\kappa$, which is  defined by 
 \begin{align}
     {\rm Mean\ SD}\propto t^\kappa. 
 \end{align} In  Figure~\ref{fig:medsd}D,  the ensemble-averaged  exponent is shown by the blue circle.  We can see that,  as the data size increases,
 fluctuations characterized by the mean SD exhibit a crossover from  super-diffusion ($\kappa=\frac{2}{\alpha}$) to normal diffusion ($\kappa=1$).
 For the median SD, by contrast, we find that its diffusion exponent $\kappa$  can be computed reliably without any significant dependence on the size of the data set (orange triangles in Figure~\ref{fig:medsd}D). For example, under the parameter setting  in Figure~\ref{fig:medsd}D, given a date set of 320 trajectories, the diffusion exponent $\kappa_{\text{estimated}}$ of the median SD falls within  the interval $ [1.45, 1.57]$ with probability $\sim 68\%$. This  in turn predicts   $\alpha_{\text {estimated}}:=\frac{\kappa_{\text{estimated}}}{2} \approx 1.32\pm 0.05$, which is close to the actual value $\alpha=1.3$.

\subsection{Fixation probability}

\begin{figure}[th]\centering
   \includegraphics[width=8.5cm]{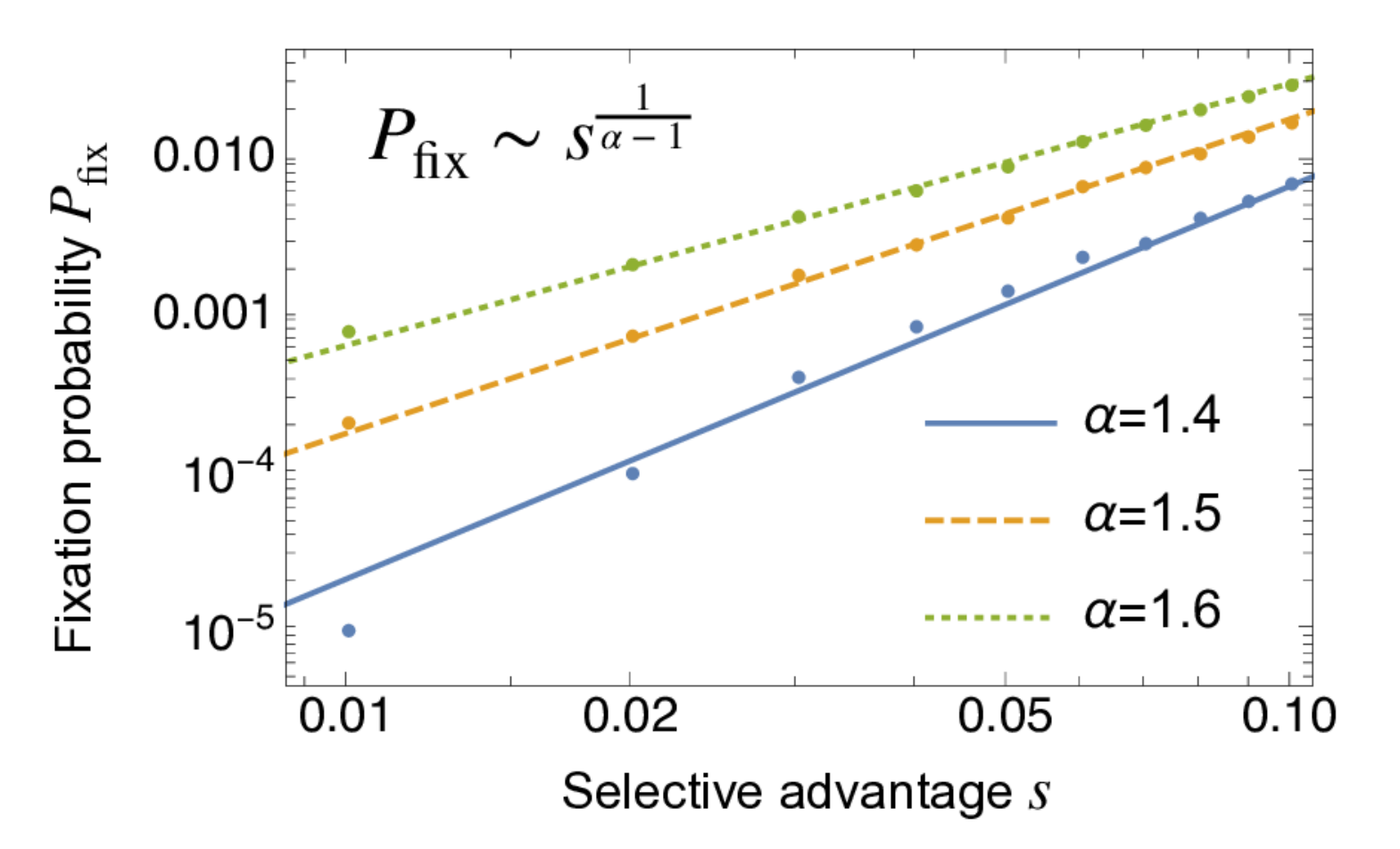}
  \caption{The fixation probability $P_{\text{fix}}$  as a function of  selective advantage $s$. The lines are the expectations from the scaling argument in Equation~\ref{Pfix}. The population size is $N=10^8$.} 
 \label{fig:Pfix}
\end{figure}

Next, we examine the effect of natural selection on the fixation probability of beneficial mutations. 
We consider a mutant  with positive selective advantage $s>0$ arising in a monoclonal population.  The fixation probability $  P_{\text{fix}}(s)$ of a single mutant depends on the  parameter $\al$ of the offspring  distribution.  In the Wright-Fisher model (or equivalently, $\al\geq 2$),  the fixation probability can be obtained using a diffusion approximation and is given by  $P_{\text{fix}}(s) = \frac{1-e^{-2s}}{1-e^{-2Ns}}$, which becomes $P_{\text{fix}}\approx  2 s$ when $N s\gg 1$ and $s$ is small. 
When $\al=1$, an analytic result has been recently obtained in \citep{hallatschek2018selection}, which can be approximated as $P_{\text{fix}}(s) \sim \frac{1}{N^{1-s}}$. For the intermediate case, $1<\al<2$, we find that the fixation probability is given by
\begin{align}
    P_{\text{fix}}(s) \sim s^{\frac{1}{\al-1}}. \label{Pfix0}
\end{align}
See Figure~\ref{fig:Pfix} for the numerical results.  Note that since $P_{\text{fix}}(s)\rightarrow \frac{1}{N}$ in the neutral limit independently of $\al$, these results hold for  sufficiently strong selection, $1\ll N s^{\frac{1}{\al-1}}$.

As Equation~\ref{Pfix0} shows, for a fixed  population size and selective advantage, the fixation probability becomes smaller as $\alpha$ decreases. Intuitively, this is because, for smaller $\alpha$, the success of fixation 
in catching a ride on a jackpot event depends more  strongly on luck than on  fitness differences.

\subsection{Site frequency spectrum}
The site frequency spectrum (SFS) is often used as a convenient summary of the genetic diversity within a population. Theoretically, the SFS is defined  in the infinite alleles model \citep{Kimura1969-un} as the density $f_{\text{SFS}}(x)$ of neutral derived alleles in the population (namely, $f_{\text{SFS}}(x)dx$ is the number of derived alleles in the frequency window $[x-\frac{dx}{2},x+\frac{dx}{2}]$). 

\begin{figure}[t]
\centering
  \includegraphics[width=7.5cm]{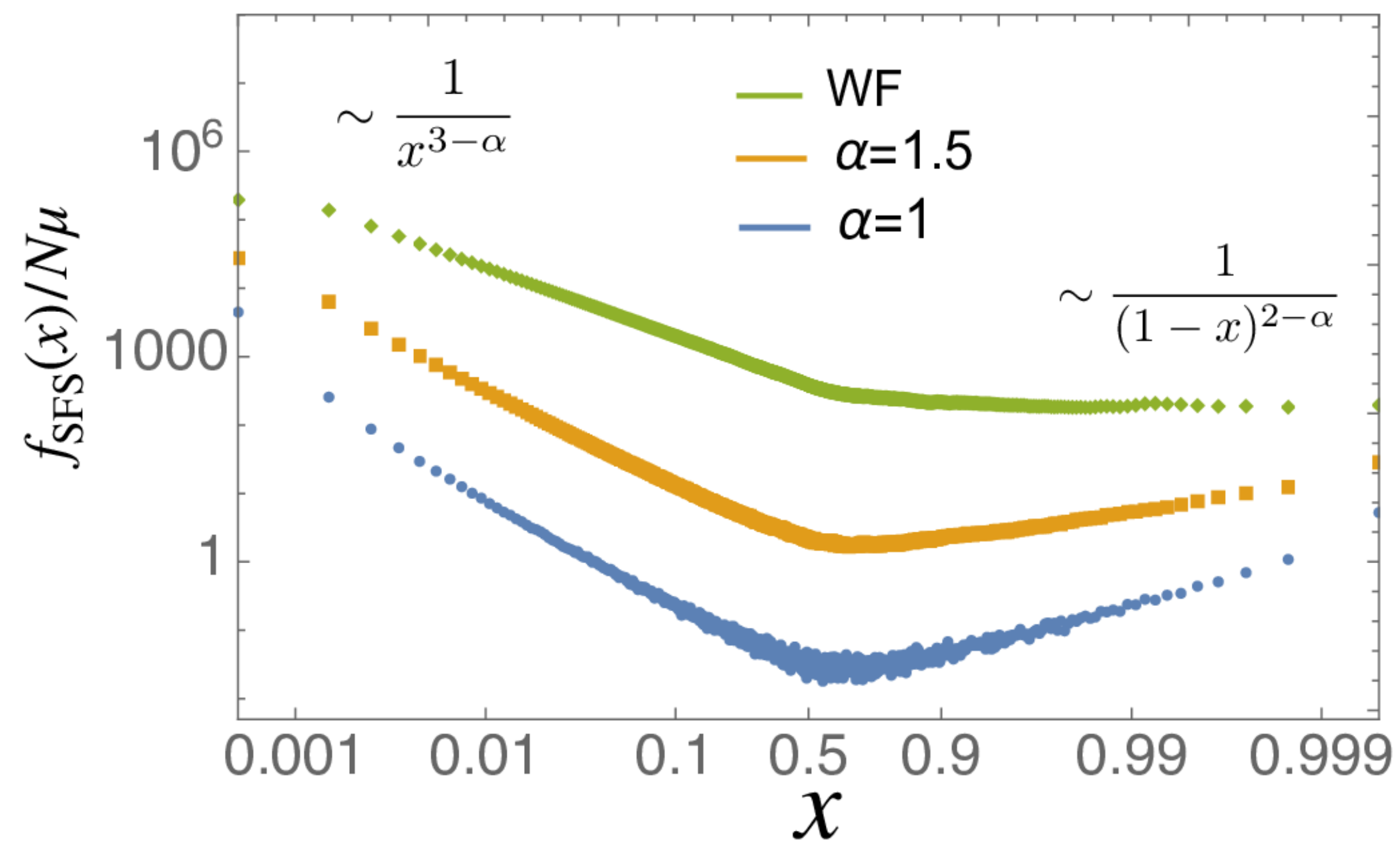}
  \caption{The site frequency spectrum for different values of $\alpha$ and fixed population size $N=10^5$.
  When $1<\alpha<2$, the rare-end spectrum and the frequent-end spectrum are $\propto \frac{1}{x^{3-\alpha}}$ and $\propto \frac{1}{(1-x)^{2-\alpha}}$, respectively (see also Figure~\ref{fig:sfs_scaling}). 
  }
 \label{fig:sfs}
\end{figure}

Figure~\ref{fig:sfs}  shows numerical plots of the SFS for  $\alpha=1,1.5$, and the Wright-Fisher model.
In the standard Wright-Fisher model, the SFS is proportional to $1/x$, which decreases monotonically as $x$ increases.
By contrast, when  offspring numbers are broadly distributed (when $\alpha < 2$), the SFS is non-monotonic with a somewhat surprising uptick towards the fixation boundary. When $\alpha=1$,  the analytic understandings of  asymptotic behaviors near both boundaries are well-established: $f_{\text{SFS}}(x)$ is proportional to $\frac{1}{(x \log x)^2}$ near $x\sim 0 $ and $- \frac{1}{(1-x)\log (1-x)}$ near $x \sim 1$, respectively \citep{Neher2013-sl, kosheleva2013dynamics} (see also Appendix E).

For the intermediate case $1<\alpha<2$, the rare-end behavior of the SFS has been analytically studied. From a backward approach (the $\Lambda$-coalescent), the authors in \citep{berestycki2014asymptotic} showed 
\begin{align}
\underset{n\rightarrow\infty}{{\rm lim}}\frac{\zeta^{(n)}_i}{n^{2-\alpha}} \propto \frac{\Gamma(i+\alpha-2)}{i!}.\label{sfs_ber}
\end{align}
 Here, $n$ is a sample size and $\zeta^{(n)}_i$ is the number of sites at which variants appear $i$ times in the sample (see \citep{berestycki2014asymptotic}  for the proportionality constant of the right-hand side of Equation~\ref{sfs_ber}). By using  Stirling's approximation in Equation~\ref{sfs_ber}, we have
\begin{align}
     f_{\text{SFS}}(x)\propto \frac{1}{x^{3-\alpha}} \quad {\rm \ when}\ x\ll 1. \label{rare_sfs}
 \end{align}

\begin{figure}[t]\centering
 \includegraphics[width=8.5cm]{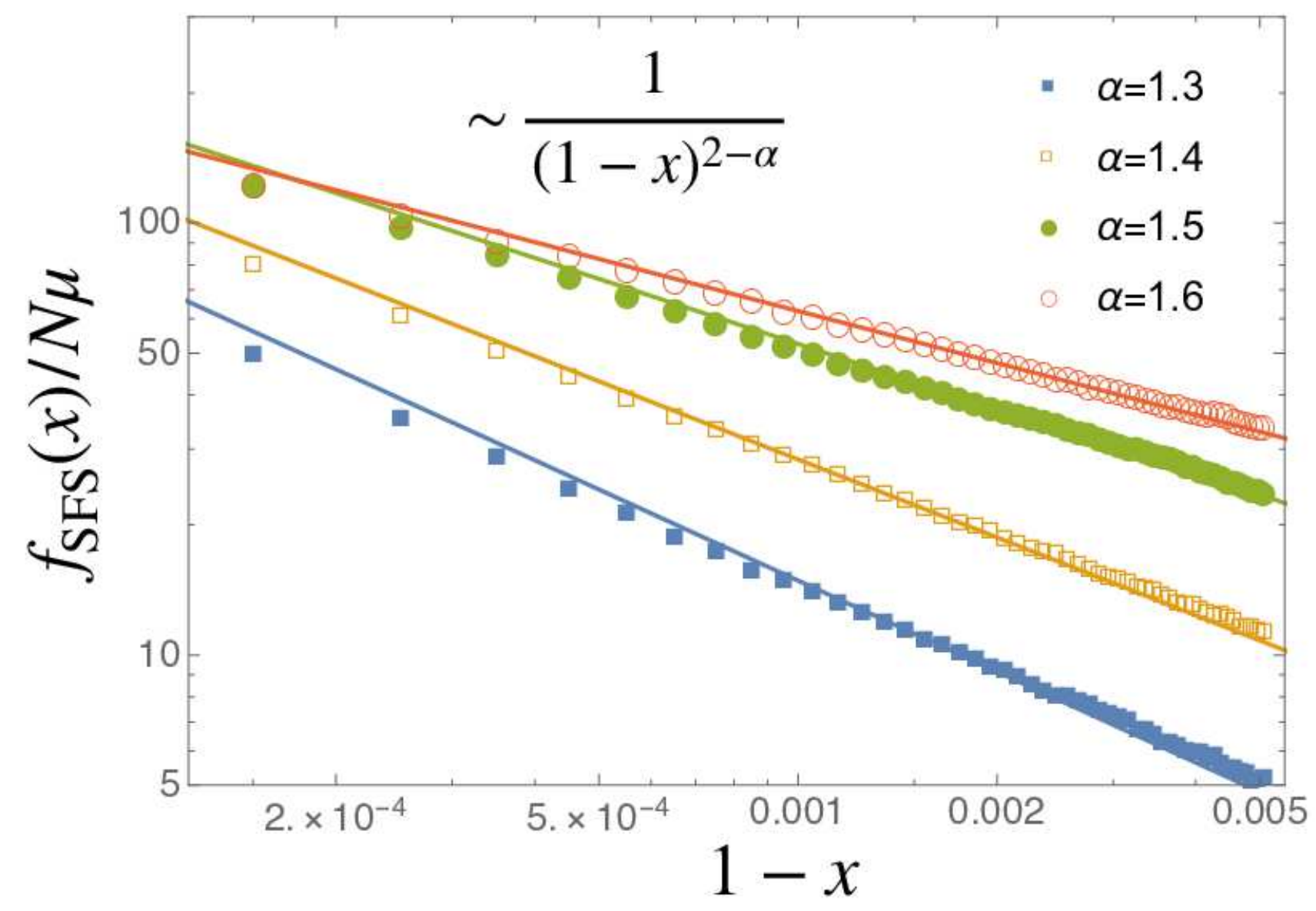} \caption{The SFS near $x=1$  for $\alpha=1.3,\ 1.4,\ 1.5,\  1.6$ (circle, squares). The horizontal axis is $1-x$. The  solid lines are drawn assuming $f_{\text{SFS}}(x)\propto \frac{1}{(1-x)^{2-\alpha}}$. $N=10^6$ is used. 
 }
\label{fig:sfs_scaling}
\end{figure}

Equation~\ref{sfs_ber}, cannot be used for high-frequency variants, because the number of times the variants appear ($i$ in Equation~\ref{sfs_ber}) is kept finite in taking the limit of the sample size $n$.  To the best of our knowledge, a precise behavior at the high-frequency end for $1<\alpha<2$ has not been reported. As shown in Figure~\ref{fig:sfs_scaling}, we find  that  the asymptotic form of the uptick of $f_{\text{SFS}}(x)$ is given by
 \begin{equation}
f_{\text{SFS}}(x) \propto  \frac{1}{(1-x)^{2-\alpha}} \quad  ({\rm for}\ 1-x\ll 1).
\end{equation}

\subsection{Mutation-drift balance}
\begin{figure*}[t]%
\includegraphics[width=.65\textwidth]{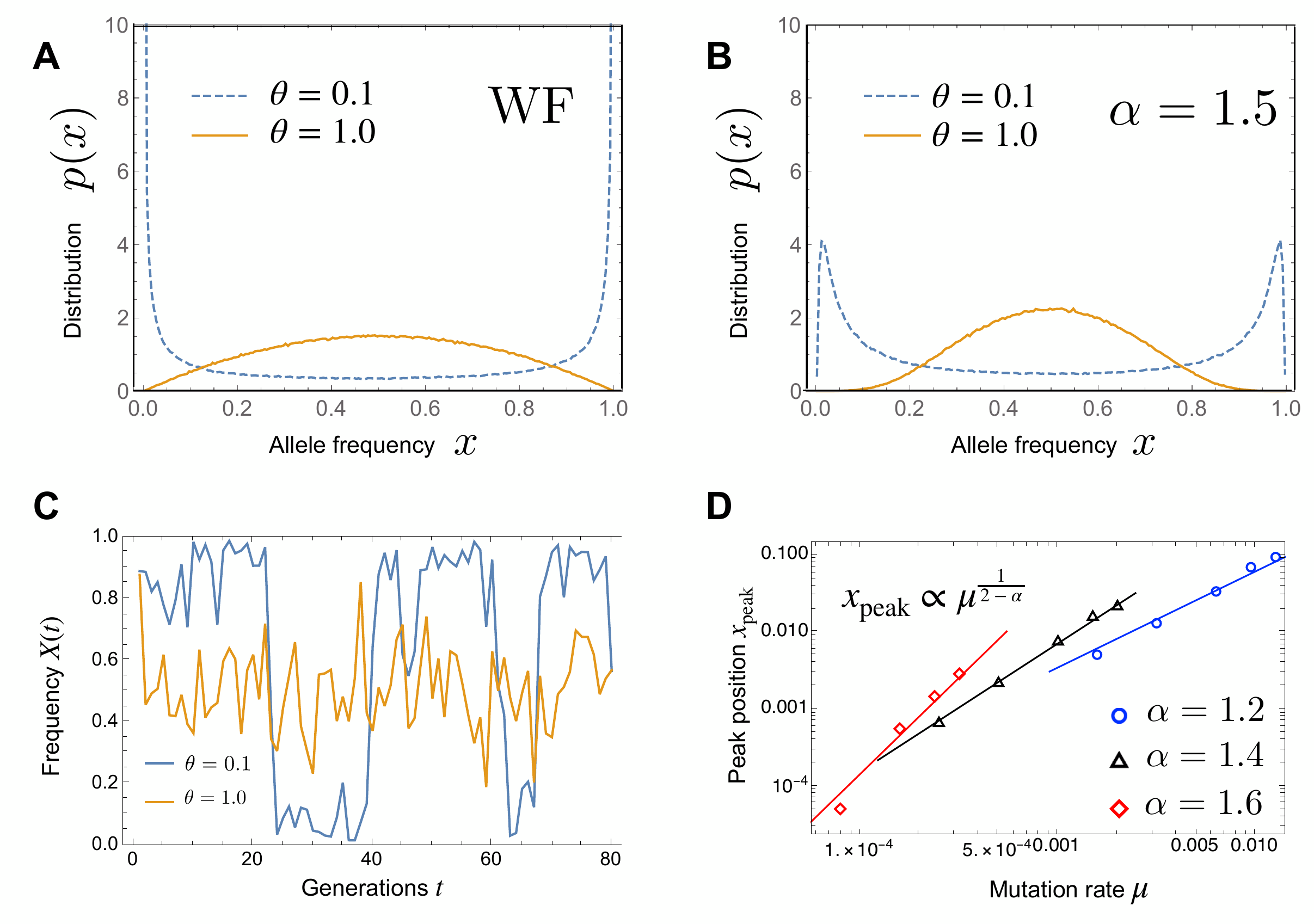}%
\hspace{0.2cm}
\begin{minipage}[top]{.23\textwidth}%
\vspace{-7cm}
\caption{ ({\bf A}) Stationary distribution of the allele frequency in  the Wright-Fisher model,  when the mutation rate is small ($\theta=0.1$) and large ($\theta=1.0$).   ({\bf B}) Stationary distribution for $\alpha=1.5$,  when the mutation rate is small ($\theta=0.1$) and large ($\theta=1.0$). ({\bf C}) The time-series of the allele frequency in the case of $\al=1.5$,  when the stationary distribution is bimodal ($\theta=0.1$)  and  unimodal ($\theta=1.0$). ({\bf D}) The position of the peak near $x=0$ of the stationary distribution versus the mutation rate $\mu$. $N=10^4$ is used.
}
\label{fig:macro_stst}
\end{minipage}%
\end{figure*}
A broad offspring  distribution also affects the stationary distribution of allele frequency when mutations and genetic drift are balancing one another. For simplicity, we consider symmetric reversible mutations between two neutral allele types. We denote the scaled mutation rate (per unit time in the continuous description)  as $\theta = T_c \mu$, where  $\mu$ denotes the mutation rate per generation. In the Wright-Fisher model, it is known that the stationary distribution is given by \citep{kimura1954stochastic}
\begin{align}
    P_{\text{WF}}(x) \propto x^{2\theta-1}(1-x)^{2\theta-1}.\label{WFdist}
\end{align}
There  is a critical value $\theta^{\text{WF}}_c=\frac{1}{2}$: When  $\theta>\theta^{\text{WF}}_c$, the  distribution in Equation~ \ref{WFdist} has a single peak at the center $x=\frac{1}{2}$;  when $\theta<\theta^{\text{WF}}_c$, it has a U-shaped distribution, where the density is increasing monotonically from the center to the boundaries. 

Figure~\ref{fig:macro_stst}A and B show the numerical results of the stationary distributions for the Wright-Fisher model and $\alpha=1.5$,  respectively. When $1\leq \al<2$,  while  a critical value of the mutation rate $\theta_c$ exists as in the Wright-Fisher model, there is a qualitatively different feature:  For a small mutation rate $\theta<\theta_c$, the stationary distribution is not a U-shaped but an M-shaped distribution with two peaks near the boundaries.  Note that the M-shaped distribution indicates a stochastic switching behavior, as illustrated  in  Figure~\ref{fig:macro_stst}D) (the blue curve). As shown in  Figure~\ref{fig:macro_stst}D, the peak positions are approximately given up to prefactors by
\begin{align}
 x_{\text{peak}},\ 1-x_{\text{peak}} & \approx \theta^{\frac{1}{2-\alpha}}.
 \label{xpeak0}
\end{align}


In  Appendix H, we show that the M-shaped stationary distribution persists even in the presence of  natural selection, provided that selection is weaker than the sampling bias at the peaks of the distribution.

A similar M-shaped distribution was observed for the EW process in  \citep{Der2014-xy}, wherein  moments of the stationary distribution were extensively studied. However, the origin of the M-shaped distribution remained unclear. Below, using scaling arguments, we explain why the bimodal distribution arises in our case.

\section{Analytical Arguments}
\subsection{Limiting process, transition density,  and time-dependent effective bias}


We now provide analytical arguments for the observations made in the simulations described in the first part of this paper. Our discussion starts with an exact but somewhat unwieldy description of the allele frequency dynamics. We then show how exact short-time and intermediate time asymptotics can be derived and used to rationalize the sampling bias and the scaling laws discovered above.

The allele frequency dynamics can be fully characterized by the transition probability density $w_N(y|x)$ that the mutant frequency changes from $x$ to $y$ in one generation. Since one generation consists of random offspring contributions to the seed pool and binomial sampling from the seed pool, 
we have 
\begin{align}
	w_N(y|x) = \int d M \int d W\,  P^{\text{MUT}}(M;x N)P^{\text{WT}}(W;(1-x) N)\non\\
\times {\rm Pr_{binom.}}(y N ,N, \frac{M}{M+W}).
\label{w_N}
\end{align}
Here, $P^{\text{MUT}} (M;x N)$ is the probability density  that  the sum of $x N$   random mutant offspring numbers takes the value $M$,  $P^{\text{WT}}(W;(1-x) N)$ is that for  the wild type, and $\rm Pr_{binom.}$ is the probability of getting $y N$ successes in $N$ trials with success probability $\frac{M}{M+W}$. First, we will focus on the neutral case, for which $P^{\text{MUT}}$  and $P^{\text{WT}}$ are the same function, i.e., $P^{\text{MUT}}(\cdot)=P^{\text{WT}}(\cdot)$. 

While the resampling distribution $w_N$ may in general behave in complex ways, it  has few options in the large $N$ limit. These 
constraints emerge from two asymptotic simplifications. First, since $M$ and $W$ are the sums of many random variables, $P^{\text{MUT}}$  and $P^{\text{WT}}$ tend to stable distributions as described by the generalized central limit theorem \citep{Gnedenko1968-gh,Uchaikin2011-pp} (see also Appendix A for a brief description of the  theorem). Second, the fluctuations associated with binomial sampling become negligible compared with those induced by offspring number contributions to the seed pool, provided that the offspring  distribution is sufficiently broad, i.e., $\alpha\leq 2$.
 Thus, we can replace ${\rm Pr_{binom.}}(y N ,N, \frac{M}{M+W})$ with a  Dirac delta function, $\delta(y - \frac{M}{M+W})$. By using these facts and evaluating the integral in Equation~\ref{w_N} (see Appendix B for  details), we obtain a simple analytical expression of $w_N(y|x)$, which is valid in the large $N$ limit:
When $\alpha=1$ \citep{hallatschek2018selection}, 
\begin{align}
w_N(y|x)=\frac{1}{\log N}\, \frac{x(1-x)}{(x-y)^2}.
\label{wn_a=1}
\end{align}
When $1<\alpha<2$,
\begin{align}
w_N(y|x) = \begin{cases}
N^{1-\al}  C_\alpha \, x(1-x)\frac{  (1-y)^{\alpha -1} }{(y-x)^{\al+1}} \quad {\rm when}\ x<y \\
N^{1-\al} C_\alpha \, x(1-x) \frac{ y^{\alpha -1}}{(x-y)^{\al+1}}    \quad   {\rm when}\ x>y,
\end{cases}
\label{wN_mag}
\end{align}
where $C_\alpha\equiv \alpha\left(\frac{\alpha -1}{\alpha }\right)^{\alpha }$.

 To obtain the continuum description, we must appropriately scale the time $t$ with the population size $N$  \citep{Gardiner2009-nq}. The characteristic timescale (coalescent timescale)  $T_c$   can be read from the dependence of the transition density on $N$.   \cite{hallatschek2018selection} showed  that,  when $\alpha=1$,  the resulting limiting process is described by
 \begin{align}
\frac{\p }{\p \tau } P(x,\tau) &= - \frac{\p }{\p x} V(x) P(x,\tau)\non\\
&+ \int d x' (w(x|x') P(x',\tau) - w(x'|x) P(x,\tau) ), \label{CK_marginal}
\end{align}
where the jump kernel $w(x'|x)$ is  given by
\begin{align}
w(x'|x )= \frac{x(1-x)}{(x-x')^2}\label{w_marginal},
\end{align}
and the advection (bias) term $V(x)$ is given by 
\begin{align}
V(x) = -{\rm P.V.}\int dx' (x'-x)w(x'|x)=x(1-x)\log\frac{x}{1-x}, \label{V_marginal}
\end{align}
where P.V. denotes the Cauchy principal value. 
It is easy to check that Equation~\ref{V_marginal} satisfies 
 the neutrality condition $\frac{\p }{\p \tau }\langle X\rangle =0$. 

To develop intuition, it is useful to interpret the different terms in Equation~\ref{CK_marginal}.
 First, $V(x)$  has a form of frequency-dependent selection that enhances the major allele (with frequency $> 50\%$) and suppresses the minor allele. The apparent fitness differences between the mutant and wild type is  given by the log-ratio of their frequencies. 
 Such a selection-like effect arises because the major allele can sample the offspring number from $P_U(u)$ more deeply than the minor allele (see  \cite{hallatschek2018selection}). 
Second, in spite of this apparent bias, the neutrality of the whole process is maintained due to rare large jumps, characterized by  $w(y|x)$. This also means that the neutrality does not hold if we focus on ``typical''  trajectories (see Figure~\ref{fig:traj}). In fact, as we show in Appendix E,  the median $x_{\text{med}}$  of the mutant frequency, which is  a proxy of ``typical''  trajectories,  evolves  according to 
\begin{align}
\frac{d}{d\tau}X^{\text{med}}(\tau ) = V(X^{\text{med}}(\tau )) \quad ({\rm when }\ \alpha=1).\label{xmed}
\end{align}

When $1<\alpha<2$,  using the same reasoning as the derivation of Equation~\ref{CK_marginal} and choosing $\tau=\frac{t}{C_\alpha N^{\al-1}}$, 
we can obtain the following differential Chapman-Kolmogorov equation, 
\begin{align}
\frac{\p }{\p \tau } P(x,\tau) &= - \frac{\p }{\p x} V(x) P(x,\tau)\non\\
&+ \int_{|x'-x|>\epsilon} d x' (w(x|x') P(x',\tau) - w(x'|x) P(x,\tau) ) \label{CK_nonmarginal}
\end{align}
where
\begin{align}
w(x'|x) = \begin{cases}
  x(1-x)\frac{  (1-x')^{\alpha -1} }{(x'-x)^{\al+1}} \quad {\rm when}\ x<x' \\
  x(1-x) \frac{ x'^{\alpha -1}}{(x-x')^{\al+1}}    \quad   {\rm when}\ x>x' 
\end{cases}
\label{w_nonmarginal}
\end{align}
and 
\begin{equation}
	V(x) = -\int_{|x'-x|>\epsilon} dx'(x'-x)w(x'|x). \label{V_nonmarginal}
\end{equation}
As in Equation~\ref{CK_marginal}, the advection term guarantees the neutrality of allele frequency. 
Equation~\ref{w_nonmarginal} means that, when $x<\frac{1}{2}$, rightward  jumps occur more frequently than  leftward ones, and  this tendency reverses when $x>\frac{1}{2}$. Noting the overall minus sign in Equation~\ref{V_nonmarginal}, this in turn means that $V_{\text{eff}}$ is a bias against the minor allele (see Figure~\ref{fig:traj}), as in the case of $\al=1$. We will later show that when $x \ll 1 $, the median trajectory is initially decaying  like $\Delta X^{\text{med}} \sim  - \tau^{\frac{1}{\alpha}}$ (Equation~\ref{dx0}).

Note that, in the limit $\epsilon\to 0$, two divergencies arise in Equation~\ref{CK_nonmarginal}, one in the integral for the advection velocity in Equation~\ref{V_nonmarginal} and one in the jump integral in Equation~\ref{CK_marginal}. However, since both divergencies exactly  cancel, the entire right-hand side of Equation~\ref{CK_nonmarginal} is well-defined.
 As shown in Appendix D, Equations~\ref{CK_marginal} and \ref{CK_nonmarginal} can also be derived as a dual of the $\Lambda$-Fleming-Viot process, namely as the adjoint operator of the backward generator (e.g.,  \cite{griffiths2014lambda,etheridge2010coalescent}).

Although it is difficult to study Equation~\ref{CK_nonmarginal} analytically, it is possible to derive exact short-time and long-time asymptotics that, combined with scaling arguments, paint a fairly comprehensive picture of the ensuing statistical genetics.

\subsection{Short-time dynamics and fluctuations}
\begin{figure*}[t]
\centering
  \includegraphics[width=18cm]{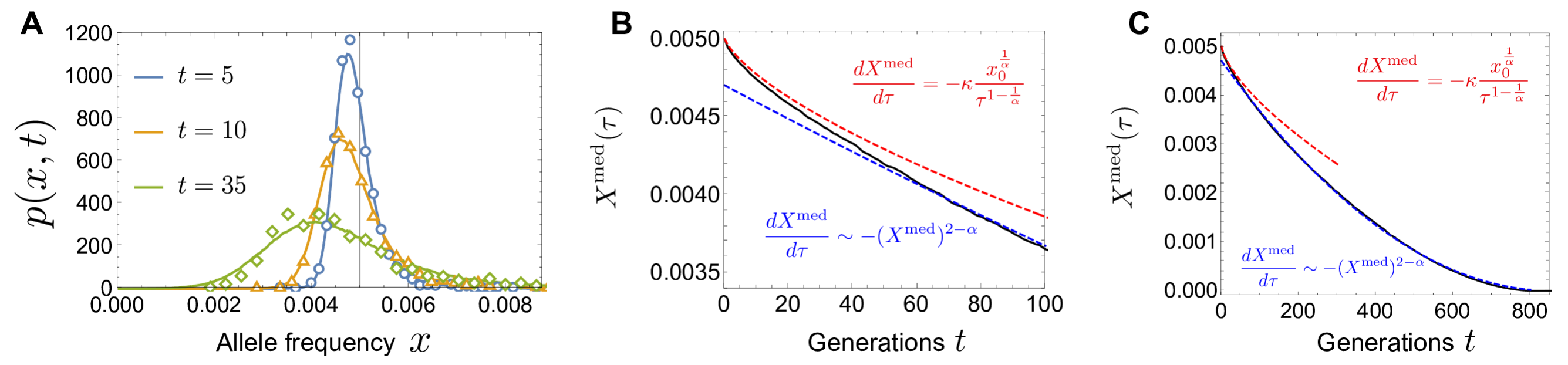}
    \caption{ (A) The allele frequency distribution $p(x,t|x_0=0.005)$ at generation  $t=5,\, 10,\, 35$,  for $\alpha=1.5$.  The solid lines denote the short-time transition densities given by Equations~\ref{dx=cz} and \ref{ch_zeta}, and the open markers denote those computed from 10000 allele frequency trajectories  in the discrete-time model. (B)  The initial dynamics of the median of the allele frequency (black). The red and blue lines denote the short-time solution in Equation~\ref{short_med} and the long-time solution in Equation~\ref{med_long}, where constants of integration and the prefactor of Equation~\ref{veff_long} are determined by fitting to the discrete-time model (black line) between $40<t<800$. (C) The overall trajectory to extinction. The color scheme is the  same as that in (B).  In (A-C), $\alpha=1.5,\ N=10^7, \ x_0=0.005$ are used.
  }
 \label{fig:xmed_short_long}
\end{figure*}

First, we describe the transition  density $P(x,\tau|x_0,\tau=0)$ of Equation~\ref{CK_nonmarginal} for small times. 
When $1<\alpha<2$, the allele frequency changes due to the deterministic bias $V(x)$ and random occurrence of jumps, sampled from the broad distribution in Equation~\ref{w_nonmarginal}. Since  the number of jump events is enormous $(\sim \frac{\tau}{\epsilon^\alpha})$ even for small $\tau$, 
the generalized central limit theorem  applies, and  $X(\tau)$ is asymptotically distributed according to  a stable distribution \citep{Gnedenko1968-gh}. For a general stable distribution, its analytical expression is  not available,  and only its characteristic function can be expressed analytically. As we show in {Appendix C}, 
the random displacement $\Delta X(\tau)= X(\tau)-x_0$ can be expressed 
as  
\begin{align}
\Delta X(\tau) = X(\tau)- x_0 = \gamma(\tau, x_0) Z. \label{dx=cz}
\end{align}
Here $Z$ is sampled from the stable distribution $p(z)$ whose  characteristic function $\langle e^{i k Z} \rangle \equiv  \int dz e^{ik z } p(z)$  is given by 
\begin{align}
\langle e^{i k Z} \rangle 
& =\exp\biggl[-|k |^\alpha \biggl(1 - i \beta(x_0)  \tan \frac{\pi \alpha}{2}  {\rm sign} (k )\biggr)\biggr],\label{ch_zeta}
\end{align}
and the scale parameter $\gamma(\tau,x)$ and the skewness parameter $\beta(x)$ are respectively  given by
\begin{align}
\gamma (\tau, x) & \equiv \tau^{\frac{1}{\alpha}}\biggl(   \frac{\pi ( x(1-x)^\alpha + (1-x)x^\alpha )  }{ 2\Gamma(\al+1) \sin \frac{\pi \alpha}{2}}
\biggr)^{\frac{1}{\alpha}},\label{bn}\\
\beta(x) & \equiv \frac{ x(1-x)^\al - x^\al(1-x)}{x^\al(1-x) + x(1-x)^\al}. \label{beta}
\end{align}
 Note that  statistical properties of $Z$  are independent of $\tau$, and $\Delta X(\tau)$ depends on $\tau$ via the scale parameter $\gamma(\tau,x_0)$.
As shown in Figure~\ref{fig:xmed_short_long}A, for small times,  the transition density  $P(x,\tau|x_0,\tau=0)$ computed from the stable distribution agrees precisely  with  numerical simulation results in the discrete-time model.
Our result can be regarded as  a counterpart of the Gaussian approximation often employed for Wright-Fisher diffusion  (see  \cite{tataru2017statistical} and the references therein).




Now, we study the mean and median of the allele frequency using the short-time expression.
The mean  does not change in time since  $\langle \Delta X(\tau) \rangle = \gamma(\tau,x_0) \langle Z \rangle =0$, which is consistent with the neutrality. On the other hand,  the median   changes as   
\begin{align}
\Delta X^{\rm med}(\tau) = \gamma(\tau,x_0) Z^{\rm med}(x_0),\label{short_med}
\end{align} 
where $ Z^{\rm med}(x_0)$ denotes the median of $Z$.  $ Z^{\rm med}(x_0)$ depends on $x_0$ via $\beta(x_0)$ (see Equation~\ref{ch_zeta}), and  $Z^{\rm med}(x_0) \lessgtr 0$ for $x_0 \lessgtr \frac{1}{2}$.  Equation~\ref{short_med} agrees with   numerical simulations in the discrete-time model, while $X(\tau)$ is  close to the initial frequency $x_0$  (see the red and black curves in Figure~\ref{fig:xmed_short_long} (B)). 
 
The scaling property $\Delta X(\tau)\propto \tau^\frac{1}{\alpha}$ in Equation~\ref{dx0} immediately follows from  Equation~\ref{short_med}, since $\gamma\propto \tau^{\frac{1}{\alpha}}$. This scaling implies that there is a time-dependent bias driving the median of the allele frequency. Differentiating Equation~\ref{short_med} with respect to time gives
\begin{align}
\frac{d}{d\tau} X^{\rm med}(\tau) = V_{\rm eff}(\tau)\label{adiabatic}
\end{align}
where  the {\it effective time-dependent bias} $V_{\rm eff}(\tau) $ is given  by 
\begin{align}\label{v_adiabatic}
V_{\rm eff}(\tau) \equiv 
\frac{\partial \gamma(\tau,x_0)}{\partial \tau }Z^{\rm med}(x_0)
.
\end{align}
Near the boundaries $x=0$ and  $x=1$, 
  $V_{\rm eff}(\tau)$   is approximately given by  
\begin{align}
V_{\rm eff}(\tau) \approx  
\begin{cases}
-k \frac{ x_0^{\frac{1}{\alpha}}}{\tau^{1-\frac{1}{\alpha}}} \quad \quad (x\ll 1)\\
+k  \frac{ (1-x_0)^{\frac{1}{\alpha}}}{\tau^{1-\frac{1}{\alpha}}}\quad (1-x\ll 1)
\end{cases},\label{veff0}
\end{align}
where $k \equiv \frac{|Z^{\rm med}(x_0=0)|}{\alpha}(\frac{\pi }{2 \Gamma(\alpha+1) \sin \frac{\pi \alpha}{2}})^\frac{1}{\alpha}$  is a positive constant.

\begin{figure}[t]\centering
  \includegraphics[width=9cm]{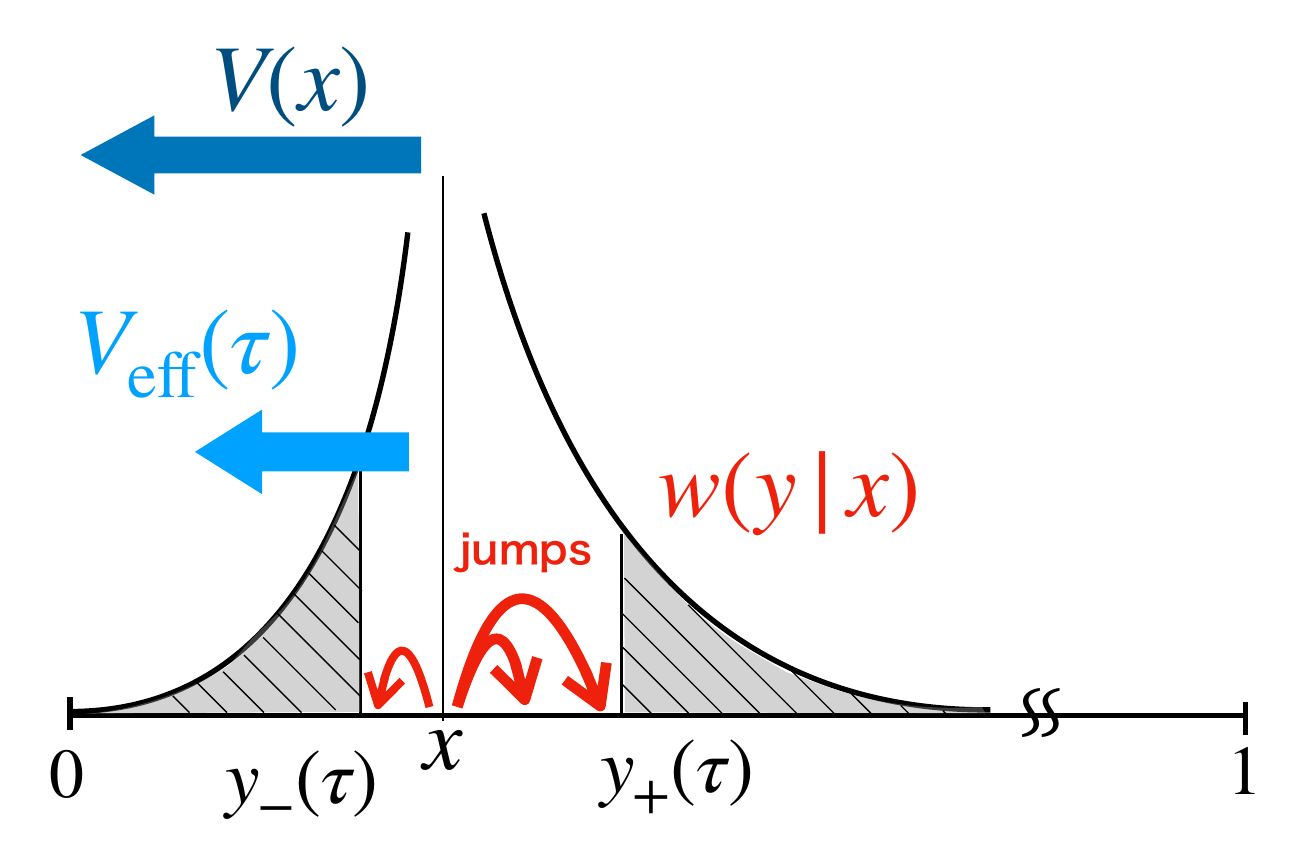}
  \caption{ Schematic explanation of the effective  time-dependent bias $V_{\text{eff}}(\tau)$. The black curve shows the  jump rate $w(y|x)$ in Equation~\ref{w_nonmarginal} when $x\ll 1$.  In a time $\tau$, small jumps within the region $[ y_-(\tau), y_+(\tau)]$ are likely to occur,  offsetting a part of the original bias $V(x)$. $V_{\text{eff}}(\tau)$ is  the  residual part of the bias.
  }
 \label{fig:w_ypm}
\end{figure}

\paragraph*{The advection term arises from a sampling bias.} Intuitively, the time-dependent bias $V_{\rm eff}(\tau)$ arises from a time-dependence of the largest sampled offspring number (Figure~\ref{fig:w_ypm}). To see this, consider a typical trajectory of the allele frequency starting from $x$. 
Up to  a short time $\tau$, only jumps from $x$ to $y\in [y_-(\tau), y_+(\tau)]$ are likely to occur,
 where $y_-(\tau)$ and $y_+(\tau)$ can be estimated from 
 \begin{align}
 &\tau \times \int_{0}^{y_-(\tau)} w(y|x)dy\sim 1,\quad \tau\times \int_{y_+(\tau)}^1 w(y|x)dy\sim 1. \label{ypmest}
 \end{align} 
These conditions give  
\begin{align}
y_{-}(\tau) \sim  \frac{x}{1+(\frac{ \tau (1-x)}{\alpha})^{1/\alpha}},\quad y_{+}(\tau) \sim  \frac{x+(\frac{  \tau x}{\alpha})^{1/\alpha}}{1+(\frac{ \tau x}{\alpha})^{1/\alpha}}.
\label{main_pm}
 \end{align}
Because these small jumps cancel a part of the  bias $V(x)$ in Equation~\ref{V_nonmarginal},  the typical trajectory is then  driven by 
  the uncanceled residual part of the bias $V(x)$,  
\begin{align}
V'_{\text{eff}} \equiv  - \int_{y\in [0,y_-(\tau)]\cup[y_+(\tau),1]}dy (y-x)w(y|x).\ \label{veff_def_old}
\end{align}
When $x\ll 1$,   
 the dominant contribution to this integral is from  $y\approx y_+(\tau)$. Using $w(y|x)\sim \frac{x}{(y-x)^{\alpha+1}}$ from the first line of Equation~\ref{w_nonmarginal} and  $y_+ -x\sim (\tau x)^\frac{1}{\alpha}$ from Equation~\ref{main_pm}, the above integral can be evaluated as $V'_{\text{eff}} \sim - \frac{x}{(y_+-x)^{\alpha-1}} \sim - \frac{x^{\frac{1}{\alpha}}}{\tau^{\frac{\alpha-1}{\alpha}}} $, which agrees with $V_{\text{eff}}$ in Equation~\ref{veff0} for $x\ll 1$ (up to the factor $\kappa$). When $1-x\ll 1$,  the dominant contribution to  $V'_{\text{eff}}$ is from   $y\approx y_-(\tau)$ and can be evaluated in a similar way, reproducing  $V_{\text{eff}}$ in Equation~\ref{veff0} for $1-x\ll 1$.

\paragraph*{Allele frequency fluctuations are inconsistent with Wright-Fisher diffusion.}
In the simulations, we found that, for $1\leq \alpha<2 $, allele frequency fluctuations are inconsistent with Wright-Fisher diffusion and characterized by super-diffusion with diffusion exponent $\frac{2}{\al}$ (see Equation~\ref{MedSD_main}).  This finding is readily explained by the short-time asymptotic in Equation~\ref{dx=cz}. Recalling statistical  properties of $Z$ are independent of $\tau$, the median SD is given by 
 \begin{align}
    {\rm Median \ SD}  = \gamma(\tau,x_0)^2 {\mathbb M}[ (Z- Z^{\rm med} )^2]\propto \tau^{\frac{2}{\alpha}}.
 \end{align}

This scaling can also be  justified  heuristically by noting that, for $1<\alpha<2$, the square displacement is  dominated by large jumps. During time $\tau$,  an allele  frequency $X(\tau)$ around $x$  typically jumps to $y_\pm$ given in  Equation~\ref{main_pm}. When $\tau\ll 1$, it is easy to see $|y_\pm-x|\sim \tau^{\frac{1}{\al}}$ with  $x$-dependent prefactors. Because the median SD is dominated by the largest displacements,  it can be evaluated as
 \begin{align}
     \text {Median SD} \sim  (y_\pm(\tau)-x)^2 \sim \tau^{\frac{2}{\alpha}},
 \end{align}
  where $\tau = \frac{t}{T_c}\ll 1$ is assumed.

\subsection{Long-time dynamics and extinction time}
Above, we saw that at short times, allele frequencies carry out an unconstrained Levy flight. This random search process, however, gets distorted as soon as the allele frequency starts to get in reach of one of the absorbing boundaries. Interestingly, the dynamics then enters a universal intermediate asymptotic regime that controls both the characteristic extinction time as well as establishment times and fixation probabilities.

To see this, let us consider the extinction dynamics of a trajectory starting from a small frequency $x_0 \ll 1$ (Figure~\ref{fig:medsd}). At short times, we can apply the short-time asymptotics in Equations~\ref{adiabatic}, \ref{veff0}. We  expect Equations~\ref{adiabatic}, \ref{veff0} to break down  when the displacement $\Delta X^{\rm med}(\tau)$ computed from Equation~\ref{adiabatic} becomes comparable to $x_0$, which occurs at $\tau\sim \tau_{\rm short} \equiv x_0^{\alpha-1}$. By taking a coarse-grained view, the  rate of the frequency change in  $\tau_{\rm short}$ is roughly given by
\begin{align}
 \frac{\Delta X^{\text{med}}}{\tau_{\rm short}} \sim -x_0^{2-\alpha}. 
\end{align}
This suggests that, in a long timescale ($\tau \gtrsim  \tau_{\rm short}$), the median frequency decreases as
\begin{align}
\frac{d}{d\tau} X^{\text{med}} (\tau) = \tilde V_{\rm eff}(X^{\text{med}}\left(\tau)\right) \quad ({\rm for}\ X\ll 1), \label{med_long}
\end{align}
where, up to a prefactor,  the frequency-dependent bias $\tilde  V_{\rm eff}(X)$ is given by
\begin{align}
\tilde V_{\rm eff}(X) \sim -  X^{2-\alpha}.\label{veff_long}
\end{align}
 In Figure~\ref{fig:xmed_short_long}C, it is numerically shown that the long-time trajectory  $X^{\text{med}}(\tau> \tau_{\rm short})$   is consistent with Equation~\ref{med_long}. 
  By solving Equation~\ref{med_long}, the median trajectory goes to extinction  at $\tau_{\text{ext}}\sim  x_0^{\al-1}$ (Equation~\ref{tauext0}), in agreement with our simulations (Figure~\ref{fig:medsd}). Note that, for $1-x\ll 1$, the bias in Equation~\ref{veff_long} is replaced  by $\tilde V_{\rm eff}(X)\sim + (1-X)^{2-\alpha}$.

\begin{figure}[t]
 \centering
  \includegraphics[width=9cm]{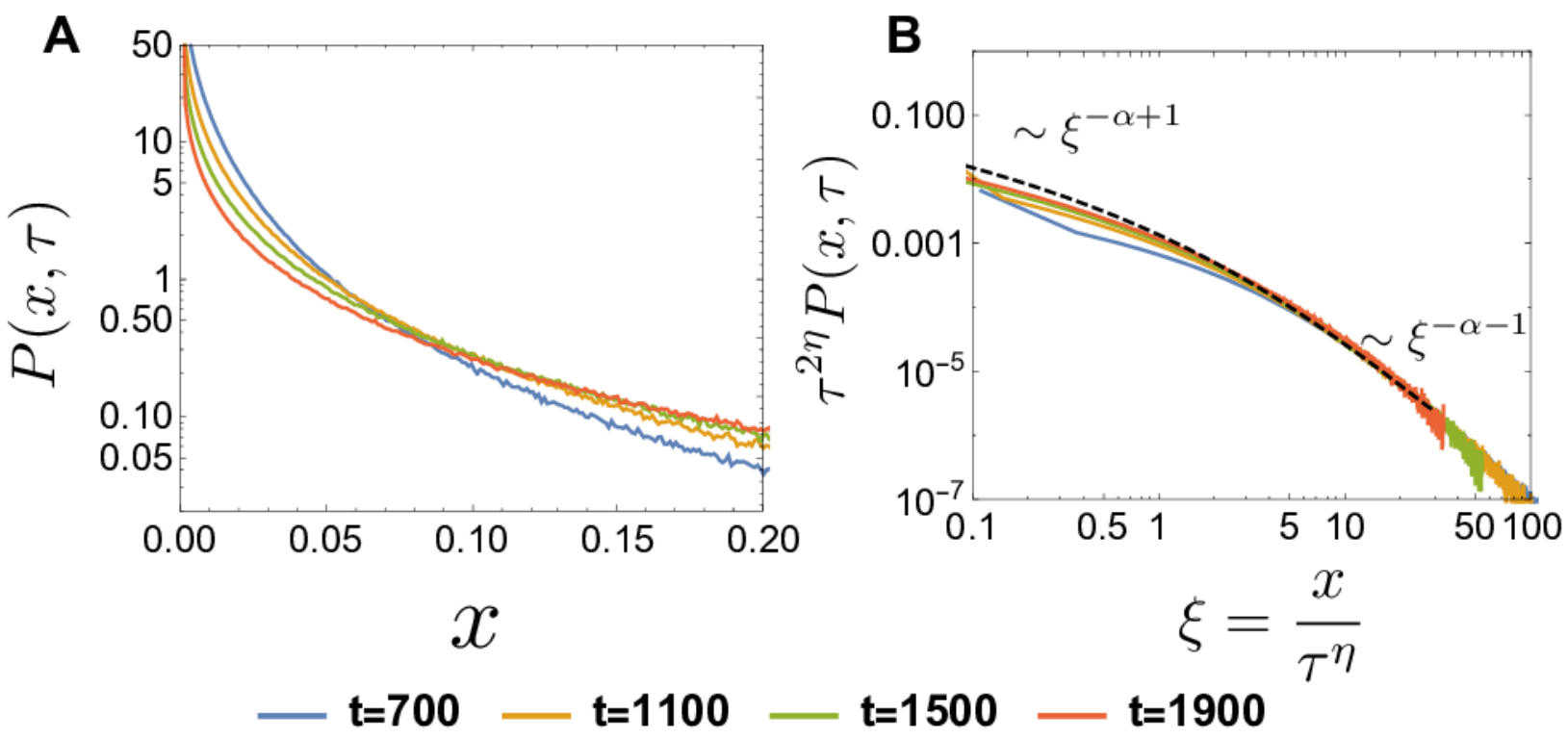}
  \caption{(A) Log plot of $P(x,\tau|x_0=0.01)$ at generations $t=700,1100,1500,1900$ computed from the discrete-time model.
  $N=10^7$ and $\alpha=1.5$. (B) Log-Log plot of $\tau^{2\eta} P(x,\tau|x_0=0.01)$  versus $\xi=x/\tau^{2\eta}$, where $\eta=(\alpha-1)^{-1}$, at $t=700,1100,1500,1900$ (solid curves). The dashed curve represents the analytic result of $g(\xi)$ (see Appendix C).
  The curves $\tau^{2\eta} P(x,\tau|x_0=0.01)$ at the different time points collapse into $g(\xi)$, supporting the scaling ansatz in  Equation~\ref{P_scaling}.
  }
 \label{fig:P_scaling}
\end{figure}

Importantly, Equations~\ref{med_long} and \ref{veff_long} can also be rigorously justified from a scaling ansatz for the  transition density. After some time,  $P(x,\tau|x_0)$ spreads broadly  over the region  $x\ll 1$ with a peak at  $x=0$ (Figure~\ref{fig:P_scaling}A). 
As shown in  Figure~\ref{fig:P_scaling}B, $P(x,\tau)$ is consistent with the following scaling ansatz;
\begin{align}
P(x,\tau)\sim \tau^{-2 \eta} g(\xi)\quad ({\rm for} \ x\ll 1),\label{P_scaling}
\end{align}
 where $\eta\equiv \frac{1}{\alpha-1}$ and $g(\xi)$ is a function of $\xi \equiv \frac{x}{\tau^\eta}$. Up to an overall constant,  $g(\xi)$ can be determined analytically and  expressed as an infinite series (see Appendix C).  Note that  the $\tau$-dependent factor in Equation~\ref{P_scaling} is motivated from the fact that the extent over which the distribution spreads increases like $\tau^\eta$. Equation~\ref{P_scaling} implies that, conditional on establishment at $\tau$, the median frequency increases as $X^{\rm med}(\tau)|_{\rm establish}\sim \tau^\eta$.  Then, Equation~\ref{veff_long} follows by evaluating the bias  in  Equation~\ref{veff0} at 
$\tau \sim (X^{\rm med })^\frac{1}{\eta}$ and at $X^{\rm med }$, instead of at $x_0$.

As a consistency check of the exponent $\alpha-1$ in Equation~\ref{tauext0}, we consider two solvable, extreme cases. First, in the limit  $\al\rightarrow 2$,  the dependence on $x_0$ in Equation~\ref{tauext0} becomes linear. In the Wright-Fisher model, the  mean extinction time can be obtained analytically by solving the backward equation $-1=x(1-x)\frac{\p^2 \tau_{\text{ext}}(x)}{\p x^2}$ (see, for example, \cite{Karlin1981-vz}). The solution is proportional to $x_0$  with a logarithmic correction, $\tau_{\text{ext}}\approx  - x_0\log x_0$.  Second, when $\al \rightarrow 1$, the mean extinction time  no longer depends on $x_0$. We can obtain this explicitly,  by solving Equation~\ref{xmed}: Using $V(X)\simeq X\log X$   when $X\ll 1$, the solution  is given by $\log \frac{\log X(\tau)}{\log x_0} = \tau$. Therefore, if we approximately define the mean extinction time $\tau_{\text{ext}}$  as  $X(\tau_{\text{ext}}) \approx  \frac{1}{N}$, we obtain  $\tau_{\text{ext}} \approx \log \frac{\log N}{-\log x_0}\approx \log \log N$, which is to leading order independent of $x_0$ if $x_0$ is taken to be of order one. 

\subsection{Natural selection and fixation probability}
One important advantage of the forward-time perspective is that we account for natural selection by introducing an appropriate bias favoring of the beneficial variant. Suppose that the mutant type has a selective advantage  $s>0$, such that the average offspring number of mutants is increased by a factor of $1+s$ relative to the wild type. In time-rescaled Chapman-Kolmogorov equation, this adds the term $\sigma  x(1-x) $, where $\sigma = T_c s$,  into the advection $V(x)$ of Equation~\ref{CK_nonmarginal}.

\begin{figure}[t]\centering
  \includegraphics[width=8cm]{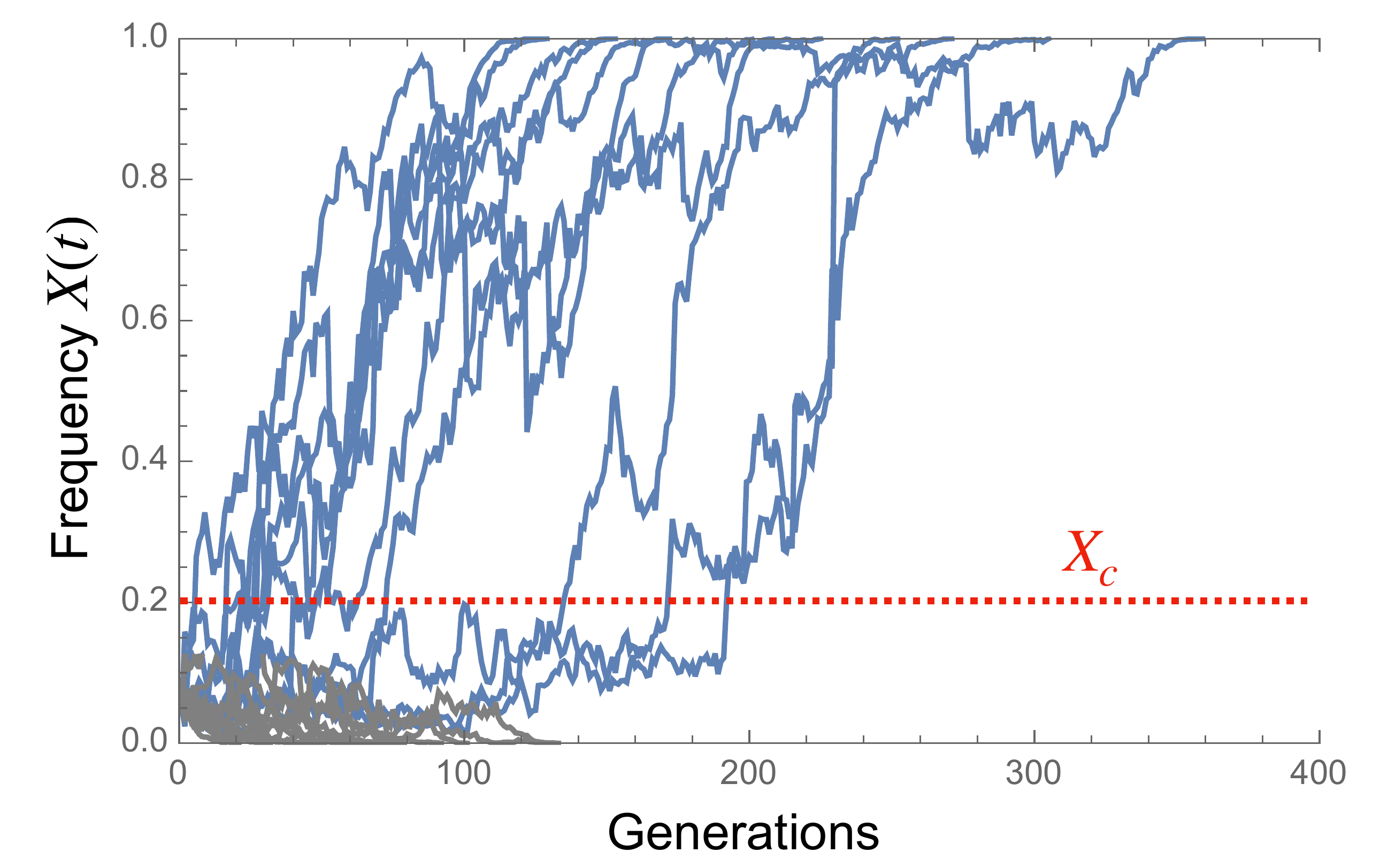}
  \caption{Example of trajectories of the  frequency of the beneficial allele, starting from $x_0=0.05$. $\al=1.5, s=0.03$ and $N=5000$. Fixed trajectories are colored in blue and extinct ones in gray. Here, the crossover point $X_c$ can be estimated as  $X_c\sim 0.2$ (, assuming that the proportional constant in Equation~\ref{xc} is one). 
   Once a trajectory reaches the crossover point,  it becomes fixed in high probability.
  }
 \label{fig:Pfix_traj}
\end{figure}

\begin{figure}[t]\centering
  \includegraphics[width=8.5cm]{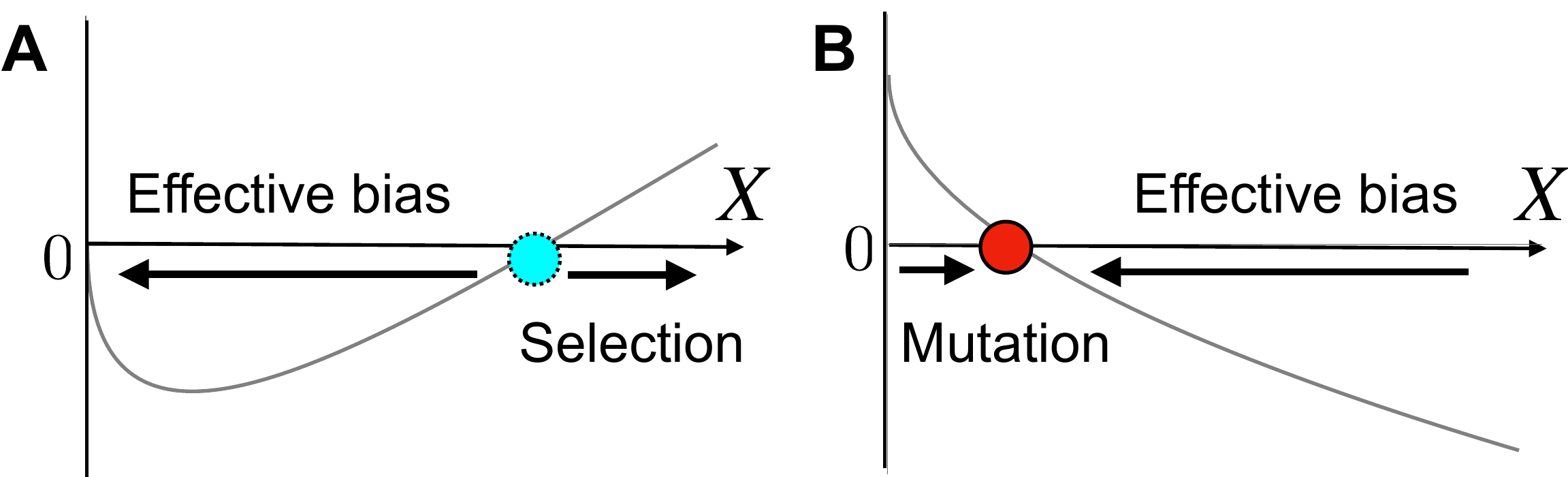}
  \caption{(A) The crossover from the effective bias to genuine selection. $V(X)\sim - C X^{2-\alpha} + \sigma X$ is plotted, where $C$ is a positive coefficient and $\sigma>0$. Deterministically, an unstable point exists at $x\sim X_c$.
(B)  The balance between the effective bias and mutation. $V(X)\sim - C X^{2-\alpha} + \theta$ is plotted. Deterministically, a stable point exists at $X\sim \theta^{1/(2-\alpha)}$.
  }
 \label{fig:balance}

 \end{figure}

 The key observation underlying the argument below is that  when $X$ is sufficiently small, the selection force is negligible compared to the bias  $\tilde  V_{\rm eff}(X)$ in Equation~\ref{veff_long} because while the former is approximately linear in $X$, the latter is sublinear. If the frequency happens to grow and reach a certain value $X_c$, the genuine selection begins to  dominate over the bias, and the trajectory fixes with high probability (see Figure~\ref{fig:Pfix_traj} for example trajectories and  Figure~\ref{fig:balance})A). By using Equation~\ref{veff_long}, the crossover point $X_c$ can be estimated from balancing selection with the sampling bias,
\begin{align}
\sigma \ X  =-\tilde V_{\text{eff}}(X ) \sim  X^{2-\al},\label{crossover}
\end{align} 
which gives
\begin{align}
X_{c} \sim \sigma^{-\frac{1}{\al-1}} = \frac{1}{N}s^{-\frac{1}{\al-1}}.\label{xc}
\end{align}

 The fixation probability $P_{\text{fix}}$ can be estimated by using the neutral fixation probability in a population of size $\approx N X_c$, because  the dynamics are essentially neutral for $X  \ll X_c$, and the trajectory grows almost deterministically for $X>X_c$. Thus, the fixation probability 
 is approximately given by 
 \begin{align}
P_{\text{fix}}\sim \frac{1}{N X_c} =s^{\frac{1}{\al-1}},
\label{Pfix}
\end{align}  
which is valid  for $\frac{1}{N}\ll s^{\frac{1}{\al-1}}$. Equation~\ref{Pfix} reproduces our simulation results in Figure~\ref{fig:Pfix} for $1<\alpha<2$ and, as $\alpha \rightarrow 2$, also reproduces the known result of the Wright-Fisher model, $P^{\text{WF}}_{\text{fix}}\sim 2 s$ (up to a prefactor).

\subsection{Site frequency spectrum}
  { By using the time-dependent effective bias, we can also estimate the behavior of the SFS $f_{\text{SFS}}(x)$ for frequent and rare variants. 
   While the SFS is theoretically defined in the infinite alleles model, it can be computed from our  biallelic framework (\cite{ewens1963diffusion}): $f_{\text{SFS}}(x)\Delta x$ is defined as the expected number of neutral derived alleles in the frequency interval $[x-\frac{\Delta x}{2},x-\frac{\Delta x}{2}]$ in a sampled population (here, the whole population). 
Because new mutations are assumed to arise uniformly in time, the SFS for unlinked neutral loci is  given by the product of the total mutation rate $\mu N $ and the mean sojourn time, namely, the average time  an allele  spends in the frequency interval $[x-\frac{\Delta x}{2},x-\frac{\Delta x}{2}]$ until fixation or extinction.}

First, we consider the low-frequency end, $x\ll 1$, of the SFS (see \cite{Cvijovic2018-nc} for a similar argument). 
Since the SFS is proportional to the sojourn time,   trajectories whose maximum frequencies are $x$ or slightly larger than $x$  dominantly contribute to the SFS $f_{\text{SFS}}(x)$ at $x$. Since  these trajectories typically go extinct due to the bias, and we can roughly estimate their sojourn times at $x$  as the inverse of ``velocity'', $|\tilde V_{\text{eff}}(x)|\sim x^{2-\alpha}$  {in Equation~\ref{veff_long}}. 
Since the probability that a trajectory  grows above a frequency $x$  is roughly  given by $\sim 1/(N x)$,  the SFS is proportional to
\begin{align}
   \frac{1}{N x \tilde V_{\text{eff}}(x)} \propto \frac{1}{x^{3-\alpha}} \quad ({\rm for}\ x\ll1). 
\end{align}

Similarly, for the high-frequency end of the SFS, only the trajectories that grow above $x$ can contribute to $f_{\text{SFS}}(x)$. Typically, these trajectories go to fixation due to the bias $\tilde V_{\text{eff}}(x) \sim (1-x)^{2-\alpha}$. Therefore, the SFS is proportional to
\begin{align}
   \frac{1}{N x \tilde V_{\text{eff}}(x)} \propto \frac{1}{(1-x)^{2-\alpha}} \quad ({\rm for}\ 1-x\ll1). 
\end{align}

The effect of the genuine selection on the SFS can also be studied by using the effective bias. See Appendix F. 

\subsection{Bimodality of stationary distribution}
Now, we turn to explaining the bimodality observed at mutation-drift balance. We found that, when the mutation rates are small, the stationary allele frequency distribution is not a  U-shaped, as expected from Wright-Fisher dynamics, but M-shaped, as shown in Figure~\ref{fig:macro_stst}. The M-shaped distribution arises from the balance between the mutational force and the effective bias (see Figure~\ref{fig:balance}B).
In the Chapman-Kolmogorov equation, the mutational force  is given by
\begin{align}
    -\theta x + \theta (1-x)  \approx  
    \begin{cases}
    + \theta \quad (x \ll 1)\\
   - \theta \quad (1-x\ll 1),
    \end{cases}
\end{align}
which pushes the frequency toward the center  $x=\frac{1}{2}$.
On the other hand, the effective  bias, $\tilde  V_{\text{eff}}(x) \approx -x^{2-\alpha}$ for $x\ll1$ and $\tilde V_{\text{eff}}(x) \approx(1-x)^{2-\alpha}$ for $1-x\ll 1$, pushes a trajectory toward the closer boundary. Therefore, the positions where these two forces balance  are approximately  given by
\begin{align}
 x_{\text{peak}} & \approx c \theta^{\frac{1}{2-\alpha}}, 1-c \theta^{\frac{1}{2-\alpha}},\label{xpeak}
\end{align}
where $c$ is a positive constant. 
If $\theta$ is sufficiently small, we can always find the balancing points.
The presence of these two balancing points means that we can think of the allele frequency dynamics  as a two-state system, essentially analogous to a super-diffusing particle in a double-well potential (see  Figure~\ref{fig:macro_stst}C for a realization of trajectories). This explains the bimodal shape of the stationary distribution.

Finally, we remark that, even in the presence of natural selection, the balancing positions are still determined from the mutation-effective bias balance provided that $\theta\ll 1$: while the effective bias and the mutational term are sub-linear and constant  respectively, the selection term $\sigma x(1-x) $ is linear in $x$ when $x\ll 1$. Thus, when $\theta$ is sufficiently small, 
the magnitude of the selection term around $x=c \theta^{\frac{1}{2-\alpha}}, 1-c\theta^{\frac{1}{2-\alpha}}$ is negligible, and the peak positions are given by  Equation~\ref{xpeak}.

\section{Discussion}

In this study, we analyzed the effect of power law offspring distributions on the competition of two mutually exclusive alleles. Our main reason to consider such broad offspring distributions is that they often emerge in evolutionary scenarios that inflate the reproductive value~\citep{barton2011relation} of a small set of founders. For example, range expansions blow up the descendant numbers of the most advanced individuals in the front of the population, an effect that has been called gene surfing~\citep{hallatschek2008gene}. Likewise, continual rampant adaptation boosts the descendant numbers of the most fit individuals. The resulting allele frequency dynamics becomes asymptotically similar to that of a population with scale-free offspring distributions.

In the case of narrow offspring distributions, which is  predominant assumption in population genetics, it is usually an excellent approximation to describe the allele frequency dynamics by a biased diffusion process, which forms the basis of powerful inference frameworks~\citep{tataru2017statistical}. If the offspring distribution is broad, however, allele frequency trajectories are disrupted by discontinuous jumps, resulting from so-called jackpot events - exceptionally large family sizes drawn by chance from the offspring distribution. Our goal was to find an analytical and intuitive framework within which we can understand the main features of these unusual dynamics.

We found that the main counter-intuitive features can be understood and well-approximated from a competition of selection and mutations with a time-dependent emergent sampling bias,  $V_{\text{eff}}(\tau)$. The sampling bias favors the major allele and arises, because the sub-population carrying the major allele typically samples deeper into the tail of the offspring distribution than the minor allele fraction. 


In the remainder, we first summarize the unusual population genetic patterns that can be explained by the action of these effective forces. We then discuss how broad offspring dynamics could be detected in natural populations and what its implications are for the dynamics of adaptation. Finally, we demonstrate that these dynamics are also ubiquitous in populations with narrow offspring distributions, when mutational jackpot are possible. Therefore, we believe our theoretical framework may be taken as a general null model for populations far from equilibrium.

\subsection{Unusual dynamics}
We found that the sampling bias effectively acts like time- and frequency-dependent selection. In the absence of true selection, $V_{\text{eff}}(x,\tau)$ drives the major allele to fixation, first rapidly and than gradually slowing down with time and proximity to fixation. The slowing down of the sampling bias near fixation also leads to an excess of high-frequency alleles, given continual influx of neutral mutations. This generates a high-frequency uptick in the site frequency spectrum, which is characteristic of the tail of the offspring distribution.  In mutation-drift balance, the allele frequency distribution is M-shaped, in contrast to the U-shape expected from Wright-Fisher dynamics. The peaks reflect the balance of the mutational and sampling bias. 

Non-neutral dynamics depends on whether the genuine selection force dominates over the sampling bias. The sampling bias tends to dominate near extinction or fixation, and wanes near 50\% frequency. A de-novo beneficial allele will not be able to fix unless it overcomes, by chance, the switch-point frequency at which genuine selection becomes stronger than the sampling bias. Finally, fluctuations in typical trajectories are getting stronger over time. 
As a consequence, allele frequencies super-diffuse: fluctuations grow with time more rapidly than under regular Wright-Fisher diffusion.

\subsection{Detecting dynamics driven by broad offspring distributions}
 The time-dependent over-dispersion 
  is most readily detected by plotting the median square displacement as a function of time (see {Figure~\ref{fig:medsd}B}). Testing deviations in this statistics are an attractive avenue for detecting deviations from Wright-Fisher diffusion because the signal is strong for intermediate allele frequencies, which can be accurately measured by population sequencing. By contrast, the time-dependent bias vanishes when an allele has 50\% frequency. So, the detection of the sampling bias requires accurate time series data of low frequency variants, which is difficult to obtain given sequencing errors.

It is clear that a single super-diffusing but neutral allele would not abide by the diffusive Wright-Fisher null model  and thus might be falsely considered as an allele under selection. But importantly, allele super-diffusion has an impact even on statistics that sum over many unlinked loci. This is significant for inference methods, for instance to detect polygenic selection, which argue that trait values follow a diffusion process, if not for an underlying Wright-Fisher dynamics of the allele frequencies then because they sum over many independent allele frequencies~\citep{berg2014population}. However, $\alpha<2$ dynamics breaks both of these arguments. In particular, sums of many unlinked loci tend to non-Gaussian distributions (so called alpha-stable distributions). Hence, for traditional inference methods based on Wright-Fisher diffusion or standard central limit theorem~\citep{tataru2017statistical}, an underlying super-diffusion process should be ruled out.

If time series are not available, broad offspring numbers can also be detected from the site frequency spectrum (SFS)~\citep{Neher2013-sl}. A tail-tale sign of the sampling bias is a characteristic uptick at the high-frquency tail of the SFS, which is difficult to generate by demographic variation~\citep{Neher2013-sl}. As we have shown, the shape of the uptick is characteristic of the tail of the offspring distribution (the parameter $\alpha$).


\subsection{Implications for the dynamics of adaptation}
We found that the fixation probabilities quite sensitively depends on the broadness $\alpha$ of the offspring distribution (Equation~\ref{Pfix}). Accordingly, the dynamics of adaptation, which  ultimately depends on the fixation of beneficial variants, should change quantitatively. To estimate these modifications, we consider an asexual population of constant size $N$  with a broad offspring  distribution with $1<\al<2$, wherein  beneficial mutations occur at the  rate $\mu_{\rm B}$. For low mutation rates, mutations sweep one after the other but when mutation rate are sufficiently high, multiple mutations occur and most mutations are outcompeted by fitter mutations. Such a situation is known as clonal interference. 

We can study the effect of the exponent $\alpha$ on the adaptation dynamics  quantitatively by repeating the argument in \citep{Desai2007-sm}, wherein the variance of offspring numbers is assumed to be narrow. As discussed in Appendix G,
 clonal interference should occur if
\begin{align}
\mu_{\rm B} N s^{\frac{2-\alpha}{\alpha-1}} \ln (N s^{\frac{1}{\al-1}}) \gtrsim 1\quad {\rm  (clonal
\ interference) },\label{CI}
\end{align}
where $s>0$ is the fitness effect of a mutation, which we assume to be constant. The rate $R$ of adaptation  is given by
\begin{align}
    R\sim \begin{cases}
    \mu_{\rm B} N s^{\frac{\al}{\al-1}} \quad (\text{successive\ selective\ sweeps})\\
    \frac{2 s^2 \ln (N s^{\frac{1}{\al-1}})}{(\ln \frac{s}{\mu_{\rm B}})^2}\quad (\text{clonal\ interference })
    \end{cases}
\label{vel}
\end{align}
Note that  the second line in Equation~\ref{vel} reproduces Equation~5 of \citep{Desai2007-sm} in the limit $\al \rightarrow 2$.
Thus, the rate of adaptation depends only weakly (logarithmically) on $\alpha$ in the clonal interference regime, even though the condition for clonal interference in Equation~\ref{CI} depends on $\al$ quite sensitively,  .

\subsection{Emergence of skewed offspring distributions in models of range expansions}

\begin{figure*}[t]
\centering
  \includegraphics[width=17cm]{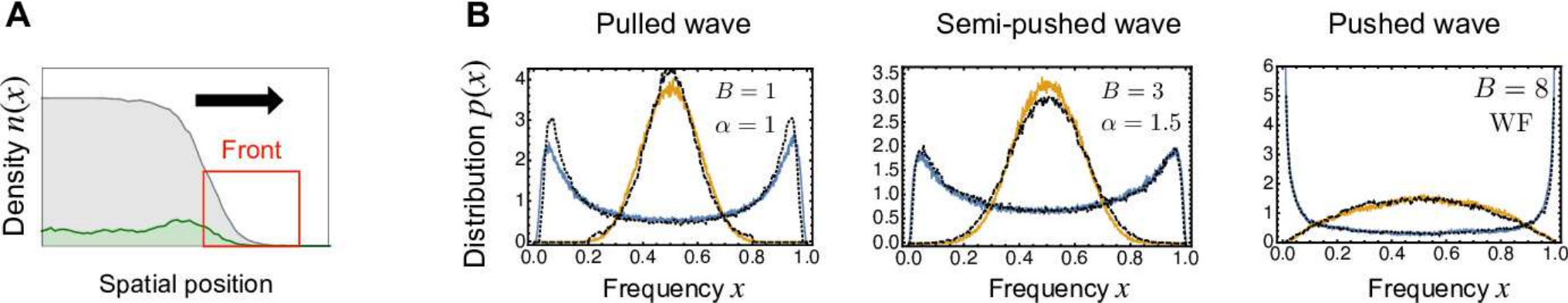}
  \caption{ ({\bf A}) The model of a range expanding  population with two neutral alleles (green and gray). A broad offspring  distribution arises dynamically in the front region.  ({\bf B}) Stationary distributions of the allele frequency when mutation rate $\theta$ is small (blue) and when $\theta$ is large (orange). The wiggling lines (blue/orange) are the numerical results in the traveling wave model, while the dotted lines (black)  are those in the macroscopic model.  The parameters of the Allee effect $B$ are $B=1$ (left), $3$ (middle), and $8$ (right). See Appendix I for the details of the implementation of the simulation and other parameter values.
  }
 \label{fig:wave_stst}
\end{figure*}

Our study can be regarded as an analysis of the population genetics induced by power-law offspring distributions. The main reason to consider these scale-free offspring distributions is that they quite generally \textit{emerge} in models of stochastic traveling waves~\citep{Birzu2018-wq}. Such models are ubiquitous in population genetics because they describe a wide range of evolutionary scenarios, including range expansions, rampant asexual and sexual adaptation as well as Muller's ratchet~\citep{brunet2007effect,Neher2013-sl,desai2013genetic,kosheleva2013dynamics,schweinsberg2017rigorous,Birzu2018-wq}. Our analysis should therefore apply most directly to these evolutionary scenarios, which we now demonstrate using a simple model of a range expansion. We end by discussing the question of whether some of our results may also arise in scale-rich offspring distributions.

Ref.~\citep{Birzu2018-wq} argued that any exponent $1\leq \alpha \leq 2$ can emerge in a simple model of range expansions that incorporates a tunable level of cooperativity between individuals (Figure~\ref{fig:wave_stst}A). The  model can be described by a generalized stochastic Fisher-Kolmogorov equation
\begin{align}
\frac{\partial n}{\partial t} = D \frac{\partial^2 n}{\partial x^2} + r(n)n +{\rm noise},\label{Birzu}
\end{align}
for the  time-dependent population density $n(x,t)$ at position $x$ in a linear habitat and time $t$. The growth rate $r(n)$ is assumed to be density-dependent,
with 
\begin{align}
r (n)=r_0 (1-\frac{n}{K})(1+B\frac{n}{K}),
\end{align}
where the parameter $B\geq 0$ accounts for co-operativity among individuals, which is also called an Allee effect.
As discussed in \cite{hallatschek2018selection}, lineages in the region of the wave tip are diffusively mixed within the timescale $\tau_{mix}\sim \frac{1}{r} \ln^2 K \sqrt{\frac{D}{r}}$. This implies that, in this microscopic model, resampling from an offspring  distribution roughly occurs every $\tau_{mix}$ generations. In \cite{Birzu2018-wq,birzu2020genealogical}, it was argued that depending on the strength of the Allee effect, the offspring  distributions corresponding to any of the three distinct classes of the beta coalescent process can arise; namely, the Bolthausen-Sznitman coalescent when $B<2$, the beta coalescent with $1<\al<2$ when $2<B<4$, and the Kingman coalescent when $B>4$.

To demonstrate  clearly that our present study can serve as a macroscopic analysis of the traveling model, we introduce reversible mutations in the traveling wave model and measured  the mutant frequency of the first {$N\sim \frac{K}{k}$ individuals from the edge of the front. Here, $k$ is the spatial decay rate, i.e., $n\sim e^{-k \tilde x}$ where $\tilde x$ is the coordinate comoving with the expansion}. This definition of the mutant  frequency is reasonable because  only  the wave front has a skewed offspring distribution due to the founder effect.  In Figure~\ref{fig:wave_stst}B,  for $B=1$ (left), $3$  (middle), and $8$ (right),    the frequency distributions in the traveling wave model are shown
 when the mutation rate is small (orange jagged line) and when it is large (blue jagged line).
 The corresponding distributions in the macroscopic model  are shown by black dotted lines. The stationary distributions in the traveling wave model agree well with those in the macroscopic model. Especially, the transition from the M-shaped or U-shaped distribution to the monomodal distribution is consistently reproduced  in the traveling wave model. These results underscore  the correspondence between the traveling wave  with the Allee effect and the beta coalescent process.
 
The above-described correspondence suggests that the spatial area occupied by one allele type in a range expansion should behave statistically like the time-integral over the allele frequency in the Cannings model. In  the context of  adapting  (non-spatial) populations,  this quantity  describes the  total number of mutational opportunities of a mutant lineage \citep{Desai2007-sm, weissman2009rate, neher2011genetic}.
As presented in Appendix J,  the distribution of the time-integrated frequency exhibits a scaling behavior, that depends on the offspring distribution sensitively.  While a full discussion is beyond the scope of this paper, we expect that the distribution of areas  serves as a useful observable to distinguish different prototypes of traveling waves~\citep{Birzu2018-wq}.

\textit{Broad offspring distributions with a scale:} While scale-free offspring distributions often emerge over an intermediate time scale ($\tau_{mix}$ in the above traveling wave model), there are also species that over single generations show broad offspring numbers and violate Wright-Fisher diffusion. For such species, it may be more natural to consider offspring distribution with a characteristic scale. In `sweepstake' reproduction~\citep{Eldon2006-ko}, a fixed and finite fraction of the population is replaced at every sweepstake event (specified by the parameter $\Psi$ in \cite{Eldon2006-ko}). Because $\Psi$ sets a characteristic scale in offspring numbers, power law relationships for 
the median of allele frequencies as well as frequency fluctuations
cannot be expected, which we confirm in Appendix K. Nevertheless, the qualitative features of a sampling bias can be recognized quite clearly for sweepstake reproduction as well. 

Either type of model ultimately is an approximation to true offspring distributions, and it depends on the situation, which one to use. As we argued, the beta-coalescent along with the forward-in-time model described in this paper is the natural choice for range expansions, rapid adaptive process or other scenarios where the reproductive value of a chosen few are highly inflated.

\section*{Acknowledgements}
This work is in part supported by RIKEN iTHEMS Program. Research reported in this publication was supported by the National Institute of General Medical Sciences of the National Institutes of Health under award R01GM115851, a National Science Foundation CAREER Award (\#1555330), a Simons Investigator award from the Simons Foundation (\#327934), and JSPS KAKENHI (Grant Number JP19K03663). We express our sincere thanks to  Benjamin H. Good, Daniel B. Weissman,  Jiseon Min,  Joao Ascensao, Michael M. Desai, and Stephen Martis for their helpful discussions and comments.

\bibliography{broadoffspring.bib}

\onecolumngrid
\appendix

\setcounter{equation}{0}
\section{Generalized central limit theorem}

Here, we briefly summarize the generalized central limit theorem  \citep{ Gnedenko1968-gh, Uchaikin2011-pp}. 
Suppose that each random number $u_i$ is sampled from  the Pareto distribution $P(u)=\frac{\alpha}{u^{\alpha+1}}\ (u\geq 1)$ and consider the shifted and rescaled random variable $\zeta$;
\begin{align}
\zeta = \frac{\sum_{i=1}^n u_i - a_n}{b_n},\label{zeta}
\end{align}
where $a_n$ and $b_n$ are 
\begin{align}
a_n &= 0, \quad  b_n = \bigl(\frac{\pi n}{ 2 \Gamma(\alpha) \sin \frac{\pi \alpha}{2}}\bigr)^{1/\alpha} \quad {\rm for} \  0<\alpha<1,\non\\
a_n &= n\log n, \quad b_n = \frac{\pi}{2} n \quad {\rm for} \ \alpha=1,\non\\
a_n &= \frac{\alpha}{\alpha-1} n, \quad b_n = \bigl(\frac{\pi n}{ 2 \Gamma(\alpha) \sin \frac{\pi \alpha}{2}}\bigr)^{1/\alpha}\quad {\rm for} \  1<\alpha<2,\non\\
a_n &=\frac{\alpha}{\alpha-1} n=2n, \quad b_n=(n \log n)^{1/2}\quad {\rm for} \  \alpha=2.
\label{anbn}
\end{align}
It is well-known that the distribution of  $\zeta$ is well-approximated by the $\alpha$-stable distribution, which we denote as $P_\alpha(\zeta)$. While an explicit expression of $P_\alpha(\zeta)$ is not available in general,  the characteristic function is given by 
\begin{align}
\langle e^{is \zeta} \rangle &=  \int d\zeta e^{i s \zeta } P_\alpha(\zeta)\non\\
& \sim  
\begin{cases}
\exp\bigl[- |s|^\alpha \bigl(1+i\, {\rm sgn} 
(s) \frac{2}{\pi} \log |s|)\bigr] \quad \quad (\alpha=1)\\
\exp\bigl[- |s|^\alpha \bigl(1-i\, {\rm sgn} 
(s) \tan \frac{\pi \alpha}{2}\bigr)\bigr] \quad \quad \ (\alpha \neq 1)
\end{cases}, \quad {\rm for } \ n\rightarrow \infty.
\label{ch_fn} 
\end{align}

\section{The transition density of an allele frequency $w_N(y|x)$ and  the asymptotic   dynamics for lager $N$} 
Allele-frequency change in a generation is characterized by the transition density $w_N(y|x)$, which is the probability distribution of the allele frequency $y$ at the next generation given the current allele frequency $x$. When $N$ is lager, the asymptotic  dynamics can be described by a time-continuous differential Chapman-Kolmogorov equation, which  is defined by  an advection velocity $V(x)$, diffusion coefficient $D(x)$, and jump kernel $w(y|x)$ \citep{Gardiner2009-nq}. The triplet is obtained from the transition density $w_N(y|x)$ as follows:
 \begin{align}
    w(y|x)&= \lim_{N\rightarrow \infty} \frac{w_N(y|x)}{\delta t_N}\non\\ 
     V(x)&= \lim_{N\rightarrow \infty} \frac{1}{\delta t_N} \int_{|y-x|<\epsilon} (y-x) w_N(y|x) dy \non\\
      D(x)&= \lim_{N\rightarrow \infty} \frac{1}{\delta t_N} \int_{|y-x|<\epsilon} (y-x)^2 w_N(y|x) dy, 
 \end{align}
where $\delta t_N$ is an $N$-dependent timescale, corresponding to one generation measured in units of the coalescent timescale. 
In the following, we derive the transition density $w_N(y|x)$ and the asymptotic dynamicsfor general $\alpha$ by using a similar computational technique used in \cite{hallatschek2018selection}, wherein the case of $\alpha=1$ is studied extensively. 

As mentioned in the main text, when $\alpha \leq 2$, the binomial sampling error is negligible for large $N$ compared to the stochasticity coming from broad offspring number fluctuations, and we can replace the binomial distribution  in Equation~\ref{w_N} of the main text  with the Dirac delta function;
\begin{align}
w_N(y|x) =\langle \delta(y-\frac{M}{M+W})\rangle_{M,W} =\langle \int_{-\infty}^{+\infty}  \frac{d\sigma}{2\pi }e^{i(y-\frac{M}{M+W})\sigma}\rangle_{M,W}.
\end{align}
Here $\langle \cdot\rangle_{M,W}$ means the average over $M = \sum_{i=1}^{x N} u_i$ and $W=\sum_{i=1}^{(1-x) N} v_i$. 
Using the variable $s=\frac{\sigma}{M+W}$, we can rewrite $w_N$ as
\begin{align}
w_N(y|x)& =\langle \int_{-\infty}^{+\infty}   \frac{d s}{2\pi  } (M+W)e^{-i s( M-y (M+W))}\rangle_{M,W}\non\\
& =\partial_y \int  \frac{d s}{2\pi i s}\langle e^{-i s M (1-y)+i s W y)}\rangle_{M,W}\non\\
& =\partial_y W_N(y|x).\label{app:dywn}
\end{align}
Here, 
\begin{align}
W_N(y|x) = \int_{-\infty}^{+\infty}\frac{ds}{2\pi i s } \Phi(-s(1-y);xN)\Phi(sy ;(1-x)N)
\end{align}
with
\begin{align}
\Phi(s ;n) = \langle e^{i s \sum_{i=1}^{n} u_i }\rangle . 
\end{align}

To use the properties of  the $\alpha$-stable distributions in Appendix A, we further rewrite $W_N(y|x)$ as follows:
\begin{align}
W_N(y|x) &=  \int_{-\infty}^{+\infty}\frac{ds}{2\pi i s } \Phi(-s(1-y);xN)\Phi(sy ;(1-x)N)\non\\
& =   \int_{-\infty}^{+\infty}\frac{ds}{2\pi i s } \langle e^{-i s(1-y) \sum_{i=1}^{xN} x_i}\rangle \langle e^{i s y  \sum_{i=1}^{(1-x)N} x_i} \rangle \non\\
& =   \int_{-\infty}^{+\infty}\frac{ds}{2\pi i s } \biggl\langle e^{-i s(1-y) ( b_{xN} \zeta + a_{xN})}\biggr\rangle_\zeta 
\biggl\langle e^{i s y  ( b_{(1-x)N} \zeta' + a_{(1-x)N})}\biggr\rangle_{\zeta'}\non\\
& =    \int_{-\infty}^{+\infty}\frac{ds}{2\pi i s }
e^{-is  (1-y) a_{xN} + is y a_{(1-x) N}}
 \biggl\langle e^{-i s(1-y) b_{xN} \zeta }\biggr\rangle_\zeta \biggl\langle e^{i s y   b_{(1-x)N} \zeta'} \biggr\rangle_{\zeta'}.\label{WN_before}
\end{align}
When $N$ is large, the quantities in the two brackets in the last line can be approximated by the characteristic functions of $\alpha$-stable distribution, Equation~\ref{ch_fn}, with $s\rightarrow -s (1-y)b_{xN}$ and $s\rightarrow s y b_{(1-x)N}$, respectively. Thus, when $\alpha\neq 1$, Equation~\ref{WN_before} can be computed as 
\begin{align}
W_N(y|x) & =   \int_{-\infty}^{+\infty}\frac{ds}{2\pi i s }e^{-is  (1-y) a_{xN} + is y a_{(1-x) N}}\non\\
&\quad \quad \quad\times e^{-|s|^\alpha (1-y)^\alpha b_{xN}^\alpha (1+ i {\rm sgn}(s) \tan\frac{\pi \alpha}{2}
	)} e^{-|s|^\alpha y^\alpha b_{(1-x)N}^\alpha (1- i {\rm sgn}(s) \tan\frac{\pi \alpha}{2}
	)}\non\\
& =   \int_{-\infty}^{+\infty}\frac{ds}{2\pi i s }
e^{-|s|^\alpha \{  (1-y)^\alpha b_{xN}^\alpha+ y^\alpha b_{(1-x)N}^\alpha \}}\non\\
& \quad \quad \quad \times 
e^{-is  (1-y) a_{xN} + is y a_{(1-x) N}} e^{-i |s|^\alpha  {\rm sgn}(s) \tan\frac{\pi \alpha}{2}
	\bigl( 
	(1-y)^\alpha b_{xN}^\alpha  - y^\alpha b_{(1-x)N}^\alpha
	\bigr)}\non\\
& =  \int_0^\infty \frac{ds}{\pi  s }
e^{-s^\alpha \{  (1-y)^\alpha b_{xN}^\alpha+ y^\alpha b_{(1-x)N}^\alpha \}}\non\\
& \quad \times 
\sin\biggl[
{-s \left( (1-y) a_{xN} - y a_{(1-x) N}  \right)}
 - s^\alpha   \tan\frac{\pi \alpha}{2}
\bigl( 
(1-y)^\alpha b_{xN}^\alpha  - y^\alpha b_{(1-x)N}^\alpha
\bigr)\biggr].
\label{WNgeneral}
\end{align}
In the following, we evaluate the integral expression of $W_N(y|x)$ and compute the transition density $w_N(y|x)$ from Equation~\ref{app:dywn}.

\subsection*{When $\alpha<1$}\ \\
 By using Equation~\ref{anbn},
\begin{align}
a_n =0, \quad b_n^\alpha  =\pi (2\Gamma(\al)\sin \frac{\pi \al}{2}) n  \equiv c_\al n,
\end{align}
we have
\begin{align}
W_N(y|x) =&\int_0^\infty \frac{ds}{\pi s} e^{-s^\al c_\al N ((1-y)^\al x + y^\al (1-x))}\non\\
&\times \sin [-s^\al c_\al N \tan \frac{\pi \al}{2} ((1-y)^\al x -y^\al (1-x))].
\end{align}
By setting $N c_\al s^\al  = \sigma$, $W_N(y|x)$ becomes
\begin{align}
W_N(y|x) &=\frac{1}{\al} \int_0^\infty \frac{d \sigma }{\pi \sigma} e^{-\sigma ((1-y)^\al x + y^\al (1-x))}\sin [-\sigma \tan \frac{\pi \al}{2} ((1-y)^\al x -y^\al (1-x))] \non\\
& = -\frac{\tan ^{-1}\left(\frac{\tan \left(\frac{\pi  \alpha }{2}\right) \left(x (1-y)^{\alpha }-(1-x) y^{\alpha }\right)}{(1-x) y^{\alpha }+x (1-y)^{\alpha }}\right)}{\pi  \alpha }.
\end{align}
By differentiating it with respect to $y$, we obtain
\begin{align}
w_N(y|x)& = \frac{ x(1-x) \sin (\pi  \alpha ) ((1-y) y)^{\alpha -1}}{\pi  \left(x^2 (1-y)^{2 \alpha }+(1-x)^2 y^{2 \alpha }+2 x (1-x)  \cos (\pi  \alpha ) ((1-y) y)^{\alpha }\right)}.
\end{align}
Note that this does not depend on $N$, which is consistent with the fact that the coalescent time is $\mathcal O(N^0)$ when $\alpha<1$. 

\subsection*{When $1<\alpha<2$}\ \\
 By using Equation~\ref{anbn},
\begin{align}
a_n &= \frac{\al}{\al-1}n\non\\
b_n^\al &= B_\al n, \ {\rm where} \ B_\al \equiv \frac{\pi}{2 \Gamma(\al) \sin \frac{\pi \al}{2}},
\end{align}
Equation~\ref{WNgeneral} becomes 
\begin{align}
W_N(y|x) 
&=   \int_0^\infty \frac{ds}{\pi s }
e^{- B_\al N  s^\alpha \{  (1-y)^\alpha x + y^\alpha (1-x) \}}\non\\
& \quad \quad \quad \times 
\sin[- s  \frac{\al}{\al-1} N (x-y)- s^\al   \tan\frac{\pi \alpha}{2}
B_\al N \bigl( 
(1-y)^\alpha x  - y^\alpha (1-x) 
\bigr)
].
\end{align}
By changing the variable of integration as $\sigma =N^{1/\al} s $, we have
\begin{align}
W_N(y|x) 
&=   \int_0^\infty \frac{d\sigma }{\pi \sigma }
e^{- B_\al   \sigma^\al \{  (1-y)^\alpha x + y^\alpha (1-x) \}}\non\\
& \quad \quad \quad \times 
\sin[- \sigma  \frac{\al}{\al-1} N^{1-\frac{1}{\alpha}} (x-y)- \sigma^\al  \tan\frac{\pi \alpha}{2}
B_\al  \bigl( 
(1-y)^\alpha x  - y^\alpha (1-x) 
\bigr)
].
\end{align}
By changing the variable of integration as $\sigma' =\frac{\al}{\al-1}|x-y|\sigma$ and redefining $\sigma'$ as $\sigma$, we have 
\begin{align}
W_N(y|x) =\int_0^\infty  \frac{d\sigma}{\pi \sigma} e^{-\mu_1 \sigma^\al}\sin(-{\rm sgn} (x-y) N^{1-\frac{1}{\alpha}} \sigma -\mu_2 \sigma^\alpha),
\end{align}
where
\begin{align}
\mu_1 &= B_\al (\frac{\al-1}{\al})^\al \frac{  (1-y)^\alpha x + y^\alpha (1-x) }{|x-y|^\al}\non\\
\mu_2 &=\tan \frac{\pi \alpha}{2} B_\al(\frac{\al-1}{\al})^\al \frac{(1-y)^\alpha x  - y^\alpha (1-x) }{|x-y|^\al}.
\end{align}
The transition probability $w_N(y|x)$ is given by 
\begin{align}
&w_N(y|x) = \partial_y W_N(y|x)\non\\
 &={\rm sgn} (x-y)\frac{ \partial_y \mu_1}{\pi}\int_0^\infty  d \sigma \sigma^{\al-1}e^{-\mu_1 \sigma^\al} \sin(N^{1-\frac{1}{\alpha}} \sigma+ {\rm sgn} (x-y)\mu_2 \sigma^\al)\non\\
&-\frac{ \partial_y \mu_2}{\pi}\int _0^\infty d \sigma \sigma^{\al-1}e^{-\mu_1 \sigma^\al} \cos(N^{1-\frac{1}{\alpha}} \sigma+{\rm sgn} (x-y) \mu_2 \sigma^\al).\label{app:wN_interm}
\end{align}
Consider the integral 
\begin{align}
J_\alpha = \int_0^\infty d \sigma \sigma^{\al-1}e^{-\mu \sigma^\al} e^{ i N^{1-\frac{1}{\alpha}} \sigma}
\end{align}
where $\mu = \mu_1 -i\  {\rm sgn} (x-y) \mu_2$. Then, the transition probability can be written as
\begin{align}
w_N(y|x) &={\rm sgn} (x-y)\frac{ \partial_y \mu_1}{\pi} {\rm Im} J_\al  - \frac{ \partial_y \mu_2}{\pi} {\rm Re} J_\al.  \label{wNJ}
\end{align}
From  Watson's lemma, the integral $J_\al$ can be  expressed as a series expansion;
\begin{align}
J_\alpha= \sum_{m=1}^{\infty}\frac{1}{N^{m(\al-1)}}e^{i \frac{\pi}{2}  m\al}(-\mu)^{m-1}\frac{\Gamma(m\al)}{\Gamma(m)}. \label{Jal}
\end{align}
By substituting Equation~\ref{Jal} into Equation~\ref{wNJ} and writing $\mu = |\mu| e^{i \theta}$, we obtain 
\begin{align}
w_N(y|x) &=\sum_{m=1}^\infty \frac{(-|\mu|)^{m-1}}{N^{m(\al-1)}}
\frac{\Gamma(m\al)}{\pi\Gamma(m)} \non\\ &
\biggl[ {\rm sgn}(x-y) {\partial_y \mu_1}{} \sin(\frac{\pi}{2}  m\al  +(m-1)\theta) -{\partial_y \mu_2}{}\cos( \frac{\pi}{2}  m\al +(m-1)\theta)  \biggr]
\end{align}
The leading order ($m=1$) is given by
\begin{align}
w_N(y|x) &=\frac{\Gamma(\alpha)}{N^{\al-1}}\left({\rm sgn} (x-y)\frac{ \partial_y \mu_1}{\pi} \sin\frac{\pi \alpha}{2} -\frac{ \partial_y \mu_2}{\pi} \cos\frac{\pi \alpha}{2}\right)\non\\
& = \begin{cases}
N^{1-\al} \alpha\left(\frac{\alpha -1}{\alpha }\right)^{\alpha }  \, x(1-x)\frac{  (1-y)^{\alpha -1} }{(y-x)^{\al+1}} \quad {\rm when}\ x<y \\
N^{1-\al} \alpha \left(\frac{\alpha -1}{\alpha }\right)^{\alpha } \, x(1-x) \frac{ y^{\alpha -1}}{(x-y)^{\al+1}}    \quad   {\rm when}\ x>y. 
\end{cases}
\label{jump_ck_app}
\end{align}
 Equation~\ref{w_nonmarginal} in the main text can be obtained by introducing the continuous time $\tau \equiv t/(C_\alpha N^{\alpha-1})$ where $C_\alpha\equiv \alpha\left(\frac{\alpha -1}{\alpha }\right)^{\alpha }$. Equation~\ref{V_nonmarginal} follows from the neutrality  $\frac{d}{dt}\langle x\rangle =0$. 
Note that the expansion of Equation~\ref{Jal} is possible only when  $|x-y|$ is finite, i.e., when $|x-y|>\epsilon$ where $\epsilon$ is an  $N$-independent positive constant. Although $w_N(y|x)$ in \ref{jump_ck_app} diverges as $|x-y|\rightarrow 0$, this divergence is not a problem, because  the jump term of the  asymptotic dynamics in Equation~\ref{CK_nonmarginal} can be obtained  from $w_N(y|x)$ for $|x-y|>\epsilon$  (see \cite{Gardiner2009-nq}).


\subsection*{When $\alpha=2$}\ \\

 $a_n$ and $b_n$ are given by 
\begin{align}
a_n &=\frac{\alpha}{\alpha-1} n=2n, \quad b_n=(n \log n)^{1/2}.
\end{align}

Equation~\ref{WNgeneral} then becomes 
\begin{align}
W_N(y|x) 
&= \int_0^\infty \frac{ds}{\pi s} e^{-s^2\{ (1-y)^2 x N  \log xN  + y^2 (1-x)N\log (1-x)N \}}
\times \sin\bigl(- 2s N(x-y) \bigr)\non\\
&= \int_0^\infty \frac{ds}{\pi s} e^{-s^2\left\{ 
(-2 x y+x+y^2) N\log N + \left( (1-y)^2 x \log x + y^2(1-x) \log(1-x)\right)N
 \right\}}
\times \sin\bigl(- 2s N(x-y) \bigr).\non
\end{align}
By changing the variable of integration as $\sigma =  (N\log N)^{\frac{1}{2}}s$, 
\begin{align}
W_N(y|x) &= \int_0^\infty \frac{d\sigma}{\pi \sigma} e^{-\sigma^2\left\{ 
(-2 x y+x+y^2)  + \left( (1-y)^2 x \log x + y^2(1-x) \log(1-x)\right)({\log N})^{-1}
 \right\}}
\times \sin\bigl(- 2 (\frac{N}{\log N})^{\frac{1}{2}}\sigma (x-y) \bigr)\non\\
&\approx \int_0^\infty \frac{d\sigma}{\pi \sigma} e^{-\sigma^2\
(-2 x y+x+y^2)  
 }
\times \sin\bigl(- 2 (\frac{N}{\log N})^{\frac{1}{2}}\sigma (x-y) \bigr)\non\\
&=-\frac{1}{2} \text{erf}\left((x-y) \sqrt{\frac{N}{\log (N) \left(-2 x y+x+y^2\right)}}\right),
\end{align}
where $\text{erf}(x)$ is the Gauss error function
\begin{align}
\text{erf}(x) = \frac{1}{\sqrt{\pi}} \int_{-x}^x e^{-t^2} dt.
\end{align}
By differentiating $W_N(y|x)$ with respect to $y$, we have
\begin{align}
w_N(y|x)=(\frac{N}{\log N})^{\frac{1}{2}}
\frac{1}{\sqrt{\pi}}\frac{
(1-x) x }{(-2 x y+x+y^2)^{3/2}} e^{-\frac{N (x-y)^2}{\log N  \left(-2 x y+x+y^2\right)}}.
\end{align}
Suppose that $\epsilon$ is a sufficiently small but finite constant. For $|x-y|<\epsilon$, $w_N$ can be approximated as
\begin{align}
w_N(y|x)&=(\frac{N}{\log N})^{\frac{1}{2}}
\frac{1}{\sqrt{\pi}}\frac{
1 }{(x(1-x))^{1/2}} e^{-\frac{N (x-y)^2}{\log N  \left(x(1-x)\right)}}\non\\
&= \frac{1}{\sqrt{2\pi \sigma^2}}e^{-\frac{ (x-y)^2}{2 \sigma^2}}.
\end{align}
where $2 \sigma^2 = \frac{\log N}{N} x(1-x)$. From the symmetry $y-x \rightarrow -(y-x)$ of $w_N(y|x)$, the advection term is zero. The diffusivity $D$ is given by 
\begin{align}
D = \frac{1}{\delta t_N} \int_{|x-y|<\epsilon}  dy \ (x-y)^2 w_N(y|x) =\frac{1}{\delta t_N}  \sigma^2  =\frac{1}{\delta t_N} \frac{\log N}{N} \frac{1}{2}x (1-x) = \frac{1}{2} x(1-x),
\end{align}
where we have introduced the natural timescale as $ {\delta t_N} =\frac{\log N}{N} $ and used the integral approximation
\begin{align}
 \int_{-\epsilon}^\epsilon d\Delta\  \Delta^2 \frac{1}{\sqrt{2\pi\sigma^2}} \exp(-\frac{\Delta^2}{2\sigma^2}) = \frac{1}{\sqrt{2\pi}}\sigma \left(\sqrt{2 \pi } \sigma  \text{erf}\left(\frac{\epsilon }{\sqrt{2} \sigma }\right)-2 \epsilon  e^{-\frac{\epsilon ^2}{2 \sigma ^2}}\right) \approx 
 \sigma^2.
 \end{align}
Finally, the jump kernel asymptotically vanishes on the time scale $\delta t_N$, 
\begin{align}
w(y|x) = \lim_{N\to\infty}  \frac{w_N(y|x) }{\delta t_N},
\end{align}
because for fixed $x,y$ with $|x-y|>\epsilon$, $w_N(y|x)$ becomes exponentially small as $N$ becomes large. 

Thus, in the large-$N$ limit, $\alpha=2$ corresponds to Wright-Fisher diffusion for a population of effective size 
\begin{align}
N_e &=N \log(N).
\end{align}

\subsection*{When $\alpha>2$}\ \\
In this case, since the Pareto distribution $P(u)=\frac{\alpha}{u^{\alpha+1}} \ (u\geq 1)$ has finite mean $a=\frac{\alpha}{\alpha-1}$ and finite variance $b^2=\frac{\al}{(\al-1)^2(\al-2)}$, and the large $N$ limit of the allele frequency dynamics should be described by the Wright-Fisher diffusion process. To confirm this more generally,  we consider a general distribution with finite mean and variance, namely, consider that each individual's offspring number $u_i$ is sampled from a distribution with mean $a$ and variance $b^2$. Then, from the central limit theorem, the shifted and rescaled variable 
\begin{align}
\zeta = \frac{\sum_{i=1}^n x_i -  a_n}{b_n }, {\rm where}\ a_n =a n, \ b_n = \sqrt{n }b, 
\end{align}
obeys the normal distribution $\mathcal N(0,1)$. Its characteristic function  is given by $
f(s)= \exp(- \frac{1}{2} s^2)
$. Thus, we have

\begin{align}
W_N(y|x) &\approx   \int_{-\infty}^{+\infty}\frac{ds}{2\pi i s }
 e^{-is  (1-y) a_{xN} + is y a_{(1-x) N}}
f(-s(1-y)b_{xN})f(syb_{(1-x)N} )\non\\
& = \int_{-\infty}^{+\infty}\frac{ds}{2\pi i s }
 e^{-is  (1-y) a x N + is y a (1-x)N}
f(-s(1-y)\sqrt{xN}b)f(sy\sqrt{(1-x)N}b )\non\\
&= \int_{-\infty}^{+\infty}\frac{ds}{2\pi i s }
 e^{-is  (1-y) a x N + is y a (1-x)N}
e^{-\frac{1}{2}s^2(1-y)^2{xN}b^2}e^{-\frac{1}{2}s^2 y^2{(1-x)N}b^2}.
\end{align}
By setting $\sigma = N^{1/2} s$, 
\begin{align}
W_N(y|x) &= \int_{-\infty}^{+\infty}\frac{d\sigma}{2\pi i \sigma }
 e^{-i   a (x-y) N^{1/2}{ \sigma }}
e^{-\frac{1}{2}b^2\sigma^2 ((1-y)^2{x} +y^2{(1-x)})}\non\\
&   = \int_{0}^{+\infty}\frac{d\sigma}{\pi  \sigma }
 \sin({-   a (x-y) N^{1/2}{ \sigma }})
e^{-\frac{1}{2}b^2\sigma^2 ((1-y)^2{x} +y^2{(1-x)})}\non\\
&   =-\frac{1}{2}   {\rm  erf}\left(\frac{a \sqrt{N} (x-y)}{\sqrt{2}b \sqrt{ \left(-2 x y+x+y^2\right)}}\right).
\end{align}
Thus, we obtain
\begin{align}
w_N(y|x) =\partial_y W_N(y|x) = \sqrt{N}\sqrt{\frac{1 }{2\pi }} \gamma x(1-x)\frac{ \exp \left(-\frac{\gamma^2 N (x-y)^2}{2  \left(-2 x y+x+y^2\right)}\right)}{ \left(-2 x y+x+y^2\right)^{3/2}}.
\end{align}
where $\gamma \equiv a/b$. For the Pareto distribution, $\gamma =( \frac{\al}{\al-1})/\sqrt{\frac{\al}{(\al-1)^2 (\al-2)}} = \sqrt{\alpha(\alpha-2)}$.

For $|x-y|>\epsilon$, $w_N(y|x)$ becomes exponentially small as $N$ becomes large, and so the jump term does not exist in the asymptotic dynamics; $w(y|x)=0$. 
For $|x-y|<\epsilon$, we can approximate $w_N(y|x)$ as
\begin{align}
w_N(y|x) &= \sqrt{\frac{ N\gamma^2 }{2\pi x(1-x)}} { \exp \left(-\frac{\gamma^2 N (x-y)^2}{2  x (1-x)}\right)}\non\\
&  = \frac{1}{\sqrt{2\pi \Sigma^2 }} e^{-\frac{(x-y)^2}{2 \Sigma^2}},
\end{align}
where $\Sigma^2  = \frac{ x(1-x)}{\gamma^2 N }$. 
From the symmetry $y-x \rightarrow -(y-x)$ of $w_N(y|x)$, the advection is zero. Finally, the diffusion  is evaluated as 
\begin{align}
\int_{|x-y|<\epsilon}  dy (x-y)^2 w_N(y|x)=\Sigma  \left(\Sigma \  \text{erf}\left(\frac{\epsilon }{\sqrt{2} \Sigma }\right)-\sqrt{\frac{2}{\pi }} \epsilon  e^{-\frac{\epsilon ^2}{2 \Sigma ^2}}\right)\approx \Sigma^2 =\frac{x(1-x)}{\gamma^2 N}.
 \end{align}
Thus,  by re-scaling time as  $\tau = \frac{t}{\gamma^2 N}$, we obtain 
\begin{align}
D= {x(1-x)},
\end{align}
which corresponds to the Wright-Fisher diffusion of a population of effective size $N_e =N \gamma^2 = N {\alpha(\alpha-2)}$. Notice  $N_e\rightarrow 0$ as $\alpha\rightarrow 2$, indicating that the concept of the effective population size breaks down when the variance of the offspring  distribution diverges.

\section{The transition density for the differential Chapman-Kolmogorov equation for $1<\alpha<2$}
Here we derive the short-time transition density given in  Equations~\ref{dx=cz} and \ref{ch_zeta} and determine $g(\xi)$ in the scaling ansatz given in  Equation~\ref{P_scaling}. 
\subsection{The short-time transition density}
Before discussing the CK equation in  Equation~\ref{CK_nonmarginal}, it is instructive to start from the simple diffusion equation,
\begin{align}
    \partial_\tau P(x,\tau) = D \partial_x^2  P(x,\tau),\label{example_fp}
\end{align}
with the initial condition $P(x,\tau=0)=\delta(x-x_0)$. 
 The solution of this initial value problem is given by 
  \begin{align}
    P(\Delta x,\tau) = \frac{1}{\sqrt{2\pi (2 D\tau) }}\exp\bigl ( -\frac{\Delta x^2}{ 2 (2 D \tau)}\bigr),\label{fp_sol}
\end{align}
which  is usually derived from the Laplace-Fourier transformation.
However, this solution can also be obtained by using the central limit theorem:  Equation~\ref{example_fp} is equivalent to  a Brownian motion where jumps $X\rightarrow X\pm a$ occur with rate $\frac{m}{2}$, where $a$ and $m$ are related with $D$ via $D=\frac{a^2 m}{2}$. Since  $n\approx m \tau$ jumps occur in time $\tau$,  the displacement is approximately given by $\Delta X(\tau) \approx \sum_{i=1}^n l_i$ where  $l_i = \pm a$.  Then, from the central limit theorem,  $\Delta X(\tau)$ is distributed according to the normal distribution with mean $n \langle l_i\rangle= 0$ and variance $n \langle l_i^2 \rangle = (m \tau)a^2  = 2 D \tau$, namely, Equation~\ref{fp_sol}. Note that, even if the diffusion constant depends on $x$,  the solution in Equation~\ref{fp_sol} (with $D\rightarrow D(x_0)$) is valid in short times.

Essentially the same argument can be applied to the CK dynamics, except that the generalized central limit theorem should be employed since the variance of jump sizes is divergent in the case of the CK dynamics.  Suppose that the initial density  is given by $P(x',\tau=0)=\delta(x'-x)$ (for notational simplicity, the subscript $0$ on $x$ is dropped). In the CK dynamics, the frequency change $\Delta X(\tau)=X(\tau) -x$ is caused by the bias $V(x)$ in Equation~\ref{V_nonmarginal} and by stochastic  jumps.  
The rate  of  a frequency-increasing jump and  that of  a frequency-decreasing jump  are given by 
\begin{align}
W_+(x) &= \int_{x+\epsilon}^1 w (x'|x)dx' = \frac{x}{\alpha}\left(\frac{1-x}{\epsilon}\right)^\alpha,\\
W_-(x) &= \int_{0}^{x-\epsilon} w (x'|x)dx' = \frac{1-x}{\alpha}\left(\frac{x}{\epsilon}\right)^\alpha,
\end{align} 
respectively. Therefore, the expected number $ n$  of jump events in time $\tau$ is  given by
\begin{align}
 n  = (W_- + W_+)\tau.
\end{align}
Because randomness in the number of jump events is negligible compared to that in jump sizes, 
it can be assumed that  exactly $n$ jumps occur in time $\tau$. 
Then, the displacement $\Delta X(\tau)=X(\tau) -x$ can be written as 
\begin{align}
\Delta X (\tau) =  V(x)\tau + \sum_{i=1}^n l_i,\label{disp_app}
\end{align}
where $l_i\in [-x,-\epsilon]\cup [\epsilon, 1-x]$ denotes the displacement due to the $i$-th jump. 
For small $\tau$,  $w(y|x(\tau'))\approx w(y|x)$ for $0<\tau'<\tau$, which means that $l_1,\ldots, l_n$ are  independent and identically distributed. From Equation~\ref{w_nonmarginal},   each $l_i$ is approximately sampled from the following power-law distribution, 
\begin{align}
P(l)= 
\begin{cases}
\frac{W_+}{W_+ +W_-} \frac{\epsilon^\alpha \alpha }{l^{\alpha+1}}\  \quad (l \in [+\epsilon,+ \infty) )\\
\quad \quad 0  \quad \quad \quad \quad \ \ (l \in (-\epsilon,+ \epsilon) ) \label{pl_app}\\ 
	\frac{W_-}{W_+ +W_-}\frac{\epsilon^\alpha \alpha }{|l|^{\alpha+1}}  \quad (l \in (-\infty, -\epsilon] ) 
\end{cases},
 \end{align}
 where the factor $\frac{W_+}{W_-+W_+}$ (resp. $\frac{W_-}{W_-+W_+}$) represents the probability that a given jump is frequency-increasing (resp. frequency-decreasing). 
 $P(l)$ is normalized as $\int_{-\infty}^{\infty} P(l)dl =1$. Note that, in Equation~\ref{pl_app}, the original range  $ [-x,-\epsilon]\cup [\epsilon, 1-x]$ of $l$  has been extended to $[(-\infty,-\epsilon]\cup [\epsilon, \infty)$. Under this modification, the variance $\langle x(\tau)^2 \rangle $ is no longer  well-defined. However, this modification does not alter  short-time properties of typical events,  because the presence of the boundaries at $x=0,1$ is not important for them.

 By noting that  $P(l)$ has a divergent variance and that   the number of jumps is $n\approx \frac{\tau}{\epsilon^\alpha}\gg 1$   even for small $\tau$ (as $\epsilon \rightarrow +0$),  the generalized central limit theorem states that  the sum $\sum_{i=1}^n l_i$ in Equation~\ref{disp_app} obeys an $\alpha$-stable distribution. The stable distribution is characterized by   $\langle l\rangle, \beta, \gamma$ given  below (see, for example, \cite{Uchaikin2011-pp}): The mean $\langle l \rangle$ is 
 \begin{align}
\langle l \rangle &= \frac{W_+  - W_-}{W_+ + W_-}\frac{\alpha }{\alpha-1} \epsilon  = 
 \frac{ x(1-x)^\al - x^\al(1-x)}{x^\al(1-x) + x(1-x)^\al}\frac{\alpha }{\alpha-1} \epsilon.
\label{mean_app} 
 \end{align}
Asymptotically,  $P(l)$ satisfies 
 \begin{align}
\int_l^\infty P(l')dl'  &= \frac{W_+}{W_- + W_+} \frac{\epsilon^\alpha}{l^\alpha} \equiv\frac{c_+}{l^\alpha} \quad (l \rightarrow \infty ), \non  \\
\int_{-\infty}^{l} P(l')dl'  &= \frac{W_-}{W_- + W_+} \frac{\epsilon^\alpha}{|l|^\alpha} \equiv\frac{c_-}{|l|^\alpha} \quad (l \rightarrow -\infty ).
 \end{align}
Note $c_- + c_+ = \epsilon^\alpha$.  The parameters $  \gamma$  and $\beta$ are determined from $c_\pm$;   
 \begin{align}
\gamma  &\equiv \biggl(\frac{\pi (c_+ + c_-) n}{ 2\Gamma(\al) \sin \frac{\pi \alpha}{2}}\biggr)^{\frac{1}{\alpha}} = \epsilon \biggl(\frac{\pi  n}{ 2\Gamma(\al) \sin \frac{\pi \alpha}{2}}\biggr)^{\frac{1}{\alpha}} = \biggl( \tau  \frac{\pi ( x(1-x)^\alpha + (1-x)x^\alpha )  }{ 2\Gamma(\al+1) \sin \frac{\pi \alpha}{2}}
\biggr)^{\frac{1}{\alpha}}\label{gamma_app}\\
\beta &\equiv  \frac{c_+ -c_-}{c_+ +c_-}  = \frac{W_+-W_-}{W_+ +W_-} = \frac{ x(1-x)^\al - x^\al(1-x)}{x^\al(1-x) + x(1-x)^\al} \label{beta_app}.
 \end{align}
Then, from the generalized central limit theorem,  the random variable,
 \begin{align}
Z  \equiv \frac{ \sum_{i=1}^n l_i - n \langle l \rangle }{\gamma}, \label{zeta_app}
 \end{align}
has the following characteristic function, 
 \begin{align}
\langle e^{i k  Z } \rangle = \int  e^{ik z} {P_Z(z)} dz \overset{\epsilon\rightarrow+0}= \exp\biggl[- |k|^\alpha \biggl(1 - i \beta \tan \frac{\pi \alpha}{2}  {\rm sign} k \biggr)\biggr].\label{ch_zeta_app}
 \end{align}

We can determine the characteristic function for  $\Delta x$, using Equation~\ref{ch_zeta_app} and the relation 
\begin{align}
\Delta X(\tau) =  V(x) \tau + \gamma Z + n\langle l \rangle, 
\end{align} 
which follows from Equations~\ref{disp_app} and \ref{zeta_app}. 
While $V(x)$ and $\langle l \rangle$ are divergent in the limit $\epsilon\rightarrow+0$,    we can show, by using Equation~\ref{mean_app} and $
V(x)=- \int_{|x-x'|>\epsilon} dx'(x'-x)w(x'|x)\approx\frac{1}{\epsilon^{\al-1}} \frac{1}{\alpha-1}\bigl(x^\al(1-x)- x(1-x)^\al  \bigr)$, that 
these divergent terms exactly cancel out each other. 
Therefore, the displacement is simplified as  
\begin{align}
\Delta X(\tau) = \gamma Z.  \label{dx_res_app}
\end{align} 
Equations~\ref{ch_zeta} and \ref{dx=cz}  in the main text are the same as Equations~\ref{dx_res_app} and \ref{ch_zeta_app} (with the replacement of $x\rightarrow x_0$).  By substituting this into Equation~\ref{ch_zeta_app}, we obtain the  characteristic function of the allele frequency $X(\tau)$; 
\begin{align}
\langle e^{i k  X (\tau)} \rangle = \int  e^{ik x' } {P(x',\tau|x_0)} dx' = \exp\biggl[i k x_0 - |\gamma(x_0) k|^\alpha \biggl(1 - i \beta(x_0) \tan \frac{\pi \alpha}{2}  {\rm sign} k \biggr)\biggr].
\end{align} 

\subsection{The scaling ansatz for the long-time transition density in Equation~\ref{P_scaling}}
Consider the initial distribution $P(x,\tau=0)=\delta(x-x_0)$ with $x_0\ll 1$. 
After some time, the distribution spreads  over the region $x\ll 1$ with a peak at the extinction boundary  $x=0$.  As presented in  Equation~\ref{P_scaling} of the main text, up to a constant prefactor, $P(x,\tau)$  takes the following form 
$$
P(x,\tau)\sim  \tau^{-2\eta} g(\xi),
$$
where $\eta=(\alpha-1)^{-1}$ and $\xi = \frac{x}{\tau^\eta}$. Here, we present an analytic argument to determine  $g(\xi)$.

Equation~\ref{CK_nonmarginal} can be rewritten as
\begin{equation}
    \frac{\partial P}{\partial \tau} = \int_{|\Delta |<\epsilon } d\Delta (f_\Delta(x-\Delta) P(x-\Delta,\tau) - f_\Delta(x) P(x,\tau) )+ \frac{\partial }{\partial x} \int_{|\Delta |<\epsilon } d\Delta (f_{\Delta}(x) P(x,\tau)),
\end{equation}
where $f_\Delta(x) \equiv w(x+\Delta|x)$ given by Equation~\ref{w_nonmarginal}. For $x\ll 1$, $f_\Delta(x)$  is approximately given by
\begin{equation}
   f_\Delta(x)  = \begin{cases}
   \frac{x}{\Delta^{\alpha+1}}\quad (\Delta>0)\\
    \frac{x(x+\Delta)^{\alpha-1}}{\Delta^{\alpha+1}}\quad (\Delta<0)
   \end{cases}.
\end{equation}
We substitute the ansatz $
P(x,\tau)\sim \tau^{-2\eta} g(\xi)
$ into the above CK equation. The left-hand side of the CK equation becomes
\begin{equation}
{\rm LHS}= -2 \eta \tau^{-2\eta -1} g(\xi) -\eta \tau^{-2\eta-1} g'(\xi) \xi, \label{ck_left}
\end{equation}
which is proportional to $\tau^{-2\eta-1}=\tau^{-\frac{\alpha+1}{\alpha-1}}$. The right-hand side is decomposed into  the integrals over $\Delta >0$ and those over $\Delta<0$. We can show that the former is proportional to $\tau^{-\frac{\alpha+1}{\alpha-1}}$, while the latter is proportional to $\tau^{-\frac{2}{\alpha-1}}$; 
 For example,  one of the  integrals over $\Delta >0$ is 
 \begin{align}
     \int_{\Delta>0} d\Delta f_\Delta(x) P(x) =
      \int_{\Delta>0} d\Delta \frac{x}{\Delta^{\alpha+1}} \tau^{-2\eta} g(\xi) = \tau^{-\frac{\alpha+1}{\alpha-1}}\int_{\delta>0} d\delta \frac{\xi }{\delta^{\alpha+1}}  g(\xi),\non
 \end{align}
while one of the  integrals over $\Delta <0$ is 
 \begin{align}
     \int_{\Delta<0} d\Delta f_\Delta(x) P(x) =
      \int_{\Delta<0} d\Delta \frac{x(x+\Delta)^{\alpha-1}}{\Delta^{\alpha+1}} \tau^{-2\eta} g(\xi)=  \tau^{-\frac{2}{\alpha-1}}\int_{\delta>0} d\delta \frac{\xi }{\delta^{\alpha+1}}  g(\xi),\non 
 \end{align}
 where we have changed the integration variable from $\Delta$ to $\delta = \frac{\Delta}{\tau^{\eta}}$. 
Since the extinction time for the initial frequency $x_0 \ll 1$ is much shorter than the coalescent timescale, we can assume $\tau \ll 1$, which implies that the integrals over $\Delta>0$  are negligible compared to those over $\Delta>0$. By evaluating the integrals over $\Delta>0$ using the scaling form of $P(x,\tau)$ and comparing them with Equation~\ref{ck_left}, we have
\begin{align}
-\eta \biggl(2g(\xi)+  \xi g'(\xi)\biggr)   =
\int_0^\infty\frac{d\delta}{\delta^{\alpha+1}} \biggl((\xi-\delta) g(\xi - \delta) \Theta( \xi -\delta) - \xi g(\xi) + \delta \frac{d}{d \xi }(\xi g(\xi))\biggr),\label{CK_chi}
\end{align}
where $\Theta(\cdot)$ is the Heaviside step function. Note that the variable of integration has been changed from $\Delta$ to $\delta = \frac{\Delta}{\tau^{\eta}}$, and  the upper bound in the integral has been extended into $+\infty$, to make the equation analytically tractable. It is convenient to express Equation~\ref{CK_chi} in terms of $\Phi(\xi) \equiv \xi g(\xi)$; 
\begin{align}
-\eta \biggr(\frac{\Phi(\xi)}{\xi} + \Phi'(\xi)\biggl)  =
\int_0^\infty\frac{d\delta}{\delta^{\alpha+1}} \biggl(\Phi(\xi - \delta) \Theta( \xi -\delta) -\Phi(\xi) + \delta \Phi'(\xi)\biggr).\label{CK_Phi}
\end{align}

The solution of the integro-differential equation in Equation~\ref{CK_Phi} can be obtained as a series expansion. 
Assume, for small $\xi$, 
\begin{align}
\Phi(\xi)  = c_1 \xi^\beta+ \cdots\label{Phibeta},
\end{align}
where $c_1$ is a normalization and the exponent of the leading term is denoted by $\beta \in (0,1)$. Here,  $\beta<1$ is required since we are considering the situation where  $P(x,\tau)$ is monotonically decreasing in $x$, while $\beta>0$ is required to make $P(x,\tau)$ normalizable. By substituting Equation~\ref{Phibeta} into Equation~\ref{CK_Phi}, we have
\begin{align}
- \frac{\beta+1}{\alpha-1}\frac{1}{\xi^{1-\beta}} + \cdots  =  \frac{\Gamma(-\alpha)\Gamma(1+\beta)}{\Gamma(1-\alpha+\beta)} \frac{1}{\xi^{\alpha-\beta}} + \cdots. 
\end{align}
Since $\frac{1}{\xi^{1-\beta}} \ll \frac{1}{\xi^{\alpha-\beta}}$ for $\xi\ll 1$, in order for the two sides to be balanced,   the coefficient $\frac{\Gamma(-\alpha)\Gamma(1+\beta)}{\Gamma(1-\alpha+\beta)}$ needs to be zero, which is possible only when $\Gamma(1-\alpha+\beta)$ diverges. Since $1<\alpha<2$ and $0<\beta<1$, we can conclude $\beta = \alpha-1$. Therefore, the leading term of $g(\xi)$ is given by
\begin{align}
    g(\xi) = \frac{c_1}{\xi^{2-\alpha}}+\cdots \quad  (\xi\ll 1 ).
\end{align}

More generally, by starting from the ansatz,
\begin{align}
\Phi(\xi) = \sum_{m=1}^\infty c_m \xi^{(\alpha-1)m},
\end{align}
 the coefficients $c_2, c_3,\cdots$ can be determined iteratively:
\begin{align}
c_{m+1} = - \frac{1 + (\alpha-1)m }{\alpha-1}\frac{\Gamma(m(\alpha-1))}{\Gamma(-\alpha) \Gamma(m(\alpha-1)+\alpha)} c_m \quad (m=1,2,\cdots ).
\end{align}
By using this iteratively, we can express $\Phi(\xi)$ as
\begin{align}
\Phi(\xi) =c_1 \sum_{m=1}^\infty (-1)^{m+1}\frac{ \Gamma (\alpha +1)   \left(\frac{\alpha }{\alpha -1}\right)_{m-1}}{\alpha (\alpha -1)^{m} \Gamma (-\alpha )^{m-1}\Gamma (m+1) \Gamma (m (\alpha -1))} \xi ^{(\alpha -1) m},\label{Phi_series}
\end{align}
where $(\frac{\alpha}{\alpha-1})_{m-1}$ is the Pochhammer symbol, $(q)_n = \Gamma(q+n)/\Gamma(q)$. The analytic expression of  $g(\xi)$ can be obtained from this  using $g(\xi)=  \frac{\Phi(\xi)}{\xi}$.   

On the other hand, for $\xi\gg 1$, we expect that $g(\xi)$ decreases in the same way  as the offspring distribution does; 
\begin{align}
   g(\xi)  \sim\frac{1}{\xi^{\alpha+1}}+\cdots \quad (\xi\gg 1 ).
\end{align}
Therefore, we expect there is a crossover point $\xi_{\rm c}$ such that $g(\xi)  \sim\frac{1}{\xi^{2-\alpha}}+\cdots$ for $\xi\ll \xi_{\rm c}$ and   $g(\xi)  \sim\frac{1}{\xi^{\alpha+1}}+\cdots$ for $\xi\gg \xi_{\rm c}$. The scaling form  for $\xi\gg \xi_{\rm c} $  can indeed be confirmed by 
 considering the following ansatz for $\Phi(\xi)$,
\begin{align}
    \Phi(\xi) =\begin{cases}
    c_1 \xi^{\alpha-1}+\cdots \quad (\xi < \xi_{\rm c}) \\ 
    c' \xi^{-\alpha'}+\cdots \   \quad \ (\xi > \xi_{\rm c}) \label{Phicrossover}
    \end{cases},
\end{align}
where $c'$ is a normalization and $\alpha'$ is an exponent to be determined. Substituting this ansatz into Equation~\ref{CK_Phi}, we can show  $\alpha' =\alpha$, leading to  $g(\xi) \sim\frac{1}{\xi^{\alpha+1}}+\cdots$ for $\xi>\xi_{\rm c}$.

\begin{figure}[htbp]
\centering
  \includegraphics[width=10cm]{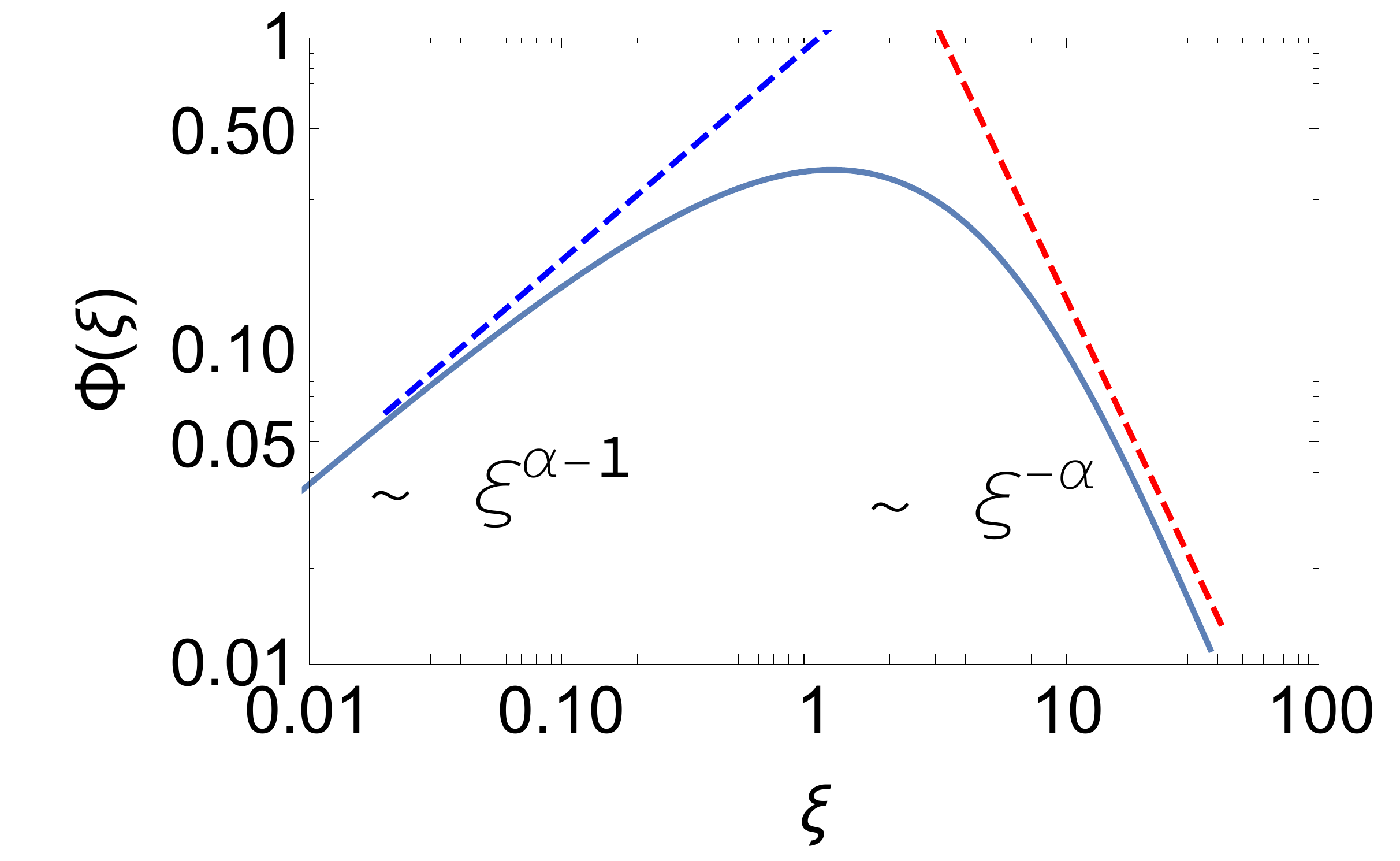}
  \caption{The infinite series in Equation~\ref{Phi_series} is  evaluated numerically by truncating at $m=150$ and using the van Wijngaarden transformation (solid line). $\alpha=1.7$ is used. The dashed blue and red lines represent the asymptotic behaviors given in Equation~\ref{Phicrossover}. }
 \label{fig:Phi}
\end{figure}

 Finally, we remark that, while Equation~\ref{Phi_series} is derived assuming  $\xi\ll 1$, the series converges for any $\xi>0$.  This indicates that the scaling form $g(\xi) \sim\frac{1}{\xi^{\alpha+1}}+\cdots$ for large $\xi$  should directly follow from a resummation of the infinite series in Equation~\ref{Phi_series}. In fact, numerical evaluation of a finite truncation of  the series indicates the crossover behavior   Equation~\ref{Phicrossover} (see Figure~\ref{fig:Phi}).

\section{From Lambda-Fleming-Viot Generator to differential Chapman-Kolmogorov equation}
In Appendix B, the jump density $w(y|x)$ is derived  from the generalized Wright-Fisher sampling, Equation~\ref{w_N} in the main text. 
Here, we present another more formal derivation of the jump density $w(y|x)$   for $1<\alpha<2$. See \cite{hallatschek2018selection} for the  case  $\alpha=1$. 

\subsection{Jump density for  general  $\Lambda$ measure}
The backward generator of the $\Lambda$ coalescent process for the biallelic model (see e.g.  \cite{griffiths2014lambda,etheridge2010coalescent}) is given by
\begin{align}
{\mathcal L} G_\tau (x|x_0) =\int_0^1\biggl( x_0 G_\tau (x| x_0 + (1-x_0)\lambda) 
-G_\tau(x|x_0) +(1-x_0) G_\tau(x| x_0 - x_0 \lambda)
 \biggr)  \frac{\Lambda(d\lambda)}{\lambda^2}.
\end{align}
This can be rewritten as a sum of  two terms:
\begin{align}
{\mathcal L} G_\tau (x|x_0)=A+B,
\end{align}
where
\begin{align}
A& = x_0 \int_0^1 \biggl(G_\tau(x|x_0 + (1-x_0)\lambda )-G_\tau(x|x_0) -(1-x_0) \lambda \partial_{x_0} G_\tau(x|x_0) \biggr) \frac{\Lambda(d\lambda)}{\lambda^2},\\
B &= (1-x_0) \int_0^1\biggl(  G_\tau(x|x_0 -x_0\lambda)-G_\tau(x|x_0) +x_0 \lambda \partial_{x_0} G_\tau(x|x_0)  \biggr)\frac{\Lambda(d\lambda)}{\lambda^2}.
\end{align}

We introduce the integration variable $x'\equiv x_0 + (1-x_0)\lambda$ for $A$ and  $x'\equiv x_0 - x_0 \lambda$ for $B$ respectively.  By writing 
\begin{align}
 \frac{\Lambda(d\lambda)}{\lambda^2}  = \frac{l(\lambda)}{\lambda^2}d\lambda,
\end{align} 
$A$ and $B$ become
\begin{align}
A& = x_0 (1-x_0)\int_{x_0}^1 \biggl(G_\tau(x|x' )-G_\tau(x|x_0) -(x'-x_0) \partial_{x_0} G_\tau(x|x_0) \biggr) \frac{l( \frac{x' -x_0}{1-x_0})}{(x'-x_0)^2}  {dx'}, \\
B &= x_0 (1-x_0) \int_0^{x_0} \biggl( G_\tau(x|x' )-G_\tau(x|x_0) -(x'-x_0) \partial_{x_0} G_\tau(x|x_0) \biggr)\frac{l(\frac{x_0 -x'}{x_0})}{(x'-x_0)^2} {dx'}.
\end{align}

Defining the jump kernel $w(x|x_0)$ as
\begin{align}
w(x'|x_0) =
\begin{cases}
 \frac{x_0(1-x_0)}{(x'-x_0)^2} l( \frac{x' -x_0}{1-x_0} ) \quad (x' > x_0),\\
 \frac{x_0(1-x_0)}{(x'-x_0)^2} l( \frac{x_0 -x'}{x_0})  \quad (x' < x_0),
 \end{cases}
\end{align}
we can formally rewrite the generator as
\begin{align}
{\mathcal L} G_\tau (x|x_0) =V(x_0) \partial_{x_0} G_\tau(x|x_0) + 
 {\rm PV} \int_0^1 w(x'|x_0) [G_\tau(x|x') - G_\tau(x|x_0) ] dx',
\end{align}
where
\begin{align}
V(x_0)  =- {\rm PV} \int_0^1 dx' w(x'|x_0) (x'-x_0).
\end{align}

\subsection{When the measure is the Beta distribution Beta$(\alpha,2-\alpha)$:}
We take the Beta$(\alpha,2-\alpha)$ distribution as the $\Lambda$ measure, which corresponds to the descendant distribution considered in this study, $\sim 1/u^{1+\alpha}$:
\begin{align}
 \frac{\Lambda(d\lambda)}{\lambda^2} =\frac{l(\lambda) d\lambda}{\lambda^2}=\frac{ \lambda^{+1-\alpha}{(1-\lambda)^{\alpha-1}}}{B(\alpha,2-\alpha)} \frac{d\lambda}{\lambda^2}=\frac{ \lambda^{-1-\alpha}{(1-\lambda)^{\alpha-1}}}{B(\alpha,2-\alpha)} d\lambda.
\end{align}
With this measure,  $A$ and $B$ become
\begin{align}
A& = \frac{x_0 (1-x_0)}{B(\alpha,2-\alpha)}\int_{x_0}^1{dx'}\,  \biggl[G_\tau(x|x' )-G_\tau(x|x_0) -(x'-x_0) \partial_{x_0} G_\tau(x|x_0) \biggr] (x'-x_0)^{-1-\alpha}(1-x')^{\alpha-1},   \\
B &= \frac{x_0 (1-x_0)}{B(\alpha,2-\alpha)} \int_0^{x_0}\,{dx'} \biggl[ G_\tau(x|x' )-G_\tau(x|x_0) -(x'-x_0) \partial_{x_0} G_\tau(x|x_0) \biggr] (x_0-x')^{-1-\alpha}x'^{\alpha-1}.
\end{align}
Note that the integrals $A$ and $B$ are convergent for $\alpha\in (0,2)$, because, near $x' \sim x_0$, the terms inside  $[\cdots]$  are $ {\mathcal O }\bigl((x'-x_0)^2\bigr)$ and so the integrands are ${\mathcal O }\bigl(|x'-x_0|^{1-\alpha}\bigr)$. The jump kernel is given by
\begin{align}
w(x'|x_0) =
\begin{cases}
\frac{x_0 (1-x_0)}{B(\alpha,2-\alpha)} (x'-x_0)^{-1-\alpha}(1-x')^{\alpha-1} \quad (x' > x_0)\\
\frac{x_0 (1-x_0)}{B(\alpha,2-\alpha)}  (x_0-x')^{-1-\alpha}(x')^{\alpha-1} \quad (x' < x_0).
 \end{cases}
\end{align}

When $1<\alpha<2$, this density agrees with Equation~\ref{w_nonmarginal} of the main text (up to a proportionality constant). The advection is given by 
\begin{align}
V(x_0)  &=- {\rm PV} \int_0^1 dx' w(x'|x_0) (x'-x_0)\non \\
&=\frac{x_0(1-x_0)}{B(\alpha,2-\alpha)}\biggl(  \int_0^{x_0-0}dx' (x_0 -x')^{-\alpha}x'^{\alpha-1} - \int_{x_0+0}^1 dx' (x'-x_0)^{-\alpha}(1-x')^{\alpha-1} \biggr). \label{vx0}
\end{align}
Note that, when $\alpha>1$,  the limit ${\rm lim}_{\epsilon\rightarrow 0} \int_0^{x_0-\epsilon} + \int_{x_0+\epsilon}^1$ in Equation~\ref{vx0} does not exist, although this divergence is rather formal since there exists  a natural cutoff $\epsilon \sim \frac{1}{N}$ for a finite-size population.

\section{Analytic results in the marginal case $\alpha=1$}
Although  the main target of our present study is the  case of  $1<\alpha<2$, we here provide  analytical results for $\alpha=1$, which have not been derived before.
\subsection{Site frequency spectrum in the presence of genuine selection}
The transition density for $\al=1$ in the presence of natural selection is derived in \cite{hallatschek2018selection} (see \cite{kosheleva2013dynamics} for  neutral case). In $x$ space, it is given by
\begin{align}
G^\sigma(x, \tau|x_0) =\frac{1}{2\pi x(1-x)}\frac{  \sin \pi   \eta}{\cos \pi  \eta + \cosh [\eta\log\frac{e^\sigma x}{1-x} -\log \frac{e^\sigma x_0}{1-x_0}]},\label{app:gsigma}
\end{align}
where $\eta \equiv e^{-\tau}$ and $\sigma$ is the selective advantage (there is an erratum in Equation~38 in \cite{hallatschek2018selection}).

For the purpose of computing the site frequency spectrum (or, equivalently, the mean sojourn time), we set $x_0=1/N$. Since we are considering the large $N$ limit, the denominator of Equation~\ref{app:gsigma} can be rewritten as 
\begin{align}
\cos \pi \eta + \cosh [\eta\log\frac{e^\sigma x}{1-x} -\log \frac{e^\sigma x_0}{1-x_0}] &\approx \frac{1}{2} \exp [\eta\log\frac{ x}{1-x} -\sigma(1-\eta)  + \log N]\non\\
&  = \frac{1}{2} N (\frac{x}{1-x})^\eta e^{-\sigma(1-\eta)}.
\end{align}
Thus, the transition density for $x_0=\frac{1}{N}$ can be written as 
\begin{align}
G^\sigma(x, \eta|x_0=\frac{1}{N}) = \frac{ e^\sigma}{\pi N x(1-x)}\frac{\sin \pi \eta}{(\frac{x}{1-x} e^\sigma)^\eta}.
\end{align}
Near the boundaries, this can be approximated as
\begin{align}
G^\sigma(x,\eta| x_0=\frac{1}{N}) = 
\begin{cases}
\frac{e^\sigma}{\pi N x}\frac{\sin \pi \eta}{(x e^\sigma)^\eta}\quad (x \ll  1)\\
\frac{e^{\sigma }}{\pi N (1-x)}( (1-x)e^{-\sigma})^\eta {\sin \pi \eta}
\quad (1-x \ll  1).
\end{cases}
\end{align}
%

The site frequency spectrum is given by $f_{\text{SFS}}(x) = N \mu  \times t(x)$, where $\mu$ is the mutation rate per generation, and  $t(x)$ is the mean sojourn time density, which is given by
\begin{align}
t(x)& = \int_0^\infty dt\, G^\sigma(x,\tau| x_0=\frac{1}{N}) = \int_0^1 \frac{d \eta}{\eta}G^\sigma(x,\eta|x_0)\non\\
& = 
\begin{cases}
\frac{e^\sigma}{\pi N x}\int_0^1  \frac{d\eta}{\eta}\frac{\sin \pi \eta}{(x e^\sigma)^\eta}\quad (x\ll 1)\\
 \frac{e^{\sigma}}{ \pi N (1-x)}\int_0^1\frac{d\eta}{\eta} \sin(\pi \eta)\, ((1-x) e^{-\sigma})^\eta \quad (1-x \ll  1).\\
 \end{cases}\label{sojourn}
\end{align}

Next, we compute the integrals in Equatiion~\ref{sojourn}, asymptotically close to the absorbing boundaries (see Equation~\ref{sfs_approx_a=1}
 for the final results). 
To evaluate Equation~\ref{sojourn} for $x\ll 1$, we first consider the  integral,
\begin{align}
I_\epsilon &=
\int_0^1 {d\eta} \exp f(\eta).
\end{align}
When $f(\eta)$ has a sharp peak at $\eta=\eta^*$,  we  approximate this  integral as
\begin{align}
I_\epsilon \approx e^{f(\eta^*)} \sqrt{\frac{2\pi}{|f''(\eta^*)|}}.
\label{sd_approx}
\end{align}
In our case, 
\begin{align}
f(\eta) &=-\log \eta - \log( \epsilon)\, \eta + \log \sin \pi\eta
\end{align}
where $\epsilon = x e^\sigma$.
 $f(\eta)$ takes the maximum value at $\eta=\eta^* \approx 1 +\frac{1}{\log \epsilon}$ \footnote{$\eta^*$ is obtained from   
$0=
f'(\eta^*)  = -\frac{1}{\eta^*} - \log( \epsilon) + \frac{\pi }{\tan \pi \eta^*}\approx - \log( \epsilon) + \frac{\pi }{\tan \pi \eta^*}\approx  - \log( \epsilon) + \frac{1 }{1-\eta^*}.
$
}.
At $\eta=\eta^*$, $f(\eta^*)\approx  -\log \epsilon -1 + \log  \frac{-\pi}{\log\epsilon} $ \footnote{Although  the magnitudes of $-1$ and $\log  \frac{-\pi}{\log\epsilon}$  are small compared to  $-\log \epsilon $, we need to retain these two terms because $f(\eta^*)$ contributes to $I_\epsilon$ through $e^{f(\eta^*)}$.  }, and $f''(\eta^*)\approx 1-\frac{\pi^2}{\sin^2 (\frac{\pi}{\log \epsilon})} \approx {\log^2 \epsilon}$. The saddle-point evaluation in Equation~\ref{sd_approx} is precise when $\epsilon\ll 1$.  By using these expressions, $I_\epsilon$ can be evaluated as
\begin{align}
I_\epsilon   \approx \frac{ \sqrt{2\pi}}{-\log \epsilon} e^{-1}\frac{-\pi}{\epsilon \log \epsilon } = \frac{\sqrt{2\pi} \pi e^{-1}}{ \epsilon \log^2 \epsilon }.
\end{align}
By setting $\epsilon = x e^\sigma$,  we find
\begin{align}
f_{\rm SFS}(x\approx 0) \sim {\mu  }\frac{\sqrt{2\pi} e^{-1} }{(x (\log x + \sigma))^2}\sim {\mu }\frac{\sqrt{2\pi} e^{-1} }{(x \log x )^2}\propto  {\mu }\frac{1 }{(x \log x )^2}.\label{sfs0}
\end{align}

Next, to evaluate Equation~\ref{sojourn} for the high-frequency end, we consider the following integral
\begin{align}
I'_\epsilon = \int_0^1 \frac{d\eta}{\eta} \sin(\pi \eta) \epsilon^\eta.
\end{align}
When $\epsilon\ll1$, the integrand takes the maximum value at the boundary $\eta=0$. Thus, 
\begin{align}
I'_\epsilon \approx \int_0^1 {d\eta} \ \pi  \epsilon^\eta = \frac{\pi(-1+\epsilon)}{\log \epsilon} \approx \frac{-\pi }{\log\epsilon}.
\end{align}
By setting $\epsilon = (1-x)e^{-\sigma}$,  we find 
\begin{align}
f_{\rm SFS}(x\approx 1) \sim -  \mu \frac{e^{\sigma}}{ (1-x)(\log (1-x) - \sigma) } \sim -  \mu \frac{e^{\sigma}}{ (1-x)\log (1-x) }. \label{sfs1}
\end{align}

In summary,  the SFS in Equation~\ref{sojourn}  is given by  
\begin{align}
f_{\text{SFS}}(x)& \sim \begin{cases}
 {\mu }\frac{1 }{(x \log x )^2}
\quad\quad  ({\rm for }  \ x\ll 1 , e^{-\sigma})\\
-  \mu \frac{e^{\sigma}}{ (1-x)\log (1-x) } \quad ({\rm for}\ 1-x \ll 1, e^\sigma).
 \end{cases}\label{sfs_approx_a=1}
\end{align}
Note that the dependence on $\sigma$ disappears when $x\ll 1$. Figure~\ref{fig:sfs} shows the plots of the SFS. 

\begin{figure}[t]
\centering
  \includegraphics[width=10cm]{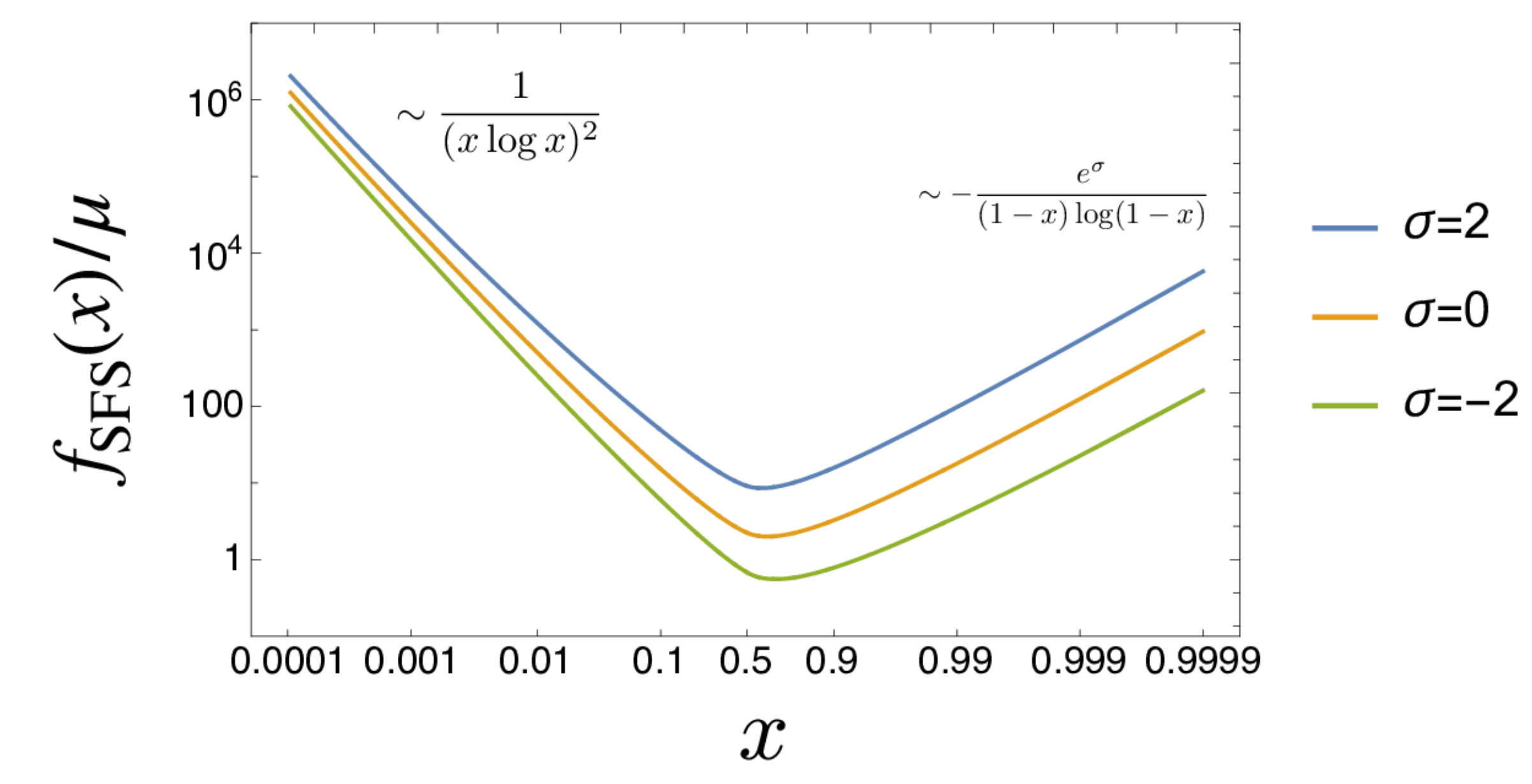}
  \caption{The SFS $f_{\text{SFS}}(x)/\mu$  when $\al=1$ for  the selective advantage $\sigma=-2, 0, 2$. $f_{\text{SFS}}(x)$ is obtained by  numerically evaluating  the exact   expression of $t(x)$ in the first line of Equation~\ref{sojourn}. As $x\rightarrow 0$,  $f(x)$ becomes independent of $\sigma$. Near $x=1$, while the magnitude of $f(x)$ depends on $\sigma$, the scaling behavior (slope in the log-log plot) does not. See Equation~\ref{sfs_approx_a=1}.}
 \label{fig:sfs_a=1}
\end{figure}

For comparison, we write the site frequency spectrum for the Wright-Fisher model ($\al \geq 2$)  (see, for example, \cite{crow1970introduction, evans2007non});
\begin{align}
f_{\text{SFS}}^{\text{WF}}(x) = \theta \frac{e^{2\sigma} (1- e^{-2 \sigma(1-x)})}{(e^{2\sigma}-1)x(1-x)}.
\end{align}
The asymptotic forms near the boundaries are given by
\begin{align}
  f_{\text{SFS}}^{\text{WF}}(x) \approx 
  \begin{cases}
  \theta \frac{1}{x}  \quad ( {\rm for} \ x\ll1) \non\\
   \theta \sigma (1+\coth \sigma )(1+(\sigma-1)(x-1)), \quad ({\rm for } \ 1-x\ll 1, |\sigma| (1-x) \ll 1) \non\\
  \end{cases}
\end{align}
where we have expanded the SFS around $x=1$ up to the sub-leading order. 
For  a sufficiently strong selection ($\sigma>1$),  the SFS increases with $x$ at the high-frequency end. However, unlike the case of $\al<2$,   the increase is not strong and the SFS approaches the constant  $\sigma (1+{\rm coth} \sigma )$ as $x \rightarrow  1$.

\subsection{Dynamics of the median of allele frequencies}
When $\alpha=1$, we can derive a simple differential equation that described the median of trajectories. 
In the logit space, the transition density is given by 
\begin{align}
G(\psi, \rho | \psi_0)  = \frac{\sin \pi \rho}{2 \pi \{ \cos \pi \eta + \cosh \left[\rho(\psi + \sigma) - (\psi_0 + \sigma)\right] \} }
\end{align}
where $\rho=e^{-\tau}$. 
The median $\Psi^{\text{med}}$ (at a given time point $\rho$) is characterized by
\begin{align}
\int_{-\infty}^{\Psi^{\text{med}}} G(\psi, \rho | \psi_0)d\psi=\frac{1}{2}.
\end{align}
From the symmetry of cosh, the median is  given by the peak of the transition density;
\begin{align}
\Psi^{\text{med}}=-\sigma + \frac{1}{\rho}(\psi_0 + \sigma).\label{psimed}
\end{align}
By differentiating Equation~\ref{psimed} with respect to $\rho$ and eliminating $\psi_0$, we obtain
\begin{align}
\frac{d}{d\rho}\Psi^{\text{med}} &= -\frac{1}{\rho^2} (\psi_0 + \sigma)= -\frac{1}{\rho} (\Psi^{\text{med}}+\sigma).
\end{align}
Noting that $\frac{d}{dt} = - \rho\frac{d}{d\rho}$, we find
\begin{align}
\frac{d}{d\tau}\Psi^{\text{med}}=\Psi^{\text{med}}+\sigma.\label{odepsimed}
\end{align}

Since the median is invariant under a coordinate transformation, the median $X^{\text{med}}$ in the $x$ space is simply related with $\Psi^{\text{med}}$ via the logit transformation, $
 \log \frac{X^{\text{med}}}{1-X^{\text{med}}}=\Psi^{\text{med}}$. 
 By differentiating this with respect to time and using Equation~\ref{odepsimed}, we obtain
\begin{align}
\frac{d}{d\tau}X^{\text{med}} = X^{\text{med}}(1-X^{\text{med}}) (\log\frac{X^{\text{med}}}{1-X^{\text{med}}}+\sigma).
\end{align}

\subsection{Allele frequency dynamics conditioned on fixation}
By using  Bayes'  theorem, the probability distribution of the allele frequency   conditioned on fixation can be written as
\begin{align}
P(x , \tau|x_0,{\rm fixation}) &= P(x, \tau,  {\rm  fixation}|x_0) \times  \frac{1}{P({\rm  fixation} |x_0)} \\
& = P(x, \tau |x_0) \times \frac{P({\rm  fixation} |x)}{ P({\rm  fixation} |x_0)}.\label{Pcond}
\end{align}
The fixation probability for the initial frequency $x_0$ is given by (see \cite{hallatschek2018selection})
\begin{align}
P({\rm  fixation}|x_0) =\frac{x_0 e^{\sigma}}{1+x_0 (e^{\sigma }-1)}.\label{Pfix_a=1x0}
\end{align}
In particular, the fixation probability of a single mutant is given by
\begin{align}
P({\rm  fixation}|x_0 =\frac{1}{N} ) \sim \frac{1}{N^{1-s}}.\label{Pfix_a=1}   
\end{align}
By using Equation~\ref{Pfix_a=1x0}, the conditioned probability in Equation~\ref{Pcond} is computed as 
\begin{align}
P(x , t |x_0,{\rm  fixation})& =\frac{1}{2\pi x(1-x)}\times \frac{x e^{\sigma }}{1+x (e^{\sigma }-1)} \times \frac{1+x_0 (e^{\sigma }-1)}{x_0 e^{\sigma }}\non\\
&\times \frac{  \sin \pi   \rho}{\cos \pi  \rho + \cosh [\rho\log\frac{e^\sigma x}{1-x} -\log \frac{e^\sigma x_0}{1-x_0}]}\non\\
& = \frac{1}{2 \pi x_0 (1-x)}\frac{1+x_0(e^\sigma-1)}{1+x(e^\sigma-1)}\frac{  \sin \pi   \rho}{\cos \pi  \rho + \cosh [\rho\log\frac{e^\sigma x}{1-x} -\log \frac{e^\sigma x_0}{1-x_0}]}.
\end{align}

\section{Site frequency spectra in presence of selection}
\begin{figure}[t]
\begin{center}
  \includegraphics[width=9cm]{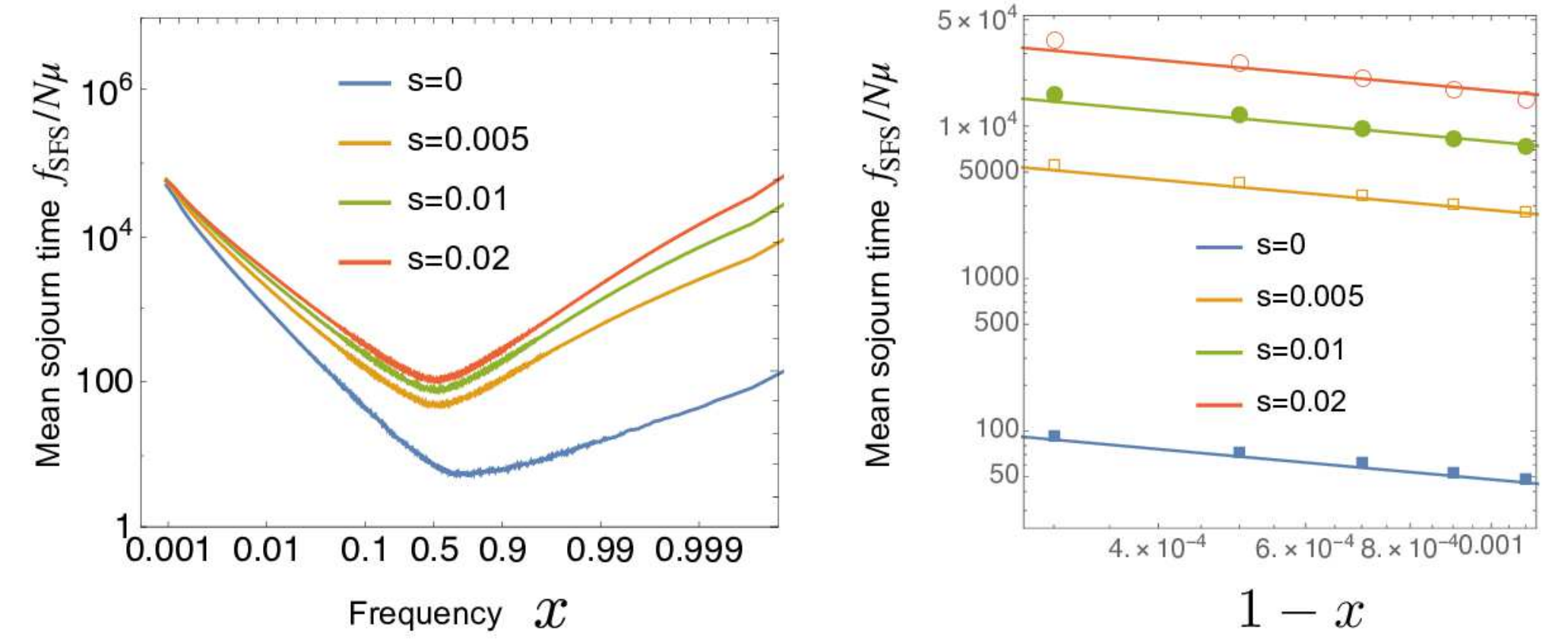}
  \caption{Left: The SFS under positive selection $s=0,\ 0.005,\ 0.01,\ 0.02$. $\al=1.5$ and $N=10^6$. Right: The SFS near $x=1$.  The straight lines are drawn assuming $SFS(x)\propto 1/(1-x)^{2-\al}$. The slope is almost independent of $s$.  
  }
 \label{fig:sfs_sel}
\end{center}
\end{figure}

Here, we  argue the effect of the genuine selection on the SFS by using the effective bias when $1<\alpha<2$. As discussed in the main text, there is a crossover point $x_c$, shown in Equation~\ref{xc}, below which the selection is negligible compared to the effective bias (see  Figure~\ref{fig:balance}). Thus, we can expect that the SFS becomes independent of the selective advantage $\sigma$ for a sufficiently small frequency $x$.
Similarly, for the high-frequency end $1-x\ll 1$, the selection is  negligible compared with the effective bias. Therefore, we expect that $f_{\text{SFS}}(x) \sim \frac{1}{V_{\text{eff}}(x)}\propto (1-x)^{-\al+2}$ even in  the presence of natural  selection. In particular, the exponent is independent of $\sigma$. Figure~ \ref{fig:sfs_sel} shows the numerical results when $\al=1.5$. As $x$ approaches $0$, the SFS becomes independent of the selective advantage $\sigma$. For frequent variants $1-x\ll 1$, the SFS can be fitted well by $(1-x)^{-\al+2}$, while the magnitude of the SFS increases with $\sigma$. A similar result can be obtained analytically when $\al=1$ (see Appendix E).

\section{Derivation of the rate of adaptation  in Equation~\ref{vel} of the main text}
Here, we conjecture the rate of adaptation for an asexual population  with a broad offspring  distribution ($1<\alpha<2$) in the clonal-interference regime, using   a self-consistency condition argument described in \citep{Desai2007-sm}. 

We assume that mutations have a fixed effect  $s$ much larger than the mutation rate $\mu_{\rm B}$ at which they arise. 
First, we consider the dynamics of the fittest sub-population that becomes established at the nose of the fitness wave. 
We can estimate the size of the sub-population when established from the establishment  probability   of a single fittest mutant;
\begin{align}
 N_{est}\sim \frac{1}{P_{\text{fix}}( qs)},
\end{align}
where $q s $ ($q\in {\mathbb N}$) is the fitness lead of the sub-population compared with the mean of the whole population, and  the fixation probability is given by  Equation~\ref{Pfix}, $P_{\text{fix}}\sim {(q s)^{\frac{1}{\al-1}}}$.  
In the time this sub-population is seeded and becomes established, the mean fitness should  increase by $s$. This implies that, after its establishment, this sub-population  will initially grow  exponentially at rate $(q-1)s$. The growth rate will slow down to $0$ when  it fixes. Therefore, the time from establishment to fixation  can be estimated as
\begin{align}
t_{\text{fix}}& \sim \frac{1}{(q-1)s/2}\ln \frac{N}{N_{est}}= \frac{1}{(q-1)s/2}\ln N P_{\text{fix}}(qs)
\end{align}
where $(q-1)s/2$ is its average growth rate between the establishment and fixation. Thus, the rate of adaptation is given by
\begin{align}
    R=\frac{(q-1)s}{t_{\text{fix}}}\sim  \frac{((q-1)s)^2 }{ 2 \ln N P_{\text{fix}}(qs)}.
    \label{v1}
\end{align}

Second, we focus on  successive events of establishments at  the edge of the fitness wave.  We define $ t_{est}$ as the mean time interval between two successive establishments.  An established sub-population grows like $n(t) \sim N_{est} e^{(q-1)st}$
, from which  the next event of establishment is produced with rate $ n(t)\mu_{\rm B} P_{\text{fix}}(qs)$. Therefore, $t_{est}$ can be estimated from
\begin{align}
\mu_{\rm B} P_{\text{fix}}(qs)  \int_0^{ t_{est}}  n(t)dt \approx 1,
\end{align}
which leads to $ t_{est} \sim \frac{1}{(q-1)s}\ln[\frac{s}{\mu_{\rm B}}]$. Since   the nose of the fitness wave advances at a speed $R=\frac{s}{ t_{est}}$, we have
\begin{align}
    R=\frac{s}{ t_{est}}\sim \frac{(q-1)s^2}{\ln \frac{s}{\mu_{\rm B}}}.\label{v2}
\end{align}

By comparing Equations~\ref{v1} and \ref{v2}, we obtain
\begin{align}
    q\sim 1+\frac{2 \ln (N P_{\text{fix}}(qs))}{\ln \frac{s}{\mu_{\rm B}}}, \quad R \sim \frac{2 s^2 \ln (N P_{\text{fix}}(qs) )}{(\ln \frac{s}{\mu_{\rm B}})^2}.\label{app:vel}
\end{align}
By substituting $P_{\text{fix}}\sim {(q s)^{\frac{1}{\al-1}}}$ into Equation~\ref{app:vel}, we obtain
\begin{align}
  q \sim  1+ \frac{2 \ln( N s^{\frac{1}{\al-1}})}{\ln \frac{s}{\mu_{\rm B}}}, \ R \sim 
    \frac{2 s^2 \ln (N s^{\frac{1}{\al-1}} )}{(\ln \frac{s}{\mu_{\rm B}})^2},\label{app:vel2}
\end{align}
where we used $\ln N q^{\frac{1}{\al-1}} \approx \ln N$. 
In the limit $\al\rightarrow 2$, the above results reproduce those  in \cite{Desai2007-sm}.

The case of $\al=1$ can be discussed in a similar way. 
 Suppose that the population is monoclonal.  The fixation probability  of a mutant  is  given by $P_{\text{fix}} \sim N^{-1+s}$ (see  Equation~\ref{Pfix_a=1}), which implies that the establishment size  is roughly given by $N_{est} \sim  N^{1-s}$.  While the timescale of establishment of a mutant is given by $(\mu_{\rm B} N P_{\text{fix}})^{-1} = (\mu_{\rm B} N^s)^{-1}$, the timescale of fixation is given by $t_{\text{fix}}\sim  \frac{1}{s}\log \frac{N}{N_{est}}\sim \log N$.   Thus,  the successive selection sweeps occur if  $(\mu_{\rm B} N^s)^{-1} \gg \log N$, or equivalently, 
\begin{align}
\mu_{\rm B} N^s \log N \ll  1 \quad ({\rm successive \ selective\  sweeps}). \label{ss_a1} 
\end{align}
By substituting $P_{\text{fix}} \sim N^{-1+s}$ into Equation~\ref{app:vel}, the rate of adaptation in the clonal-interference regime is given  by 
\begin{align}
R\sim \frac{2 s^3 \ln N }{(\ln \frac{s}{\mu_{\rm B}})^2}. \label{R1}
\end{align}
In the successive-sweeps regime, the adaptation rate is given by
\begin{align}
R =s \mu_{\rm B} N \times P_{\text{fix}}(s) \sim s \mu_{\rm B} N^s.
\end{align}

Note that  clonal interference  becomes unlikely to occur as the offspring  distribution becomes broader.  For example, when $\alpha=1$,  the population size needs to be $N \gg 10^{41}$ for $\mu_{\rm B}=10^{-4}, s=0.05$ to satisfy  $\mu_{\rm B} N^s \log N \gg  1 $.

\begin{figure*}[t]
 \centering
  \includegraphics[width=10cm]{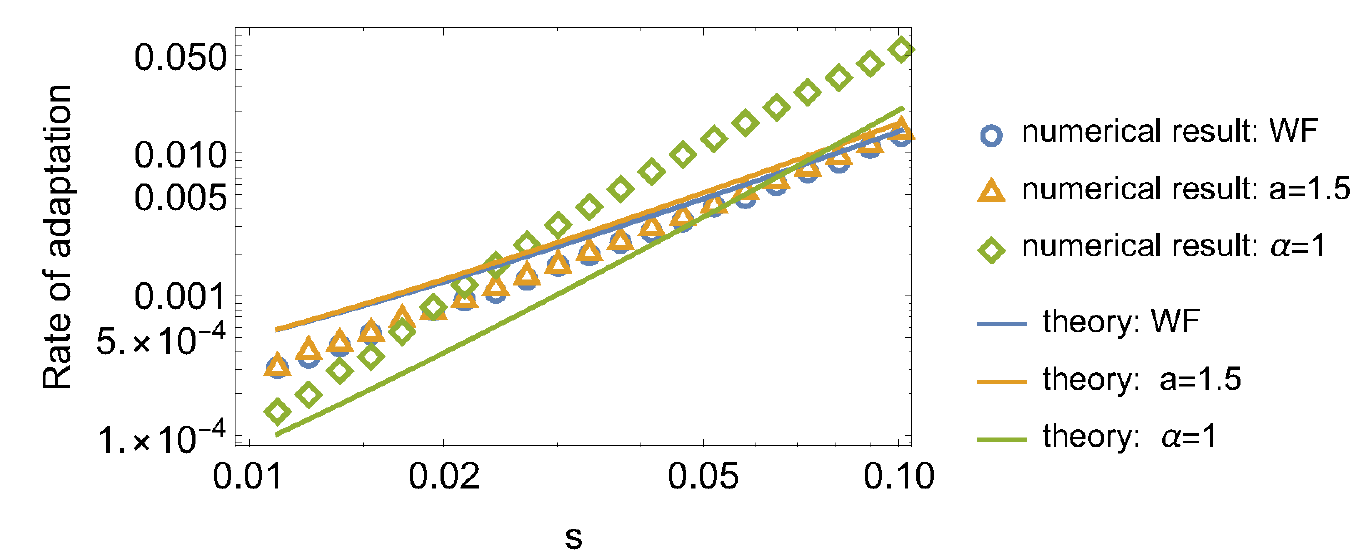}
   \caption{
   The open markers show the numerical results of $R$ as a function of $s$, while the curves show the theoretical predictions, based on the heuristic argument. The m rate of beneficial mutations  is $\mu=10^{-4}$. The population size is   $N=10^{100}$ for  $\alpha=1$,$N=10^{10} $ for $\alpha=1.5$, and $N=10^8$ for the Wright-Fisher model. 
   }
  \label{fig:rate_offsp} 
 \end{figure*}

{Figure~\ref{fig:rate_offsp} shows the numerical results of  the adaptation rate $R$ versus the selection coefficient $s$. The parameters used in the simulation are in the regime of clonal interference. 
When $1<\alpha$, $R$ is approximately proportional to $s^2$, while, when $\alpha=1$, $R$ is approximately proportional to $s^3$, which are consistent with Equations~\ref{app:vel} and \ref{R1}. However, when $\alpha=1$,  the quantitative agreement between the numerical result and the theoretical prediction is not good, and a further  investigation is needed to  validate Equation~\ref{R1}.  }

\section{Stationary distributions of traveling wave model in the presence of natural selection}
\begin{figure*}[ht]
 \centering
  \includegraphics[width=14cm]{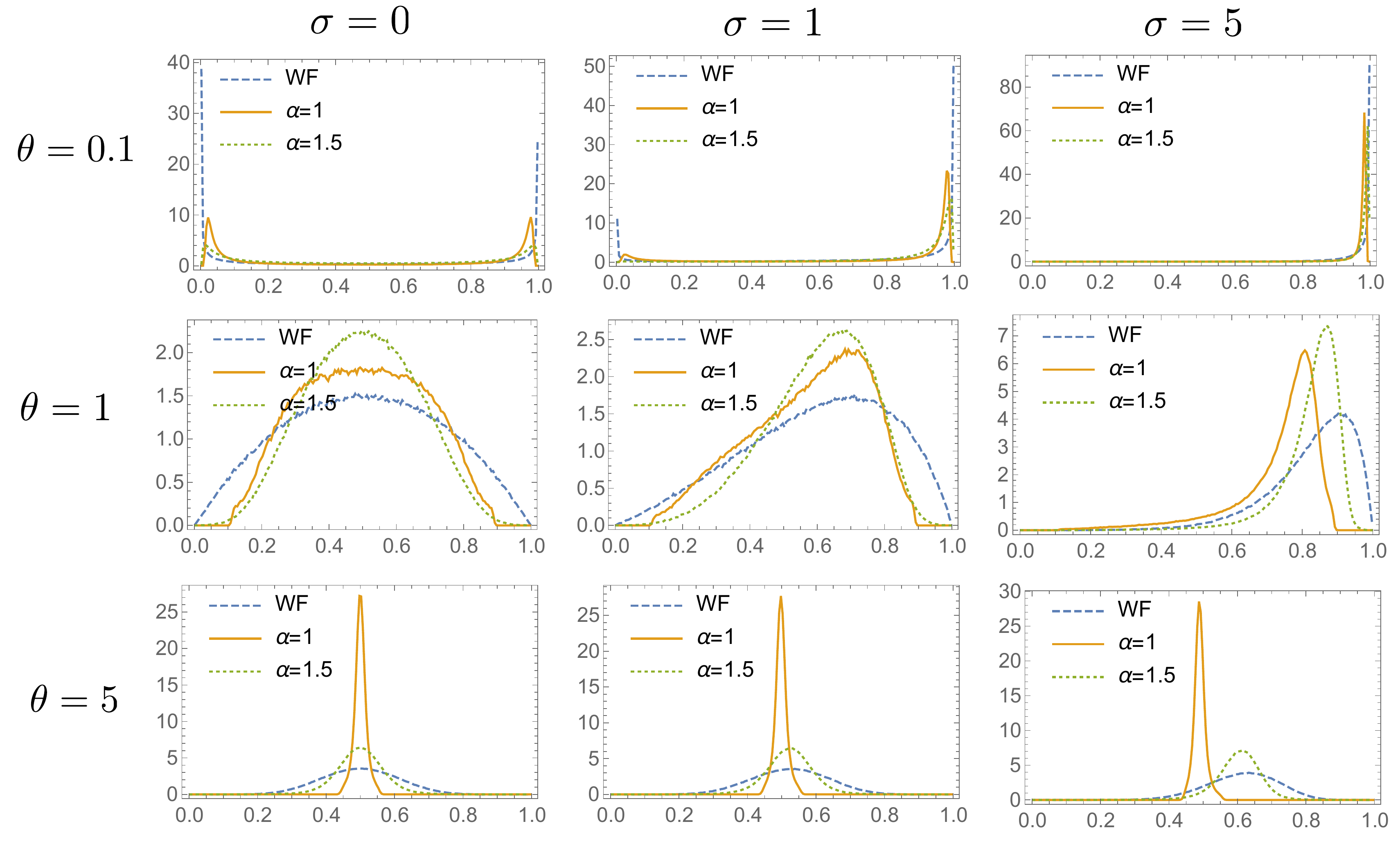}
   \caption{
   The stationary distributions of the mutant frequency for $\theta=0.1,1,5$. $\sigma=0,1,5$. $\sigma$ is the selection coefficient in the time-continuous description, $\sigma = s T_c$. 
   }
  \label{fig:stst_table} 
 \end{figure*}

In Figure~\ref{fig:wave_stst} of the main text, the mutant allele  is assumed be neutral. Here, we provide the results in the case where mutants have a fitness advantage $\sigma$ (Figure~\ref{fig:stst_table}). As in the main text, symmetrically reversible mutations  are assumed. 

\section{Numerical simulations}
Simulations are implemented in C++ with the GNU scientific library's random number generators. Results obtained from the simulations are analyzed by Mathematica. The codes are  freely available upon request.
 
\subsection{Numerical synthesis of Pareto random variables and $\al$-stable distribution}
In order to generate the mutant frequency of the gamete pool, we need to compute the sums of random Pareto variables, 
\begin{align}
M = \sum_{i=1}^{N x}u_i,\quad W  = \sum_{i=1}^{N (1-x)}v_i,\label{MW2}
\end{align}
where $u_i,v_i$ are drawn from the Pareto distribution $P_U(u)=\alpha /u^{\alpha+1}\, (u\geq 1)$.  One simple way to synthesize $u_i,v_i$ is to sample a number $r$ from the uniform distribution on $(0,1)$ and compute $r^{-\frac{1}{\al}}$. 

To generate the sums $M, N$ efficiently for large $N$ (e.g. $N  \sim 10^{6}$), we can use the generalized central limit theorem when $xN$ and $(1-x)N$ are large. In simulations,  when  $xN<100$, $M$ is generated  directly by  synthesizing $x N$ random variables $\{u_i\}$, while, when $xN\geq 100$,  $M$   is generated by sampling a random number $\zeta$  from the $\al$-stable distribution and then determining  $M=\sum_i u_i$ from Equation~\ref{zeta}. $W$ is  generated in a similar way.

After generating $M$ and $W$, the population is updated by the binomial sampling with the success probability $p=\frac{M}{M+W}$ (although  this sampling process can be omitted when $\alpha\geq 2$ since the fluctuations associated with the binomial sampling  is negligible compared to the fluctuations associated with $M$ and $N$).  Natural selection and mutations are implemented by modifying the success probability $p=\frac{M}{M+W}$ as 
\begin{align}
    \frac{p(1+s)}{p(1+s) + (1-p)}(1-\mu_{M\rightarrow W}) + \frac{(1-p)}{p(1+s) + (1-p)}\mu_{W\rightarrow M},
\end{align}
where $\mu_{W\rightarrow M}$ is the mutation rate from the wild-type to the mutant allele, and $\mu_{M\rightarrow W}$ is the mutation rate in the reverse direction.

\subsection{Site frequency spectrum}
Since the SFS is proportional to the mean sojourn time,  the SFS can be  computed numerically by generating trajectories staring with  $x_0 = \frac{1}{N}$ until fixation or extinction and measuring  how many times a trajectory visits a given frequency interval on average.

\subsection{Numerical simulation of the model of range expansion in the main text}
We first review the numerical implementation of the range expansion model with two neutral alleles without mutations \citep{Birzu2018-wq}. 
The per capita growth rate $r(n)$ with an Allee effect is given by 
\begin{align}
    r(n) = r_0 (1-\frac{n}{K})(1+B \frac{n}{K}),
\end{align}
where   $n= n_1 + n_2$ is the sum of the two population densities, and $B$ is the strength of cooperativity. In each deme, there are three types;  allele 1, allele 2, and ``empty''. At each time step, the configuration of deme $x$  is updated by  the trinomial sampling process with 
\begin{align}
    p_i =\frac{\tilde n_i  }{K(1-r(\tilde n) \tau)}\ {\rm for } \ i=1,2 \ \  {\rm and }\  p_{\rm empty} = 1-p_1-p_2, \label{p_i}
\end{align}
where $\tilde n_i$ is the population density  after migration,
\begin{align}
    \tilde n_i(t,x) = \frac{m}{2} n_i (t,x-a) + (1-m) n_i(t,x) +\frac{m}{2} n_i (t,x+a), 
\end{align}
and $\tilde n$  in the denominator of Equation~\ref{p_i} is the sum of these densities, $\tilde n = \tilde n_1 + \tilde n_2$, and $a$ denotes the width of a deme.
The expectation value of the total density $n$ after one time step is given by
\begin{align}
K \sum_{i=1,2} p_i  = \frac{\tilde n}{1-r(\tilde n)\tau} \approx \tilde n (1+r(\tilde n)\tau),\label{mean_n}
\end{align}
which explains the denominator of Equation~\ref{p_i}.
In the simulation, $a=1$ and $\tau=1$ are used. 

As in the standard Wright-Fisher model, a mutation process can be introduced  by using the success probabilities $\mathbf{p'}=(p'_1 , p'_2)^T$ given by 
\begin{align}
\mathbf{p}' = U \mathbf{p},
\end{align}
where   $\mathbf{p}=(p_1 , p_2)^T$ and  $U$ is a matrix representing  mutational transitions.
In the case of symmetrical mutations in the main text, $U$ is given by
\begin{align}
    U = \begin{pmatrix}
    1- \mu & \mu\\
    \mu & 1-\mu
    \end{pmatrix},
\end{align}

This  model serves as a microscopic description of our  (non-spatial) macroscopic model of the population with a broad offspring  distribution $p(U=u)\sim \frac{1}{u^{\alpha+1}}$. We can  argue the relation between the parameters in the two models by comparing the coalescent timescales. As established in \cite{Birzu2018-wq}, for a semi-pushed wave ($2<B<4$), the coalescent timescale  is given by 
\begin{align}
T_c^{\rm micro}\sim N^{2 \frac{\sqrt{1-\gamma(B)^2}}{1-\sqrt{1-\gamma(B)^2}}}.
\end{align}
where $\gamma(B)=\frac{v_F}{v}=2(\sqrt{\frac{B}{2}}+\sqrt{\frac{2}{B}})^{-1}$ is the ratio of  the Fisher velocity $v_F=2\sqrt{Dr_0}$ to the wave velocity $v=\sqrt{ r_0 D}(\sqrt{\frac{B}{2}}+\sqrt{\frac{2}{B}}) $. On the other hand, the coalescent timescale $T_c^{\rm macro}$ in the macroscopic description for $1<\alpha<2$ is proportional to $N^{\alpha-1}$ (see Equation~\ref{wN_mag}). By comparing the exponents, a semi-pushed wave with $B$ corresponds to the macroscopic model with 
\footnote{ Note that the definition of the parameter $\alpha_H$ in \cite{Birzu2018-wq} is different from our definition of $\alpha$. 
For $1<\alpha<2$, which corresponds to the semi-pushed wave region $-1<\alpha_H<0$,  the two definitions are related by $-\alpha_H=\alpha-1$.
 }
\begin{align}
\alpha = 2 \frac{\sqrt{1-\gamma(B)^2}}{1-\sqrt{1-\gamma(B)^2}}+1.	
\end{align}
For example, $B=3$ corresponds to $\alpha=1.5$. 
In addition, the mutation rate $\mu_{\rm micro}$ per generation in the microscopic model and the mutation rate $\mu_{\rm macro}$ per generation in the macroscopic model should be related by $\mu_{\rm micro} \times T_c^{\rm micro} \sim  \mu_{\rm macro} \times T_c^{\rm macro}$.

In the three panels (Left. Center, Right) in Figure~\ref{fig:wave_stst}B of the main text, The following parameters are used. 
\begin{itemize}
\item 	
Left: $B=1,\ \mu= (5\times 10^{-4},\ 5\times10^{-5}), \ K=28000$  for the microscopic model model, and $\alpha=1,\ \theta =(1.5,\ 0.15)$ for the macroscopic model. 
\item Center:  $B=3,\ \mu=(2\times10^{-4},\ 2\times10^{-5}),\ K=35000$  for the microscopic model, and $\alpha=1.5,\ \theta=(1.6,\ 0.16)$ for the macroscopic model. 
 \item Right: 
$B=8,\ \mu=(1 \times 10^{-5},\ 1\times 10^{-6}),\ K=57000$ for the microscopic model, and the Wright-Fisher model, $\theta =(2.4,\ 0.24)$ for the macroscopic model. 
\end{itemize}
In all of the three cases, the growth rate $\frac{r_0}{\tau}=0.01$ and the migration probability $m=\frac{2D\tau }{a^2}=0.125$ are used in the microscopic model, and the population size $N=10^5$ is used in the macroscopic model.
Note that, to compare the microscopic model with the macroscopic model, the value of the carrying capacity $K$ for each case is chosen such that the size of the front population $\frac{K}{k}$, where $k$ is the spatial decay rate of the population density\footnote{$k=\sqrt{\frac{r_0}{D}}$ for $0<B<2$, and  $k=\sqrt{\frac{r_0 B}{2D}}$ for $B\geq 2$ \citep{Birzu2018-wq}. }, approximately agrees with the population size $N=10^5$ in the macroscopic model.

\section{Areas swept by trajectories}

  \subsection{A scaling argument on  area distributions  }

\begin{figure}[t]
\centering
          \includegraphics[width=8.5cm]{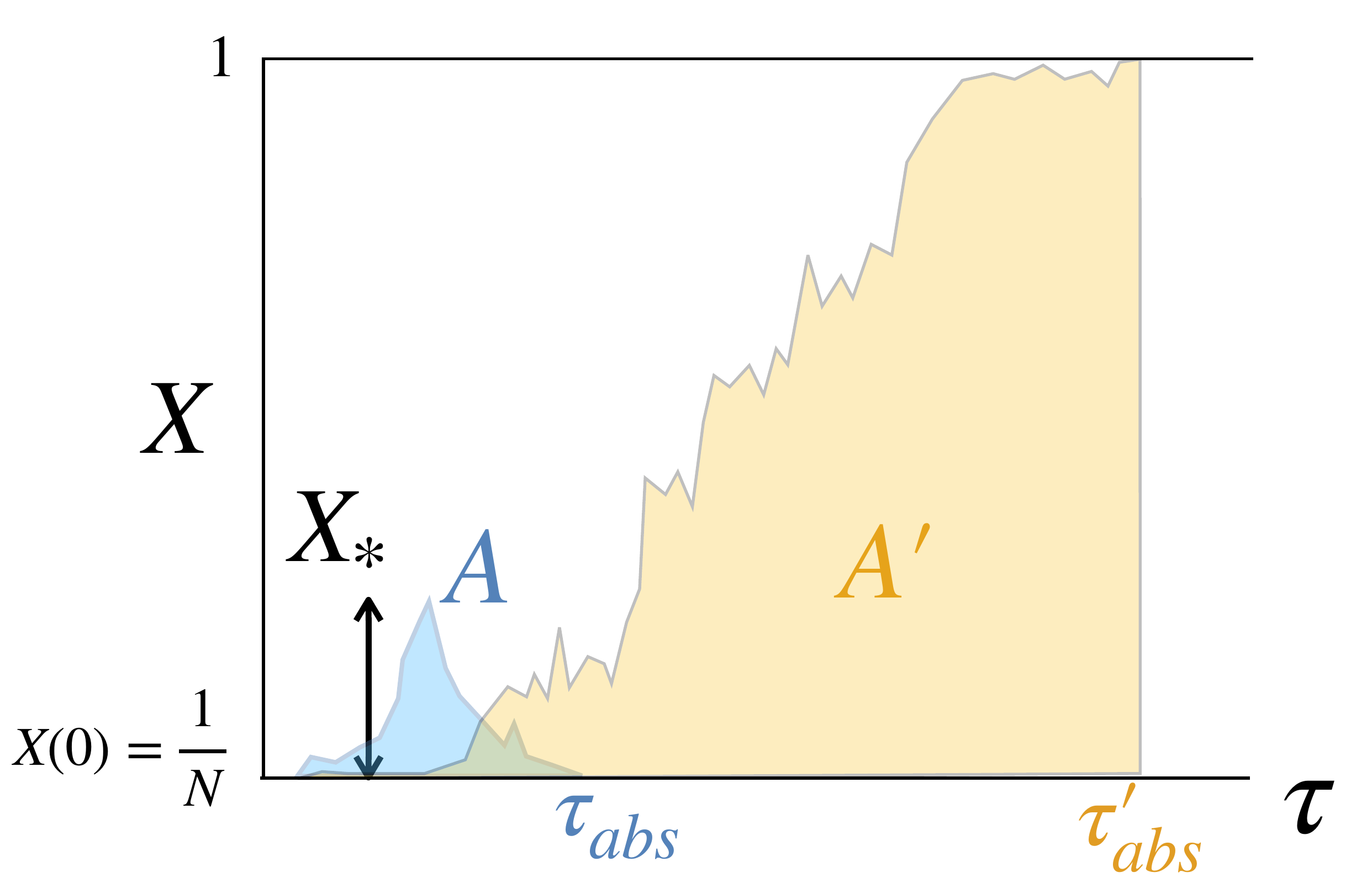}
    \caption{An area $A$ swept by a trajectory that eventually goes extinct and an area $A'$ swept by a trajectories that eventually gets fixed  are illustrated. $\tau_{\rm abs}$ and $\tau'_{\rm abs}$ are the extinction time and the fixation time, respectively.}
   \label{fig:area_sketch}
\end{figure}

Consider  frequency trajectories that depart from a single mutant $x_0=\frac{1}{N}$ and are eventually absorbed either at $x=0$ or at $x=1$. 
For each of such trajectories, we can define the area in $x-\tau$-space swept by the trajectory (see  Figure~\ref{fig:area_sketch}),
\begin{align}
 A = \int_0^{\tau_{\rm abs}} \,  x(\tau)  d\tau,\label{area}
\end{align}
where $\tau_{\rm abs}$ is the absorption time of the trajectory. While this quantity is defined for a population without spatial structure, we expect that it has a natural interpretation in a model of range expansion as a spatial integration over the mutant frequency (i.e., the abundance of the mutant type), since $\tau$ in Equation~\ref{area} is related with the spatial position of the traveling wave in the comoving frame.   

\begin{figure}[t]
\centering
    \begin{tabular}{c}
     \begin{minipage}{0.5\hsize}
        \begin{center}
          \includegraphics[width=5cm]{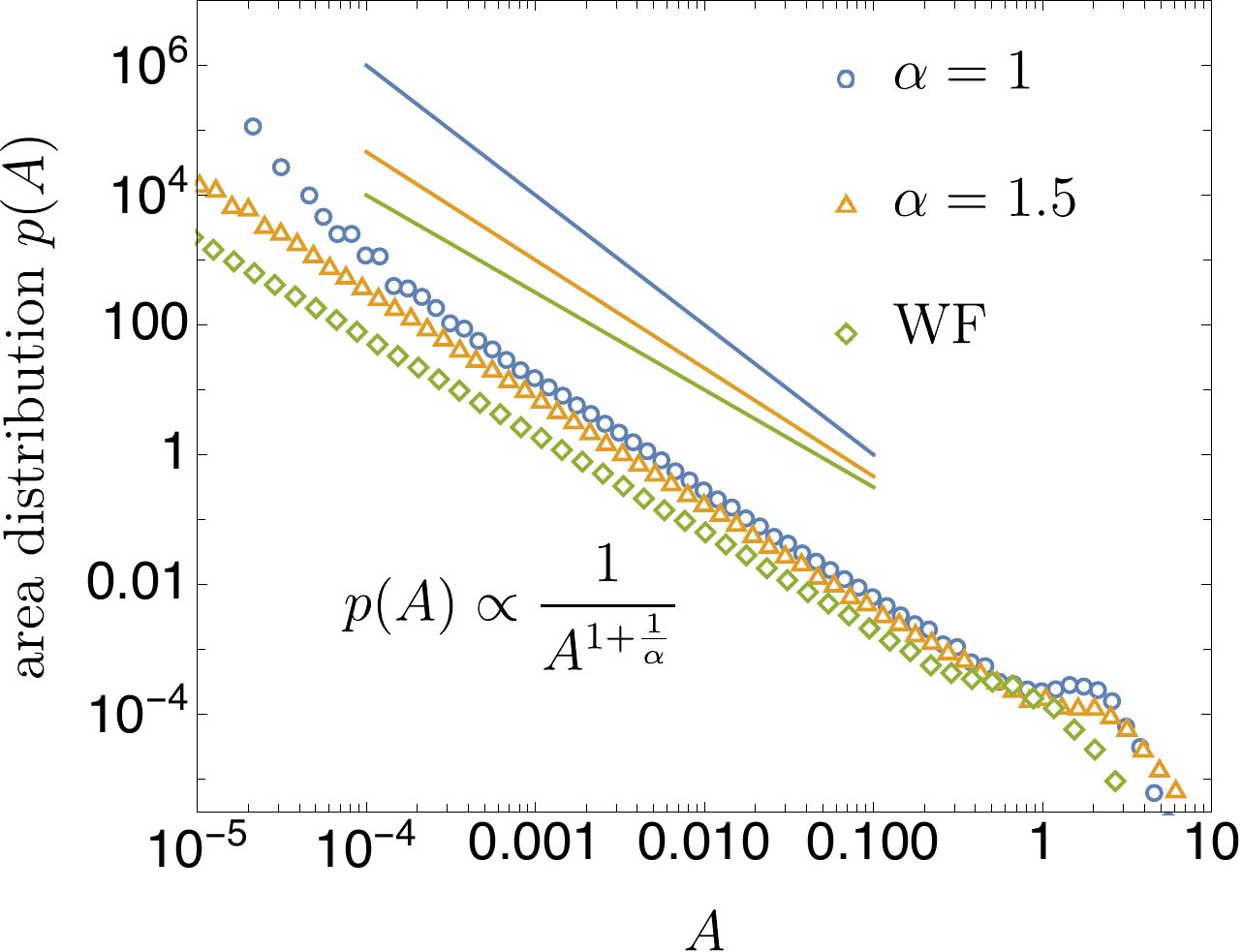}
        \end{center}
      \end{minipage}
      \begin{minipage}{0.5\hsize}
        \begin{center}
          \includegraphics[clip, width=5cm]{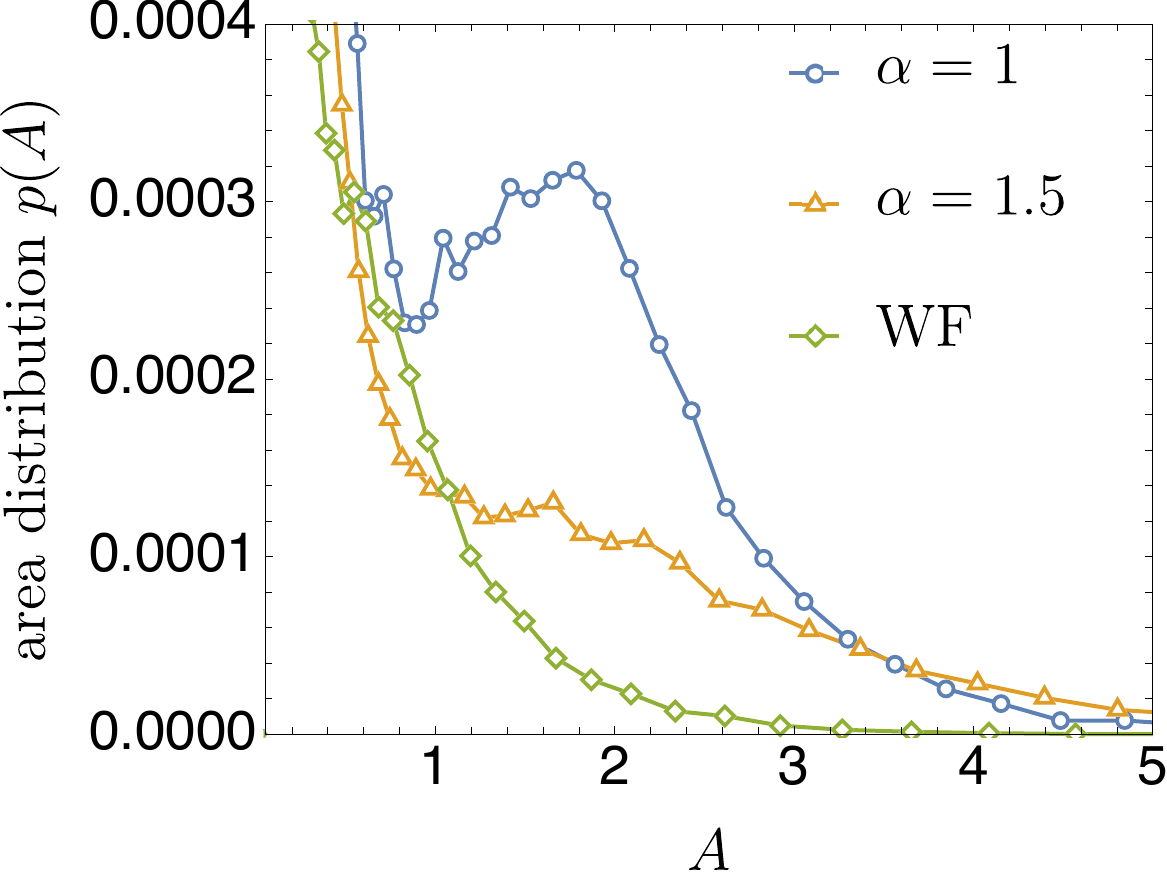}
        \end{center}
 \end{minipage}
    \end{tabular}
    \caption{ Left: The area distribution $p(A)$ for $\alpha=1,\, 1.5$ and the Wright-Fisher model. The straight lines show the scaling-argument predictions, $p(A) \propto  \frac{1}{A^{1+\frac{1}{\alpha}}}$.  $N=10^6$. Right: The tail of $p(A)$ in the large-$A$ region. }
   \label{fig:area}
\end{figure}

Here, we examine how the area $A$ defined in Equation~\ref{area} depends on the exponent $\alpha$ of the offspring  distribution. 
The left panel of Figure~\ref{fig:area} shows the numerical results of the area distribution $p(A)$  for $\alpha=1,\ 1.5$, and the Wright-Fisher model (corresponding to $\alpha\geq 2$). In a wide range of $A$, areas are distributed according to  $p(A)\sim \frac{1}{N A^{1+\frac{1}{\alpha}}}$. 

Focusing on  small areas, which correspond  to extinct trajectories,  this power-law behavior can be rationalized again from a scaling argument:  First, by using Equation~\ref{tauext0}, a trajectory whose maximum frequency is $ x_* \ll 1$  sweeps  an area roughly given by  $A\sim x_* \times \tau_{\text{ext}} \sim x_*^{\al}$ 
(see  Figure~\ref{fig:area_sketch}), i.e., $x_*\sim A^{\frac{1}{\alpha}}$. 
Second, from the neutrality, the cumulative probability ${\rm Pr}(X_*>x_*)$ that a single mutant achieves a frequency larger than $x_*$ before absorption is  estimated as  ${\rm Pr}(X_*>x_*)\sim \frac{1}{N x_*}$. Hence, the density $p(x_*)$  is  given by  $p(x_*)=-\frac{d}{dx_*}{\rm Pr}(X_*>x_*)\sim \frac{1}{N x_*^2}$. Combining these two results, we can estimate the area distribution $p(A)$ as
\begin{align}
    p(A)\sim p(x_*)\frac{dx_*}{d A}\biggr|_{x^* = A^{\frac{1}{\al}}} \sim \frac{1}{N x_*^2} (x_*)^{-\alpha+1} \biggr|_{x^* = A^{\frac{1}{\al}}}\sim \frac{1}{N } A^{-1-\frac{1}{\alpha}}.
\end{align} 
When $\alpha\rightarrow 2-0$ (Wright-Fisher limit), the distribution becomes  $p(A)\sim \frac{1}{N}A^{-\frac{3}{2}}$, which can be analytically confirmed by solving a backward diffusion equation of the Wright-Fisher diffusion (see Appendix J-2).

The numerical results indicate that, when $1\leq \alpha<2$, there is an uptick in the area distribution $p(A)$, which comes from fixed trajectories  (see the case of $\alpha=1$ in the right panel of Figure~ \ref{fig:area}). The uptick becomes less pronounced as $\alpha$ increases. For the Wright-Fisher model, we can analytically prove that $p(A)$ monotonically decreases with $A$. 


%

\subsection{Area distribution in the Wright-Fisher model}
Here, we derive an analytical result of Equation~\ref{area} for the Wright-Fisher diffusion process. 

Consider a Langevin  equation 
\begin{align}
    \frac{dX}{d\tau} =  v(X)+ \xi(\tau),
\end{align}
with $\langle \xi(\tau)\xi(\tau')\rangle= 2 D(x)\delta(\tau-\tau')$. 
Assume  the initial value $X(\tau=0)=x_0 \in (0,1)$ and the absorbing boundaries at $X=0,1$. 
For a given trajectory departing from $x_0$ and ending at either one of the boundaries, we consider the ``area'' defined by
\begin{align}
   A = \int_0^{\tau_{\rm abs}} X(\tau) d\tau\ .\label{app:area_x}
\end{align}
where $\tau_{\rm abs}$ is the absorption time. 

The area distribution  $\Pi(A;x_0)$ for a given initial condition $X(0)=x_0$ obeys a backward equation.
To show this, we discretize the dynamics;
\begin{align}
    \Delta X = v h + W
\end{align}
where $h$ denotes a short time interval and  $\langle W_i  W_j \rangle= 2 D h \delta_{i,j}$.
The transition density is given by
\begin{align}
    T(x_0+\Delta x|x_0) = \frac{1}{\sqrt{\pi (2 D(x_0) h)}} \exp(-\frac{(\Delta x - v(x_0)h)^2}{2 (2 D(x_0) h)}).
\end{align}
Note that
\begin{align}
    \langle \Delta x\rangle_{x_0} &= v(x_0) h,\non\\ 
      \langle (\Delta x)^2\rangle_{x_0} &= v(x_0) h + 2 D(x_0) h.\label{app:mean}
\end{align}

By separating a trajectory into the initial step and the remaining part, we have
\begin{align}
    \Pi(A;x_0)= \int d(\Delta x)\ T(x_0+\Delta x|x_0)\Pi(A-x_0 h;x_0+\Delta x) +  o(h),\label{app:Pi}
\end{align} 
 By Taylor-expanding $ \Pi(A-x_0 h;x_0+\Delta x)$, we have
\begin{align}
    \Pi(A-x_0 h;x_0+\Delta x)&=   \Pi(A;x_0)-\frac{\p \Pi}{\p A} x_0 h 
    +\frac{\p \Pi}{\p x_0}\Delta x\non\\
   & +\frac{1}{2}\frac{\p^2 \Pi}{\p A^2} x_0 h^2
   -\frac{\partial^2 \Pi}{\partial A\partial x_0} x_0 h \Delta x
   +\frac{1}{2}\frac{\p \Pi}{\p x_0}\Delta x^2+ \cdots \non\\
   & =  \Pi(A;x_0)-\frac{\p \Pi}{\p A} x_0 h+\frac{\p \Pi}{\p x_0}\Delta x +\frac{1}{2}\frac{\p \Pi}{\p x_0}\Delta x^2 +o(h),
\end{align}
Therefore, Equation~\ref{app:Pi} becomes 
\begin{align}
    \Pi(A;x_0)= \Pi(A;x_0)-\frac{\p \Pi}{\p A} x_0 h 
    +\frac{\p \Pi}{\p x_0}\langle \Delta x\rangle_{x_0}
    +\frac{1}{2}\frac{\p^2 \Pi}{\p x_0^2}\langle \Delta x^2\rangle_{x_0}  +  o(h).
\end{align}
By using Equation~\ref{app:mean}, we obtain 
\begin{align}
x_0 \frac{\p \Pi}{\p A} = v(x_0)\frac{\p \Pi}{\p x_0}
+ D(x_0)\frac{\p^2 \Pi}{\p x_0^2}. \label{app:bw}
\end{align}
Although, in the following, we consider the area defined by Equation~\ref{app:area_x},  it can be shown that, for the following integral,
\begin{align}
  \tilde A = \int_0^{T^*}  dt\ f(X), 
\end{align}
 the distribution $\Pi(\tilde A; x_0)$ satisfies 
\begin{align}
f(x_0) \frac{\p \Pi}{\p  \tilde A} = v(x_0)\frac{\p \Pi}{\p x_0}
+ D(x_0)\frac{\p^2 \Pi}{\p x_0^2}.
\end{align}

In the neutral Wright-Fisher model, $v(x_0)=0$ and $D(x_0)=x_0(1-x_0)$. The backward equation in Equation~\ref{app:bw} is  given by 
\begin{align}
 \frac{\p \Pi}{\p A} = (1-x_0)\frac{\p^2 \Pi}{\p x_0^2}. \label{PAwf}
\end{align}
From this equation, it follows that $\Pi(A|x_0)$  monotonically decreases with $A_0$ because the spectrum of the operator $\partial_{x_0}^2\sim (i k)^2$ is non-positive.

We can determine the area distribution $p(A)$ analytically at least for small $A$. 
We are interested in the invasion by a single mutant,   $x_0=\frac{1}{N}\ll 1$. Furthermore, for the purpose of determining the behavior for small areas, we expect that we can ignore the presence of the high-frequency boundary  $x=1$  and solve the problem on the semi-infinite line $x_0\in (0,\infty)$. Therefore, we consider the following problem:
\begin{align}
&  \frac{\p \Pi}{\p A} = \frac{\p^2 \Pi}{\p x_0^2},\non\\
& \Pi(A;x_0=0)= g(A)\non\\
& \Pi(A=0;x_0)=0\ {\ \rm for\ } x_0>0\non\\
&\underset{x_0 \rightarrow \infty}{\rm lim}  \Pi(A,x_0)=0
\end{align}
In our case, $g(A)=\delta(A)$, because the trajectory starting from $x_0=0$ has $A=0$.

For a function $ f(A)$ of $A$, we write the Laplace transformation as
\begin{align}
    \hat f(s) = \mathcal{L} [f(A)]
    = \int_0^\infty ds f(t)e^{-sA}. 
\end{align}
By take the Laplace transform with respect to $A$, we have
\begin{align}
    s \hat \Pi(s;x_0) = \frac{\p^2 \hat \Pi(s,x_0) }{\p x_0^2}, \quad \hat \Pi(s,0) =\hat g(s).  
\end{align}
The solution is 
\begin{align}
    \hat \Pi(s;x) = e^{-x_0\sqrt{s}} \hat g(s). 
\end{align}
We take the inverse of the Laplace transformation, 
\begin{align}
     \Pi(A;x_0) = {\mathcal L}^{-1} ( e^{-x_0\sqrt{s}} \hat g(s)).
\end{align}
From the convolution theorem, this is given by the convolution of $ {\mathcal L}^{-1} ( e^{-x_0\sqrt{s}})
= \frac{x_0}{2\sqrt{\pi} A^{\frac{3}{2}}}  e^{-\frac{x_0^2}{4 A}}$ and $g(A)$;
\begin{align}
    \Pi(A;x_0) = \int_0^A dA'  \frac{x_0}{2\sqrt{\pi} A'^{\frac{3}{2}}}  e^{-\frac{x^2_0}{4 A'}} g(A-A'). 
\end{align}
When $g(A) =\delta(A)$, we have
\begin{align}
    \Pi(A;x_0) =   \frac{x_0}{2\sqrt{\pi} A^{\frac{3}{2}}}  e^{-\frac{x_0^2}{4 A}}. 
\end{align}
Especially, when $x_0=1/N$, we have
\begin{align}
    \Pi(A;x_0=\frac{1}{N}) &=   \frac{1}{2\sqrt{\pi}N A^{\frac{3}{2}}}  e^{-\frac{1}{4 A N^2}}\non\\
    &  \approx  \frac{1}{2\sqrt{\pi}N A^{\frac{3}{2}}}, \label{PiA}
\end{align}
where  we have used  $e^{-\frac{1}{4 A N^2}}\approx 1$ since   only areas larger than $ x_{0}\times d\tau \sim \frac{1}{N}\times \frac{1}{N} =\frac{1}{N^2}$ are  meaningful for a finite-size population.

\section{Forward-in-time behaviors of the Eldon-Wakeley model}
Here, we present simulation results of the median allele frequency and the median and mean square displacements in the Eldon-Wakeley model \citep{Eldon2006-ko} (see also \cite{Der2012-jk}). As shown below, unlike our model,  these quantities do no exhibit sustained power-law behaviors,  because of the existence of a characteristic size $\psi$ in the offspring distribution.

We consider the neutral Eldon-Wakeley model, where the following offspring distribution $P_U(u)$ is given by (see Equation (7) in \cite{Eldon2006-ko});
\begin{align}
P_U(u)=(1-N^{-\gamma} )\delta_{u,2} +  N^{-\gamma}\delta_{u,\psi  N},
\end{align}
where $\delta_{a,b}$ is the Kronecker delta. $\psi\in(0,1)$ and the parameters characterizing how large and frequent `sweepstakes'  are.

 The limiting process as $N\rightarrow \infty$ depends on $\gamma$ (see Equation~(9) in \cite{Der2012-jk}). 
 For $\gamma>2$, the process is the same as the Wright-Fisher diffusion, while, for $\gamma<2$, it is described by a jump process whose  backward-time generator $\mathcal L^\dagger $  is given by
 \begin{align}
    L^\dagger P(x,\tau) = x P(x+ \psi(1-x),\tau) - P(x,\tau) + (1-x) P(x-\psi x, \tau),\label{ope_EW}
 \end{align}
 where the continuous time $\tau$ is related with generations $t$ by   $\tau = t/N^\gamma$. The first term  of the generator represents a frequency-increasing jump  $x\rightarrow x + \psi(1-x)$  with rate $x$, while  the last one represents a frequency-decreasing jump  $x\rightarrow x - \psi x$  with rate $1-x$.
 
\begin{figure*}[t]
 \centering
  \includegraphics[width=14cm]{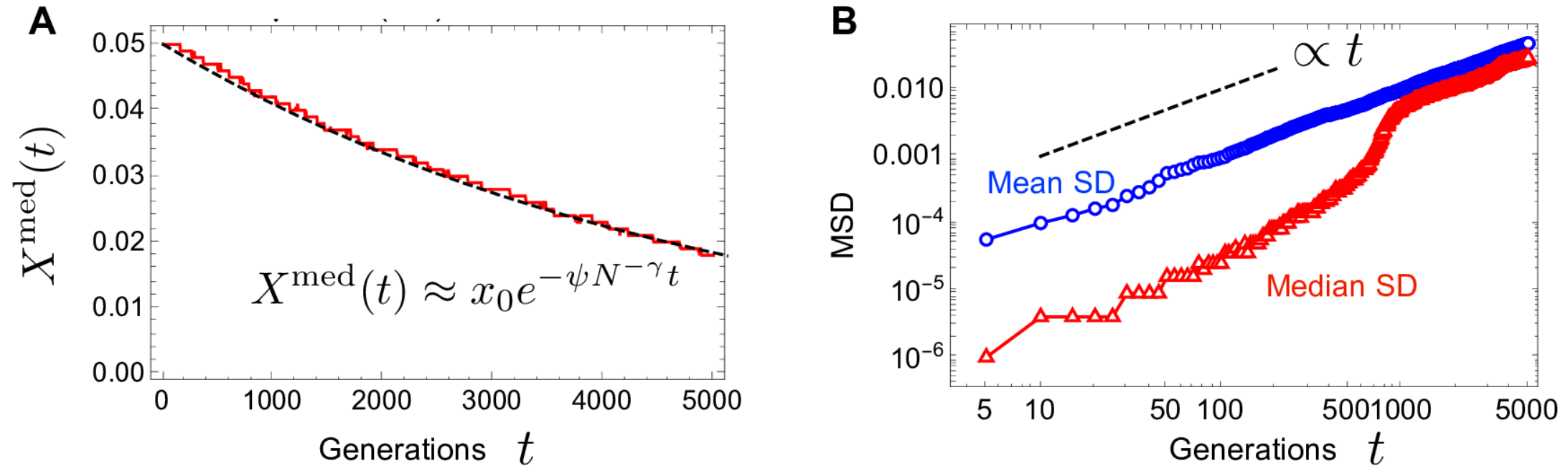}
   \caption{
   Simulation results of the Eldon-Wakeley model.
(A) The median frequency of the Eldon-Wakeley model (red solid) and $X^{\rm med}(t)=x_0 e^{-\psi N^{-\gamma }t}$ (black dashed). $N=10^3$, $\gamma =1$, $\psi =0.1 $, $x_0=0.05$.  (B) The mean and median square displacements (blue and red curves, receptively). The black dashed line  $\propto  1/t$ indicates  the expectation from the Wright-Fisher (or Moran) model.  $N=10^3$, $\gamma =1$, $\psi = 0.2$,  $x_0=0.5$.
   }
  \label{fig:EW} 
 \end{figure*}

Figure~\ref{fig:EW} shows numerical simulation results for the median of allele frequencies and the median/mean square displacements. 
The median frequency for a small initial frequency $x_0\ll 1$ is well described by $X^{\rm med} (t)=x_0 e^{-\psi N^{-\gamma} t}$ (Figure~\ref{fig:EW}A). This exponential decay can be  expected from the generator in Equation~\ref{ope_EW}; for $x \ll 1$,  frequency-increasing jumps (with rate $x$) are unlikely to occur, and an allele frequency typically decreases by  $- \psi x$  with rate $1-x \approx 1$.
Thus, the median frequency in the Eldon-Wakeley model does not exhibit a power-law behavior.

As for frequency fluctuations, while the mean SD exhibits  a normal diffusion as in  the Moran (or Wright-Fisher) model, i.e., ${\rm Mean\ SD} \propto t$, the median SD does not  exhibit a sustained power-law behavior (Figure~\ref{fig:EW}B); in a short- and long time scales, the median SD exhibits a normal diffusion (${\rm Median\ SD} \propto t$), but, for an intermediate timescale ($t\sim 500 - 1000$ generations in the figure), it increases more rapidly than expected from a normal diffusion.

\end{document}